\documentclass[%
 reprint,showpacs,preprintnumbers,
 amsmath,amssymb,
 aps,
prc,
]{revtex4-1}

\usepackage{graphicx}
\usepackage{dcolumn}
\usepackage{bm}
\begin{document}

\title{A Neutron Star is born}

\author{D\'ebora P. Menezes}
\affiliation{%
  Departamento de Fisica, CFM - Universidade Federal de Santa Catarina,\\
  C.P. 476, CEP 88.040-900, Florian\'opolis, SC, Brasil 
}%

\begin{abstract}
 A neutron star was first detected as a pulsar in 1967. It is one of
the most mysterious compact objects in the universe, with a radius of the
order of 10 km and masses that can reach two solar masses. In fact,
neutron stars are star remnants, a kind of stellar zombies (they die,
but do not disappear). In the last decades, astronomical observations
yielded various contraints for the neutron star masses and finally, in
2017, a gravitational wave was detected (GW170817).  Its source
was identified as the merger of two neutron stars coming from NGC
4993, a galaxy 140 million light years away from us. The very same
event was detected in $\gamma$-ray, x-ray, UV, IR, radio frequency and even in the
optical region of the electromagnetic spectrum, starting the new era
of multi-messenger astronomy. To understand and describe neutron stars, an
appropriate equation of state that satisfies bulk nuclear matter
properties is necessary. GW170817 detection contributed with extra
constraints to determine it. On the other hand, magnetars are the same
sort of compact objects, but bearing much stronger magnetic fields
that can reach up to 10$^{15}$   G on the surface as compared with the usual
10$^{12}$  G present in ordinary pulsars. While the description of
ordinary pulsars is not completely established, describing magnetars 
poses extra challenges.
In this paper, I give an overview on the history of neutron stars and
on the development of nuclear models and show how the description of
the tiny world of the nuclear physics can help the
understanding of the cosmos, especially of the neutron stars. 
\end{abstract}

\maketitle

\section{Introduction}
\label{Intro}

Two of the known existing interactions that determine all the conditions of
our Universe are of nuclear origin: the strong and the weak nuclear
forces. It is not possible to talk about neutron stars without
understanding them, and specially the strong nuclear interaction,
which is well described by the Quantum Chromodynamics (QCD). But
notice: a good description through a Lagrangian density does not mean
that the solutions are known for all possible systems subject to the
strong nuclear force.

\begin{figure}
\includegraphics[width=9.cm]{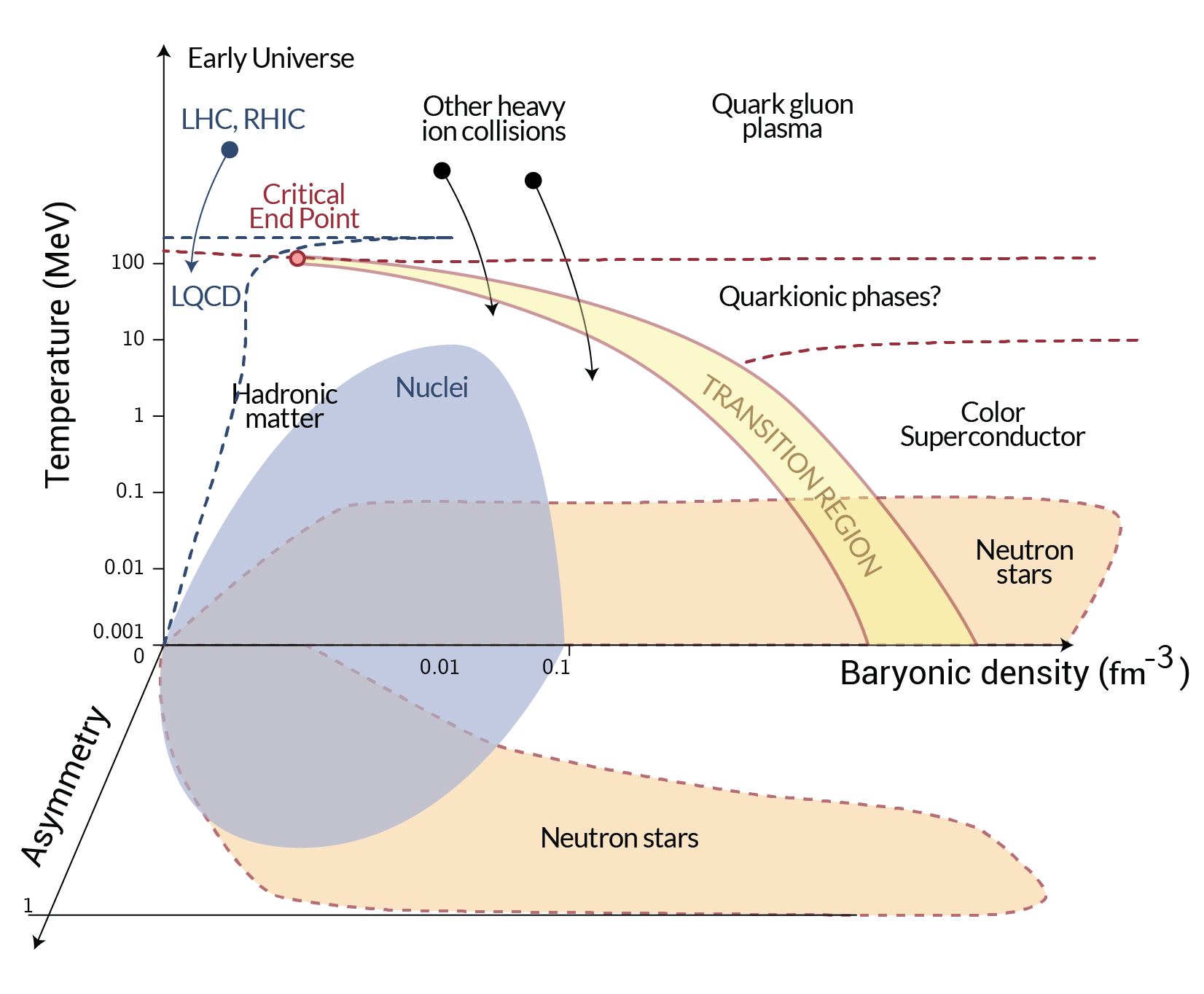} 
\caption{QCD phase diagram.  On the left of the transition region
  stands hadronic matter and on the right side, the quark gluon
  plasma. Quarkyonic phases represent a region where chiral
  symmetry has been restored but matter is still confined.  
Figure taken and adapted from \cite{annawatts}.}
\label{fig0}
\end{figure}

Based on the discovery of asymptotic freedom \cite{asym}, which
predicts that strongly interacting matter undergoes a phase transition
from hadrons to the  quark-gluon plasma (QGP) and on the possibility
that a QGP could be formed in heavy-ion collisions, the QCD phase
diagram has been slowly depicted. While asymptotic freedom is expected to
take place at both high temperatures, as in the early universe and
high densities, as in neutron star interiors, heavy-ion collisions can
be experimentally tested with different energies at still relatively
low densities, but generally quite high temperatures.
If one examines the QCD phase diagram, shown in Fig. \ref{fig0}, it
is possible to see that the nuclei occupy a small part of the diagram,
at low densities and low temperatures for different
asymmetries. One should notice the temperature log scale, chosen to
emphasise the region where nuclei exist.
Neutron stars, on the other hand, are compact objects
with a density that can reach 10 times the nuclear saturation density,
discussed later on along this paper. While
heavy ion collisions probe experimentally some regions of the
diagram, lattice QCD (LQCD) calculations explain only the low density
region close to zero baryonic chemical potential. 
Hence, we rely on effective models to advance our understanding and
they are the main subject of this paper. 

Since the beginning of the last century, many nuclear models have been
proposed. In  Section \ref{nuclear}, the first models are mentioned
and the notion of nuclear matter discussed. The formalisms that
followed, either non-relativistic Skyrme-type models \cite{skyrme} or 
relativistic ones that gave rise to the quantum hadrodynamics model
were based on some basic features described by the early models, the
liquid drop model \cite{liquiddrop} and the semi-empirical mass
formula \cite{semiempirical}.
Once the nuclear physics is established, the very idea of a neutron
star can be tackled. However, it is very important to have in mind the
model extrapolations that may be necessary when one moves
from the nuclei region shown in Fig. \ref{fig0} to the neutron star (NS)
region. A simple treatment of the relation between these two regions
and the construction of the QCD phase transition line can be seen in
\cite{Carline2}.

The exact constitution of these compact objects, also commonly
named pulsars due to their precise rotation period, is still unknown and all the
information we have depends on the confrontation of theory with
astrophysical observations. As the density increases towards their
center, it is believed that there is an outer crust, an inner crust,
an outer core and an inner core. The inner core constitution is
the most controversial: it can be matter composed of deconfined quarks 
or perhaps a mixed phase of  hadrons and quarks. I will try to comment
and describe every one of the possible layers inside a NS along this text.

NASA's Neutron Star Interior Composition Explorer
(NICER), an x-ray  telescope \cite{NICER} launched in
2017, has already sent some news \cite{NICER_news}: by monitoring the
x-ray emission of gas surrounding the heaviest known pulsar, PSR
J0740+6620 with a mass of $2.08 \pm 0.07$, it has measured
its size and it is larger than previously expected, a diameter of
around 25 to 27 km, with consequences on the possible composition of the NS core.

In this paper, I present a comprehensive review of the main nuclear physics properties that
should be satisfied by equations of states aimed to describe nuclear
matter, the consequences arising from the extrapolation necessary to describe objects
with much higher densities as neutron stars and how they can be tuned
according to observational constraints.
At the end, a short discussion on quark and hybrid stars is presented
and the existence of magnetars is rapidly outlined.
Not all important aspects related to neutron stars are treated in
  the present work, rotation being the most important one that is
  disregarded, but the interested reader can certainly use it as an
  initiation to the physics of these compact objects.

\section{Historical perspectives}
\label{history}

I divide this section, which concentrates all necessary information
for the development of the physics of neutron stars, in two parts. 
In the first one, I discuss the development of the nuclear physics
models based on known
experimental properties and introduce the very simple Fermi gas model,
whose calculation is later used in more realistic relativistic
models. The second part is devoted to the history of compact objects 
from the astrophysical point of view. 

\subsection{From the nuclear physics point of view}
\label{nuclear}

The history of nuclear physics modelling started with two very simple
models: the liquid drop model, introduced in 1929
\cite{liquiddrop} and the semi-empirical mass formula, proposed in
1935 by Bethe and Weizs\"acker \cite{semiempirical}. 

The liquid drop model idea came from the observation that the
nucleus has behavior and properties that resemble the ones of an
incompressible fluid, such as: 
a) the nucleus has low compressibility due to its internal almost
  constant density; b) it presents well defined surface;
c) the nucleus radius varies with the number of nucleons as
$R = R_0 A^{1/3}$, where $R_0 \simeq 1.2 \times 10^{-15}$ m and d) 
the nuclear force is isospin independent and saturates.

\begin{figure}
\begin{tabular}{cc} 
\includegraphics[width=6.cm]{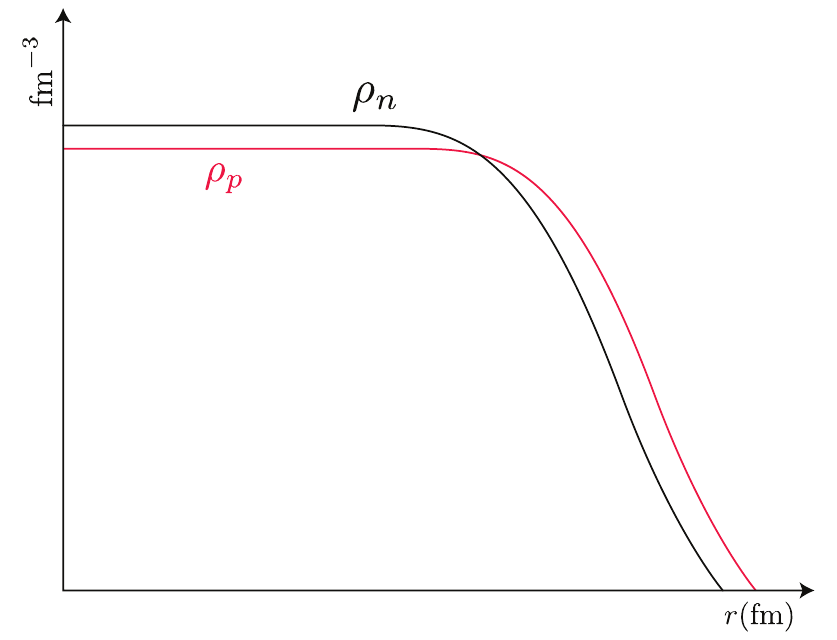}\\
\includegraphics[width=6.cm]{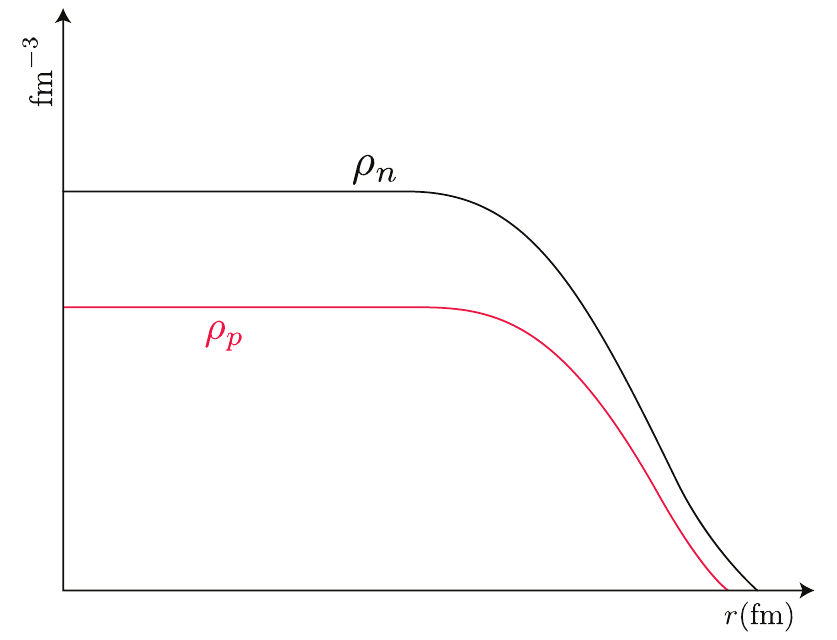} \\
\end{tabular}
\caption{Schematic representation of the nuclear densities with equal number of
  protons and neutrons (top) and a larger number of neutrons than
  protons (bottom). The proton and neutron densities depend on the
  number of nucleons so that heavier
  elements present larger densities. Typical theoretical 
densities for $^{208}$Pb are of the order
  of 0.09 fm$^{-3}$ for neutrons and  0.06 - 0.07 fm$^{-3}$ for
  protons \cite{skin2}.}
\label{fig1}
\end{figure}

Typical nuclear density profiles are shown in Fig. \ref{fig1}, from
where one can observe some of the features mentioned above, i.e.,
the density is almost constant up to a certain
point and then it drops rapidly close to the surface,
determining  the nucleus radius. The mean square radius is usually
defined as

\begin{equation}
R_{i}^{2}=\frac{\int d^{3}r\,r^2 \rho_{i}(\mathbf {r})}{\int d^{3}r\,
 \rho_{i}(\mathbf {r})}, \quad i=p,n 
\end{equation}
where $\rho_p$ is the number density of protons and $\rho_n$ the
number density of neutrons.

A nucleus with an equal number of
protons and neutrons has a slightly larger proton radius because they
repel each other due to the Coulomb interaction. A nucleus with more neutrons than
protons (as most of the stable ones) has a larger neutron radius than
its proton counterpart and the small difference between both radii is 
known as neutron skin thickness, given by \cite{skin,skin2, PREX1,PREX2}:

\begin{equation}
\theta=R_n-R_p. \label{skin}
\end{equation}

For the last two decades, a precise measurement of both charge and
neutron radii of the $^{208}$Pb nucleus has been tried at the parity radius
experiment (PREX) at the Jefferson National Accelerator Facility
\cite{Jefflab} using polarised electron scattering. The latest 
experimental results \cite{PREX2} point to $\theta=0.283 \pm 0.071$ fm and
to the interior baryon density $\rho_0=0.1480 \pm 0.0036 ({\rm exp}) \pm
0.0013 ({\rm theo})$ fm$^{-3}$. 
These quantities have shown to be important for the understanding of some
of the properties of the neutron star. I will go back to this discussion later on.

\begin{figure}
\begin{center}
\includegraphics[width=0.9\linewidth]{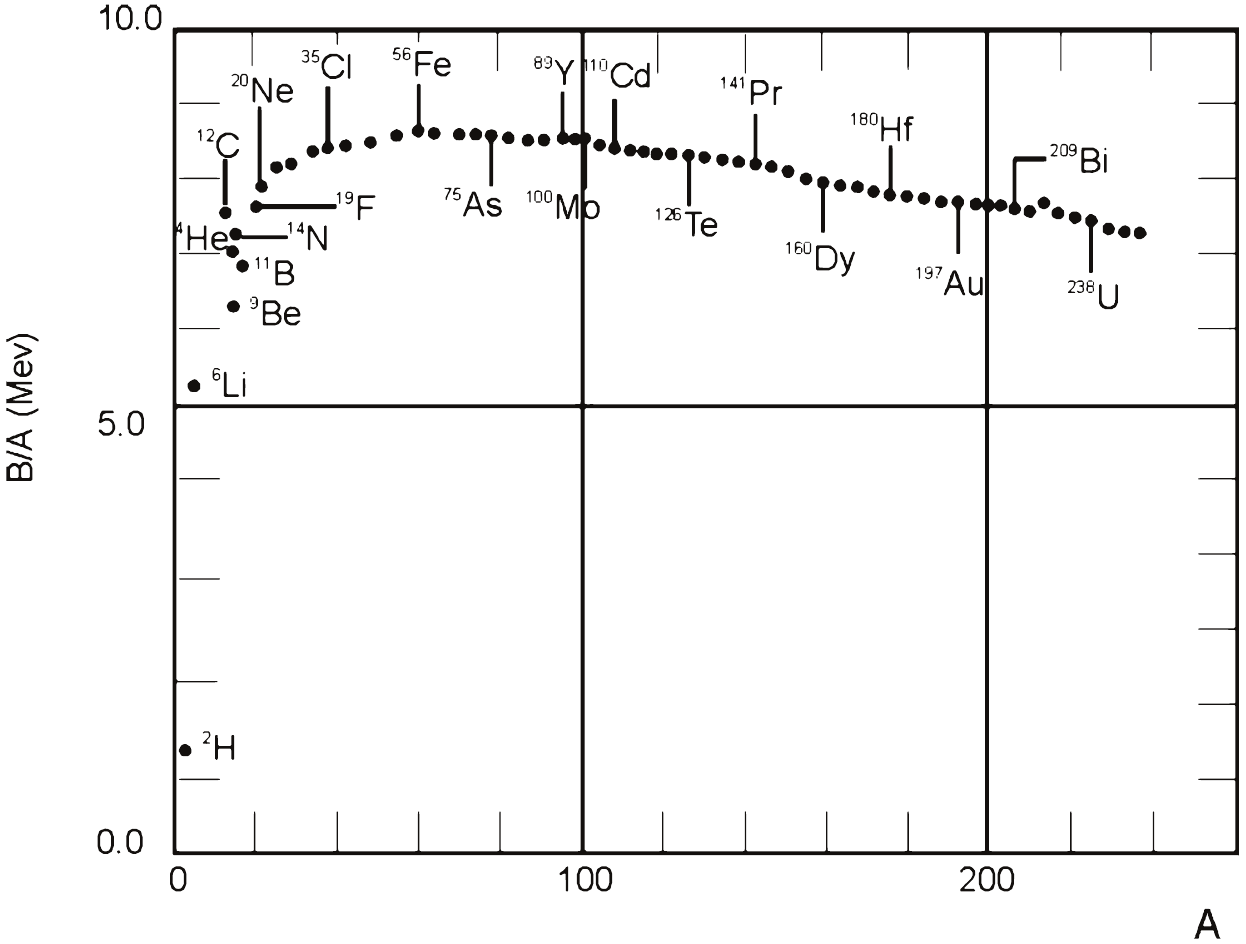}
\caption{\label{fig2} Binding energy per nucleon as a function of the
  number of nucleons.}
\end{center}
\end{figure}

The binding energy $B$ of a  nucleus $^A_ZX_N$ is given by the difference
between its mass and the mass of its constituents ($Z$ protons and $N$ neutrons):

$$
B=(Z m_p + N m_n - (m(^A_ZX)-Z m_e))c^2
$$
\begin{equation}
= (Z m(^1H) + N m_n - m(^A_ZX))c^2,
\label{ener}
\end{equation}
where  $m(^A_ZX)$ is the mass of the chemical element $^A_ZX$ 
and is given in atomic mass units.
The binding energy per nucleon $\frac{B}{A}$  is shown in
Fig. \ref{fig2}, from where it is seen that the curve is relatively
constant and of the order of  $8.5$ MeV  except for light nuclei.
The semi-empirical mass formula, which is a parameter dependent expression
was used to fit the experimental results successfully and it reads:

\begin{equation}
B(Z,A)=a_v A - a_s A^{\frac{2}{3}}- a_c e^2 \frac{Z(Z-1)}{A^\frac{1}{3}}
-a_i \frac{(N-Z)^2}{A} + \delta(A).
\label{efsem}
\end{equation}

In this equation, from left to right, the quantities refer to a volume term, a surface
term, a Coulomb term, an energy symmetry term and a pairing
interaction term \cite{livro2},\cite{ livro1}. Of course, with so many
parameters, other parameterisations can be obtained from the fitting of the data. One
possible set is $a_v=15.68$ MeV, $a_s=18.56$ MeV, $a_c \times
e^2=0.72$ MeV, $a_i=18.1$ MeV 
and

\begin{equation}
\delta=
\begin{cases}
34~~ A^{-3/4} {\rm MeV}, & {\rm even-even~ nuclei}, \cr
0, & {\rm even-odd~ nuclei},\cr
-34~~ A^{-3/4} {\rm MeV}, & {\rm odd-odd~ nuclei}. \cr
\end{cases}
\end{equation}

Although quite naive, these two models combined can explain many
important nuclear physics properties, as nuclear fission, for instance
\cite{livro1}.

\begin{figure}
\begin{center}
\includegraphics[width=8.cm]{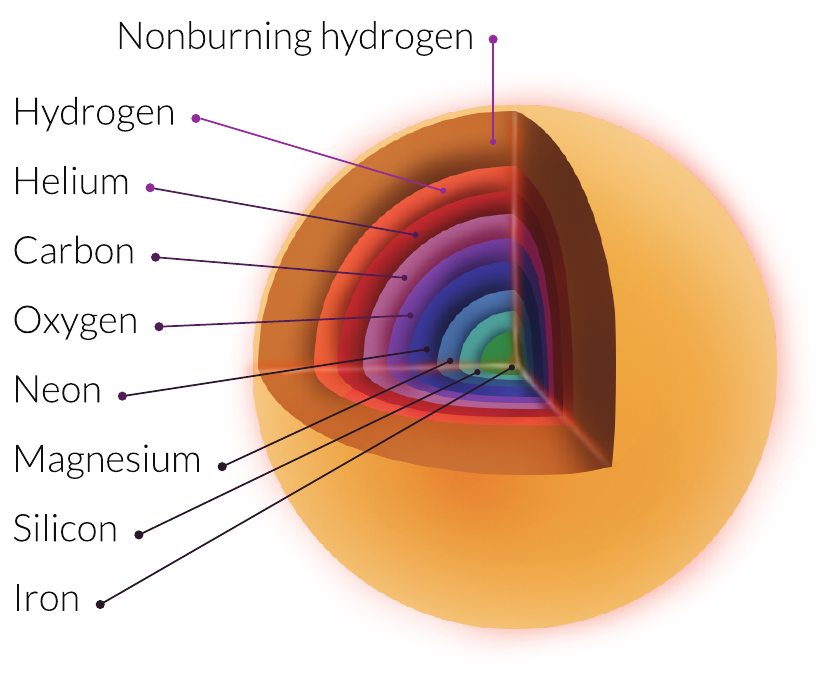}
\caption{Naive schematic representation of the possible chemical
  elements synthesised in stellar fusion.  Heavier elements are
  produced in dense stellar matter. Notice that these elements are
  produced in normal stars, not neutron stars.}
\label{fig3}
\end{center}
\end{figure}

Parameter dependent nuclear models can also explain the fusion of the
elements in the stars and the primordial nucleosynthesis with the
abundance of chemical elements in the observable universe, which is roughly the
following: 71\% is Hydrogen, 27\% is Helium, 1.8\% are Carbon to
Neon elements, 0.2\% are Neon to Titanium, 0.02\% is Lead and
only 0.0001\% are elements with atomic number larger than 60. By observing
Fig. \ref{fig2}, one easily identifies the element with the 
largest binding energy, $^{56} Fe$. Hence, it is possible to
explain why elements with atomic numbers $A \leq 56$ are
synthesised in the stars by nuclear fusion that are exothermic reactions
and heavier elements are expected to be synthesised in other
astrophysical processes, such as supernova explosions and more recently
also simulated in the mergers of compact objects. For a simplistic and
naive, but didactic idea of the stellar fusion chains, I show the possible synthesised 
chemical elements in Fig. \ref{fig3}.

\begin{figure}
\begin{center}
\includegraphics[width=8.cm]{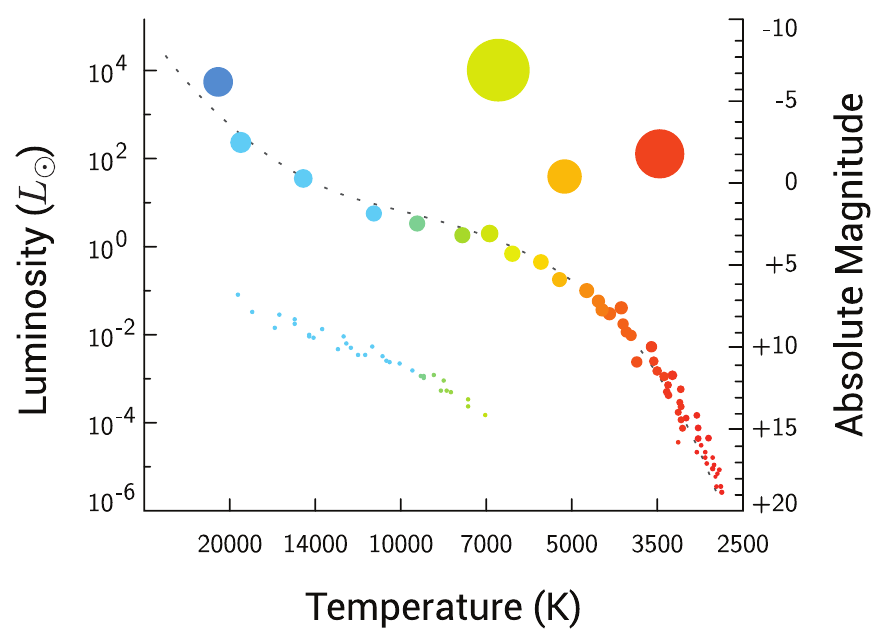}
\caption{\label{fig4} Hertzspring and Russel diagram:  luminosity (in
  terms of the Sun luminosity) as a function of the star temperature. 
 Notice that the temperature increases from right to left. 
 The yellow, orange and
  red big dots on the right top represent red giants, the blueish sequence
  on the bottom left represents white dwarfs and the central line is
  the main sequence, where the red stars are red dwarfs and the blue
  ones are blue giants.}
\end{center}
\end{figure}

After the star is born, it takes sometime fusing all the chemical
elements in its interior, until its death, which is more or less spectacular
depending on its mass. One of the most useful diagrams in
the study of stellar evolution is the Hertzsprung and Russel (HR) diagram
\cite{HR}, developed by Ejnar Hertzsprung and Norris Russel
independently in the early 1900s. According to the HR diagram, 
displayed in Fig. \ref{fig4}, the star spends most of
its life time in the central line of the diagram, the main sequence. 
Our Sun will become a white dwarf after its death, the kind of objects shown at lower
luminosities and higher temperatures, towards the left corner of the
diagram. More massive stars, with masses higher than 8 solar masses ($M_\odot$)
become either a neutron star or a black hole and these compact objects
are not shown in the HR Diagram since they do not emit
visible light waves. Moreover, neutron stars were only detected much later, as
discussed in Section \ref{astro}. For a better comprehension of the HR
diagram, please refer to  \cite{astronomytoday}. 

The main idea underlying nuclear models is to satisfy experimental
values and nuclear properties and to achieve this purpose, in almost one century of
research, they became more and more sophisticated. The
most important of these properties are the binding energy, the
saturation density, the symmetry energy, its derivatives and the
incompressibility, all of  them already explored in
the semi-empirical mass formula given in Eq. (\ref{efsem}). 
An important question to be answered is what happens when one moves
to higher densities or to finite temperature in the QCD phase diagram
shown in Fig. \ref{fig0} ?

\begin{figure}
  \begin{center}
   \includegraphics[width=9.cm]{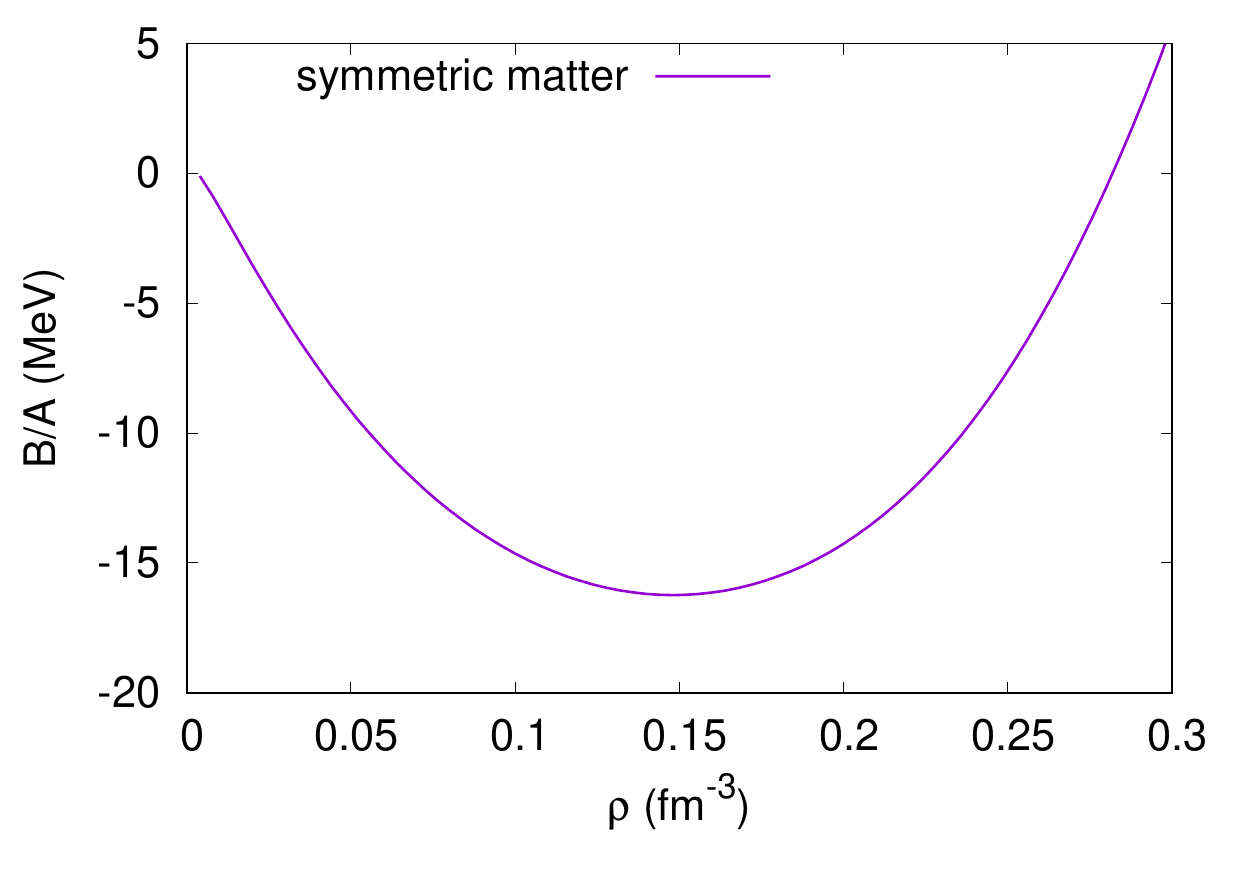}
\caption{Nuclear matter binding energy per nucleon.}
\label{ligacao}
\end{center}
\end{figure}

To better understand this point, let's discuss the concept of nuclear
matter.  This is a common denomination for an infinite matter
characterised by properties of a symmetric nucleus in its ground state,
and without the effects of the Coulomb interaction. 
If one divides eq.(\ref{efsem}) by the number of nucleons $A$, one can
see that under the conditions just mentioned, the third and forth term
disappear. If one assumes an infinite radius, $A \rightarrow \infty$
and no surface effects exist. The pairing interaction would be an
unnecessary correction. Hence, the binding energy per nucleon becomes
approximately

\begin{equation}
\frac{B(Z,A)}{A}=a_v \simeq 16~ {\rm MeV},
\end{equation}
which is what one gets for a two-nucleon system if compared with
  the average value shown in Fig.\ref{fig2}. However, the deuteron
binding energy is much smaller, around only 2 MeV. This means that
nuclear matter is not an appropriate concept if one wants to
describe  the properties of a specific nucleus, but it is rather
  useful to study, for instance, the interior of a neutron star.
Normally, it is described by an equation of
state, which consists of a set of correlated equations, like pressure,
energy and density. The equation of state that describes the ground
state of nuclear matter is calculated at zero temperature and is a
function of the proton and the neutron densities, which are the same
in symmetric matter. Useful definitions are the proton fraction
$ y_p = \frac{\rho_p}{\rho}$ and the asymmetry $\delta= \frac{\rho_n -
  \rho_p}{\rho}$, which are respectively 0.5 and 0 in the case of
symmetric nuclear matter.  In these equations $\rho=\rho_p+\rho_n$
is the total nuclear (or baryonic) density. The macroscopic nuclear energy can be
obtained from the microscopic equation of state if one assumes that

\begin{equation}
E_N = \int  {\cal E} (\rho, \delta) d^3 {\mathbf r},
\end{equation}
where ${\cal E}(\rho, \delta)$ is the energy density. Thus,

\begin{equation}
\frac{B(Z,A)}{A}=\frac{E_N}{A} -m_n = \frac{{\cal E}}{\rho} - m_n ,
\end{equation}
where $m_n = 939$ MeV is the neutron mass ($c=1$). The binding energy
as a function of the density is shown in Figure
\ref{ligacao}. We will see how it can be obtained later
  in the text. The minimum corresponds to what is generally called
  saturation density and the value inferred from experiments ranges
  between  $\rho_0 =0.148 - 0.17$ fm$^{-3}$, as mentioned earlier when
  the PREX results were given.

The pressure can be easily obtained from thermodynamics

\begin{equation}
-PV=E_N-T S - \mu~ A,
\end{equation}
or, dividing by the volume,

\begin{equation}
-P= {\cal E} - T {\cal S} - \mu~ \rho = \Omega,
\label{thermo}
\end{equation}
where $T$ is the temperature,  $\cal S$ is the entropy density,
$\mu$ is the chemical potential and $\Omega$ the thermodynamical
potential. When we take $T=0$, the expression becomes even simpler
because the term $T {\cal S}$ vanishes.

To demonstrate how a simple equation of state (EOS) can be obtained
from a relativistic model, we use the free Fermi gas and
assume that $\hbar = c=1$, known as natural units. 
 Within this model, the fermions can be either neutrons or nucleons,
 but I would like to emphasise that it is not adequate to describe
 nuclear matter properties, as will be obvious later. 
Its Lagrangian density reads:

\begin{equation}
{\cal L}_0= \bar \psi (i  \gamma^{\mu} \partial_{\mu} - m)\psi.
\label{free1}
\end{equation}
From the Euler-Lagrange equations

\begin{equation}
\partial_{\mu} \left( \frac{\partial {\cal L}}{\partial(\partial_
{\mu} \psi)} \right) - \frac{\partial {\cal L}}{\partial \psi} =0,
\label{el2}
\end{equation}
the Dirac equation is obtained:

\begin{equation}
(i \gamma^{\mu} \partial_{\mu} - m)\psi=0.
\label{dirac_apen}
\end{equation} 
Its well known solution has the form
$ \psi = \Psi (\mathbf k, \lambda) e^{ i (\mathbf k
  \cdot \vec r - E(k)t)}$, where $\Psi (\mathbf k, \lambda)$ is a
four-component spinor and $\lambda$ labels the spin. The energy can be
calculated from

\begin{equation} 
(\vec \alpha \cdot \mathbf k + \beta M)^2 \Psi (\mathbf k, \lambda) = 
E(\mathbf k)^2 \Psi (\mathbf k, \lambda),
\end{equation}

\noindent  where $\alpha = \gamma_0 \vec \gamma$
or

\begin{equation}
(\mathbf k)^2 + M^2 =  E(\mathbf k)^2, \quad  E(\mathbf k) = \pm \sqrt{\mathbf k^2 + M^2}.
\end{equation}
Moreover, one gets

\begin{equation}
< \Psi|\Psi> =\gamma \int \frac{d^3 k}{(2 \pi)^3}  ( f_+-f_-) = \rho,
\label{rhoT}
\end{equation}
where $f_\pm$ represents the Fermi-Dirac distribution for particles
and antiparticles  \cite{Greiner} . For $T=0$, $f_+$ is simply the
step function and there are no antiparticles in the system. In this case,

$$
< \Psi|\Psi> =\gamma \int \frac{d^3 k}{(2 \pi)^3}
~\theta(k_F^2 - k^2) =
$$
\begin{equation}
\frac{\gamma}{2 \pi^2} \int^{k_F}_0 k^2 dk =
\frac{\gamma~ k_F^3}{6 \pi^2} = \rho,
\label{rhofermion}
\end{equation}
with $k_F$ being the Fermi momentum and $\gamma$ the degeneracy of the
particle.  If one considers only a gas of neutrons, the degeneracy
is 2 due to the spin degeneracy. However, if one considers a gas of
nucleons,  i.e., symmetric matter with the same amount of protons and
neutrons, it is 4 because it accounts for the isospin degeneracy as well.

One can then write

\begin{equation}
< \Psi| \vec \alpha \cdot \mathbf k + \beta M |\Psi> =
\int \frac{d^3 k}{(2 \pi)^3}  \sqrt{\mathbf k^2 + M^2} (f_+ + f_-)
\label{endirac0}
\end{equation}
or

\begin{equation}
{\cal E}=\frac{\gamma}{2 \pi^2} \int {\mathbf k}^2 dk
\sqrt{{\mathbf k}^2+M^2} (f_+ + f_-).
 \label{denT}
\end{equation}
For $T=0$, it becomes

\begin{equation}
{\cal E}=\frac{\gamma}{2 \pi^2} \int^{k_F}_0 {\mathbf k}^2 dk \sqrt{{\mathbf k}^2+M^2}.
 \label{denT0}
\end{equation}
As we still do not know the value of the chemical potential in
eq.(\ref{thermo}), the pressure can be obtained from the energy momentum tensor:

\begin{equation}
T_{\mu \nu} = - g_{\mu \nu} {\cal L} +
\partial_{\nu} \psi \left(  
\frac{\partial {\cal L}}{\partial(\partial^{\mu} \psi)} \right),
\label{tem}
\end{equation}
having in mind that 

\begin{equation}
{\cal E}=<T_{00}>, \quad P=\frac{1}{3}<T_{ii}>
\label{t00}
\end{equation}
and is given by

\begin{equation}
P=\frac{1}{3} \psi^{\dagger} \left(-i \vec \alpha \cdot \nabla \right) \psi.
\label{pregas}
\end{equation}

\noindent From eq.(\ref{endirac0}) one can write

$$
< \Psi| \vec \alpha \cdot \mathbf k |\Psi> = k~\frac{\partial}{\partial k}
< \Psi| \vec \alpha \cdot \mathbf k + \beta M |\Psi> =
$$
\begin{equation}
 k~\frac{\partial}{\partial k} < \Psi| E(\mathbf k)  |\Psi>
\end{equation}
and finally
$$
\psi^{\dagger} \left(-i \vec \alpha \cdot \nabla \right) \psi =
\frac{\gamma}{ (2 \pi)^3} \int  d^3k k ~\frac{\partial E}{\partial
  k}=
$$
$$
\frac{\gamma}{2 \pi^2} \int {\mathbf k}^2 dk \cdot \frac{\mathbf
  k^2}{\sqrt{\mathbf k^2 + M^2}}, 
$$

\begin{equation}
P= \frac{\gamma}{6 \pi^2} \int dk \frac{\mathbf
  k^4}{\sqrt{\mathbf k^2 + M^2}} (f_+ + f_-),
\label{pressT}
\end{equation}
and for $T=0$,

\begin{equation}
P= \frac{\gamma}{6 \pi^2} \int_0^{k_F} dk \frac{\mathbf
  k^4}{\sqrt{\mathbf k^2 + M^2}}. 
\end{equation}

The entropy density of a free Fermi gas is given by 

\begin{equation}
{\cal S}=-\gamma \int\frac{d^3p}{(2\pi)^3}\,(f_{+} 
ln\left(\frac{f_{+}}{1-f_{+}}\right)+
ln(1-f_{+})
\end{equation}

\begin{equation}
+f_{-} ln\left(\frac{f_{-}}{1-f_{-}}\right)+ln(1-f_{-}))\;, 
\label{entropy}
\end{equation}

By minimising eq.(\ref{thermo}), the distribution functions are obtained:

\begin{equation}
\frac{\partial \Omega}{\partial f_+} =0 \rightarrow
f_+=\frac{1}{1+e^{({\cal E}-\mu)/T}}
\end{equation}
and 

\begin{equation}
\frac{\partial \Omega}{\partial f_-} =0 \rightarrow
f_-=\frac{1}{1+e^{({\cal E}+\mu)/T}}.
\end{equation}
On the other hand, the minimisation of the thermodynamical potential
with respect to the density yields the chemical potential, i.e.,

\begin{equation}
\frac{\partial \Omega}{\partial \rho} =0 \rightarrow \mu . 
\end{equation}

For $T=0$, 

\begin{equation}
\frac{\partial \Omega}{\partial k_F} =0 \rightarrow \mu = \sqrt{k_F^2
  + M^2}.
\end{equation}

In order to go back to the discussion of nuclear matter, we are
lacking exactly  the nuclear interaction and its introduction will be
seen in section \ref{rel_model}.
 
\subsection{From the compact objects point of view}
\label{astro}

I have already discussed the evolution process of a star while it
remains in the main sequence of Fig.\ref{fig4}. When the fusion
ends, it is believed that one of the possible remnants is a neutron
star. We see next how it was first predicted and then observed.

In fact, the history of neutron stars started with the observation of
a white dwarf and its description with a degenerate free Fermi gas
equation of state, as the one just introduced, but with the fermions
being electrons instead of neutrons.
In 1844, Frederich Bessel observed a very bright star that described
an elliptical orbit \cite{sirius}, known as Sirius. He proposed that Sirius was part
of a binary system, whose companion was not possible to see. In
1862, it was observed by Alvan Clark Jr. This companion, named Sirius
B, had a luminosity many orders of magnitude lower than Sirius, but
approximately the same mass, of the order of the solar mass (1
$M_\odot$). In 1914 Walter Adams concluded, through spectroscopy
studies that the temperature at the surface of both stars should be
similar, but the density of one of them should be much higher than the
density of the companion. This high density star was called white
dwarf and its properties were explained only in 1926 by Ralph Fowler
\cite{Fowler}, with the help of quantum mechanics. He claimed that the
internal constituents of the white dwarf should be responsible for a
degeneracy pressure that would compensate the gravitational
force. This hypothesis was possible since the electrons are fermions
and hence, obey the Pauli Principle. 
In 1930, Subrahmanyan Chandrasekhar calculated
the maximum  densities of a white dwarf \cite{Chandra} and subsequently
its maximum mass \cite{Chandra2} that he thought should be 0.91
M$_\odot$ due to an incorrect value of the atomic mass to charge number ratio.
It is interesting to note that the
correct Chandrasekhar limit, 1.44 M$_\odot$, was actually obtained by
Landau \cite{LandauChandra}. 

Concomitantly, Lev Landau reached the same conclusion as
Chandrasekhar and went further: he proposed that even denser objects
could exist and in this case, the atomic nuclei would overlap and the
star would become a gigantic nucleus \cite{Baym}. Landau's hypothesis is
considered the first forecast of a neutron star, although the neutrons
had not been detected yet.  Landau's paper was written in the
  beginning of 1931 but published one year later, just when the
  neutron was discovered by James Chadwick \cite{neutrons}.
The first explicit proposition of the existence of
neutron stars was made by Baade and Zwick \cite{Baade}, soon after
Chadwick's discovery.

In 1939, Toman and, independently Oppenheimer and Volkoff (TOV)
\cite{tov}, used special and general relativity to correct Newton's
equations that described the properties of a perfect isotropic fluid,
which they considered could be the interior of compact objects (white
dwarfs and neutron stars). While Tolman proposed 8 different solutions
for the system of equations, Oppenheimer and Volkoff used the equation
of state of a free neutron gas (exactly the one introduced in the last
section) and obtained a maximum mass of 0.7 M$_\odot$ for the neutron
star, what was very disappointing because it was lower than the
Chandrasekhar limit. However, soon, the limitations of this EOS
were noticed: the inclusion of the nuclear interaction could make it
harder and then generate higher masses. These calculations will be
shown in the next section.

In 1940,  Mario Shenberg and George Gamow proposed the Urca process
\cite{urca}, responsible for cooling down the stars by emitting
neutrinos, which can carry a large amount of energy with very little
interaction. 

In 1967, the first neutron star (NS)  was detected by Jocelyn Bell
and Anthony Hewish \cite{seminal}. At first, they believed they were capturing
signals from an extraterrestrial civilisation
and the booklet {\it The Little Green Men}
really existed. But they soon realised that the radio signals were
coming from a compact object with a very stable frequency (pulse) and the
object was called pulsar. 

It is worth pointing out that white dwarfs and neutron stars bear very
different internal constituents and densities. Neutron stars are much
denser. This means that general relativity is a very important
component in the study of NS, but this is not true for
white dwarfs. Hence, it would be expected that only relativistic models, as the
  ones introduced in the present text, could be used to describe
  neutron star macroscopic properties. However,
there are non-relativistic models, known as
Skyrme models  which can be used to describe NS, as far
as they do not violate causality. Moreover, some  
non-relativistic models lead to symmetry energies that
decrease too much after three times saturation density, which
is a very serious problem if we want to apply them to the study of
neutron stars, highly asymmetric systems. These problems
can be cured with the inclusion of three-body forces, what makes the 
calculations much more complicated. For a review on Skyrme models,
please see ref. \cite{skyrme}. On the other hand, relativistic models
are generally causal and always Lorentz invariant and when 
extended to finite temperature, anti-particles appear
naturally. Thus, only relativistic models are discussed in the present work.

Let's go back to  history because it continues. In 1974 Russel Hulse and  Joseph Taylor identified the
first binary pulsar PSR1913+16 \cite{PSR1913}  with a radio-telescope in Arecibo and
proposed that the system was losing energy in the form of
gravitational waves (GW), the same kind of waves foreseen by relativistic
theories.  Note that they did not detect gravitational waves
  directly but instead proved its existence via pulsar timing and were
  laureated with the Nobel prize for this discovery.
In 2015, the first GW produced by two colliding black holes
was  finally detected directly by LIGO \cite{GW15} and in 2017,
GW170817, produced by the merger of two NS
\cite{GW170817} initiated the era of multi-messenger astronomy 
\cite{multi}. These gravitational waves have become an excellent source of
constraints to the EOS used to describe neutron stars, as will be
discussed in a future section.

\section{Relativistic models for astrophysical studies}
\label{rel_model}

In Section \ref{nuclear} the EOS of a free Fermi gas was introduced
and in Section \ref{astro}, I mentioned that the EOS can
satisfactorily describe a white dwarf, as shown by Chandrasekhar
if the free Fermi gas is a gas of electrons.
However, if  the fermions are neutrons, it
cannot describe neutron stars. One important ingredient, besides the
already mentioned relativistic effects, is still missing in the
recipe: the nuclear interaction. So, let's go back to nuclear matter.

\subsection{The $\sigma-\omega$ model}
\label{sigmaomega}

This model, also known as  Walecka model \cite{walecka} or quantum
hadrodynamics (QHD-1) is based on the fact that the 
interaction inside the nucleus has two contributions: an attractive contribution, at {\it
  large} distances and a repulsive one at short distances, and
both can be reasonably well described by Yukawa type potentials and
represented by fields generated respectively by scalar and vector
mesons.
 This idea was first proposed by  Hans Peter Durr in his
Ph.D. thesis in 1956, supervised by Edward Teller, who, in 1955, also
proposed a version of the model based on classical field theory
\cite{johnson}. But the quantum version proposed by Walecka
was the one that really gained popularity and up to now it is largely
applied with different versions and extensions.  This simplified model
does not take pions into account because, as will be seen next, it  is
usually solved in a mean field approximation and in this case, the
pion contribution disappears. As the $\sigma-\omega$ model is a
relativistic model, this simpler and more common approximation is
always known as relativistic Mean Field Theory (RMF) or relativistic Hartree
approximation. 

As the name suggests, the $\sigma-\omega$ model considers that the
central effective potential for the nucleon-nucleon interaction is
given by 
$$V(r)=\frac{g_\omega^2}{4 \pi}\frac{e^{-m_\omega r}}{r} - \frac{g_\sigma^2}{4 \pi}
\frac{e^{-m_\sigma r}}{r},$$
where $r$ is the modulus of the vector that defines the relative
distance between two nucleons, the two constants $g_\sigma$ and $g_\omega$
are adjusted to reproduce the nucleon-nucleon interaction and the
meson masses are respectively $m_\sigma=550$ MeV and $m_\omega=783$ MeV. 
The interested reader, can look at the potential $V(r)$  obtained
with the coupling constants and masses used in this section in \cite{livro1}.
To obtain the binding energy that corresponds to this potential in
RMF, a Lorentz invariant Lagrangian density is necessary and it reads:
$$
{\cal L}=\bar \psi\left[\gamma_\mu(i\partial^{\mu}-g_\omega \omega^{\mu})-
(M-g_\sigma \sigma)\right]\psi
$$

\begin{equation}
+\frac{1}{2}(\partial_{\mu}\sigma\partial^{\mu}\sigma
-m_\sigma^2 \sigma^2) 
+\frac{1}{2} m_\omega^2 \omega_{\mu}\omega^{\mu} 
-\frac{1}{4}F_{\mu\nu}F^{\mu\nu},
\label{lag}
\end{equation}
where

\begin{equation}
F_{\mu\nu}=\partial_{\mu}\omega_{\nu}-\partial_{\nu}\omega_{\mu},
\end{equation}
$\psi$ represents the baryonic field (nucleons),
$\sigma$ and  $\omega^{\mu}$ represent the fields associated with the scalar
and vector mesons and $M$ is the nucleon mass, generally taken as 939 MeV.
By comparing eqs. (\ref{free1}) and (\ref{lag}), one can see that
besides the Fermi gas representing the nucleons, the latter contains
two interaction terms and kinetic and mass terms for both mesons. The
usual prescription is to use the Euler-Lagrange equations (\ref{el2})
for each field to obtain the equations of motion. They read:

\begin{equation}
\left( \partial_{\mu} \partial^{\mu} + m_\sigma^2 \right) \sigma = g_\sigma \bar \psi \psi
\label{kg},\end{equation}

\begin{equation}
\partial_{\mu} F^{\mu \nu} + m_\omega^2 \omega^{\nu} = g_\omega \bar \psi \gamma^{\nu} \psi
\label{qed2}\end{equation}
and

\begin{equation}
\left[ \gamma_\mu(i\partial^{\mu}-g_\omega \omega^{\mu})-
(M-g_\sigma \sigma)\right]\psi =0
\label{dirac2}.\end{equation}

Note that eq. (\ref{kg}) is a Klein-Gordon equation with a scalar
source, eq. (\ref{qed2}) is analogous to quantum eletrodynamics with a
conserved baryonic current ($\bar \psi \gamma^{\nu} \psi$), instead of
the electromagnetic current and eq.(\ref{dirac2}) is a Dirac
equation for an interacting (not free) gas. 

In a RMF approximation, the meson fields are replaced by their
expectation values that behave as classical fields:

\begin{equation}
\sigma\rightarrow \left<\sigma\right>\equiv\sigma_0
\end{equation}
and 

\begin{equation}
\omega_\mu\rightarrow \left<\omega_0\right>\equiv\omega_0, \quad 
\left<\omega_k \right>=0 .
\end{equation} 

The equations of motion can then be easily solved and they read:

\begin{equation}
\sigma_0 = \frac{g_\sigma}{m_\sigma^2}<\bar \psi \psi>=\frac{g_\sigma}{m_\sigma^2} \rho_s
\label{phi0}
\end{equation}

and

\begin{equation}
\omega_0 = \frac{g_\omega}{m_\omega^2}<\psi^{\dagger}
\psi>=\frac{g_\omega}{m_\omega^2} \rho ,
\label{v0}
\end{equation}
where $\rho_s$ is a scalar density and $\rho$ is a baryonic number
density. The Dirac equation becomes simply

\begin{equation}
\left[ (i \gamma_\mu \partial^{\mu}-g_\omega \gamma_0 \omega^0)-
(M-g_\sigma \sigma_0)\right]\psi =0,
\label{diracnu}\end{equation}
and

\begin{equation}
M^* = M-g_\sigma \sigma_0, \label{massaef}
\end{equation}
is the effective mass. To obtain the EOS, the recipe is the same as
already shown for the free Fermi gas, which leads to expressions for
the energy density and pressure. Assuming
$C_s^2=g_\sigma^2(M^2/m_\sigma^2)=267.1$ and $C_v^2=g_\omega^2(M^2/m_\omega^2)=195.9$,
the binding energy $E/N-M = -15.75$ MeV at the saturation density
$\rho=0.19$ fm$^{-3}$, a little bit too high.

\begin{figure}
  \begin{center}
\includegraphics[width=9.cm]{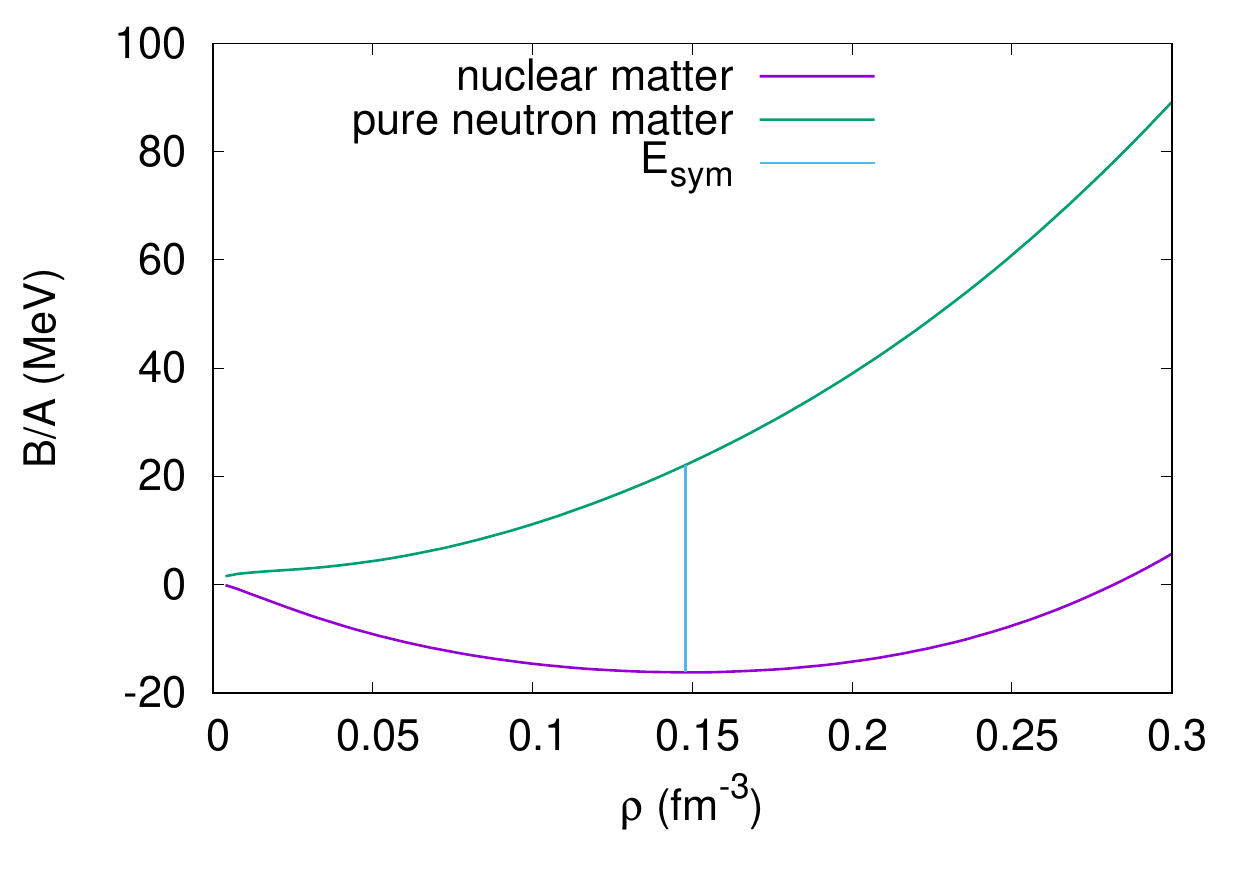}
\caption{Schematic representation of the nuclear matter symmetry energy.}
\label{simetria}
\end{center}
\end{figure}

Other important quantities directly related with nuclear matter EOS are the
symmetry energy, its derivatives and the incompressibility.  The
symmetry energy is roughly the necessary energy to transform symmetric
matter into a pure neutron matter, as shown in Fig. \ref{simetria}, i.e.,

\begin{equation}
{\cal E}(\rho, \delta) \simeq {\cal E}(\rho, \delta=0) + E_{sym}(\rho)\delta^2.
\end{equation}
Its value can be inferred from experiments and it is of the order of
30-35 MeV and it can be written as 

\begin{equation}
E_{sym}=\frac{1}{8} \left( \frac{\partial^2 ({\cal E}/\rho)}{\partial
    y_p^2} \right)_{y_p=0.5} 
= \frac{1}{2} \left( \frac{\partial^2 ({\cal E}/\rho)}{\partial
    \delta^2} \right)_{\delta=0}.
\end{equation}

It is common to expand the symmetry energy around the saturation
density in a Taylor series as

\begin{equation}
E_{sym}= J + L_0 (\frac{\rho-\rho_0}{3 \rho_0}) + \frac{K_{sym}}{2} 
(\frac{\rho-\rho_0}{3 \rho_0})^2 + {\cal{O}}(3),
\end{equation}
where $J$ is the symmetry energy at the saturation point and 
$L_0$ and $K_{sym}$ represent respectively  its slope and curvature:

\begin{equation}
L_0= 3 \rho_0 \left(\frac{\partial E_{sym}(\rho)}{\partial
  \rho}\right)_{\rho=\rho_0},
\end{equation}
and

\begin{equation}
K_{sym}= 9 \rho_0^2 \left(\frac{\partial^2 E_{sym}(\rho)}{\partial
  \rho^2}\right)_{\rho=\rho_0} .
\end{equation}

Experimental data for the symmetry energy can be inferred from
 heavy-ion collisions, giant monopole (GMR) and giant dipole
   (GDR) resonances, pygmy dipole resonances and isobaric analog 
 states. 
Accepted values for the slope until very recently lied in between
30.6 and 86.8 MeV \cite{relat,Micaela2017} and for the curvature, in between
-400 and 100 MeV \cite{Tews,Zhang}. These two quantities are correlated
with macroscopic properties of neutron stars, as will be seen later on
along this manuscript.
Based on 28 experimental and observational data, restricted bands for
the values of $J$ ($25 < J <35$ MeV) and $L_0$ ($25 < L_0 < 115$ MeV)
were given in \cite{bao} . More recently, results
obtained by the PREX2 experiment \cite{PREX2} point to a different band,
given by $L_0=106 \pm 37$ MeV \cite{PiekarewiczPREX}. If confirmed,
this result rehabilitates many of the already ruled out EOS and points
to a much larger than previously expected neutron star radius, also
discussed later on in the present paper. 

Another important quantity is the incompressibility, already mentioned
when the liquid drop model idea was introduced. It is a measure of the stiffness 
of the EOS, i.e., defines how much pressure a system can support and
it is calculated from the relation

\begin{equation}
K_0 = 9 \left( \frac{\partial P}{\partial \rho} \right)_{\rho=\rho_0,~ y_p=0.5}
\end{equation}
and ranges between 190 and 270 MeV \cite{relat,Micaela2017}.
These values can be inferred from both theory and experiments. 

I will go back to the importance of these nuclear matter bulk
properties and their connection with neutron star properties later
on.

\subsection{Extended relativistic hadronic models}
\label{rmf}

I next present one example of a
complete Lagrangian density that describes baryons interacting among each other by 
exchanging scalar-isoscalar ($\sigma$) , vector-isoscalar ($\omega$), 
vector-isovector ($\rho$) and scalar-isovector ($\delta$) mesons:

\begin{eqnarray}
\mathcal{L} = \mathcal{L}_{\rm nm} + \mathcal{L}_\sigma +
\mathcal{L}_\omega
+ \mathcal{L}_\rho + \mathcal{L}_{\delta} + \mathcal{L}_{\sigma\omega\rho},
\label{dl}
\end{eqnarray}
where \cite{Agrawal}

\begin{widetext}

\begin{eqnarray}
\mathcal{L}_{\rm nm} & =& \overline{\psi}(i\gamma^\mu\partial_\mu - M)\psi 
+ g_\sigma\sigma\overline{\psi}\psi 
- g_\omega\overline{\psi}\gamma^\mu\omega_\mu\psi 
- \frac{g_\rho}{2}\overline{\psi}\gamma^\mu\vec{\rho}_\mu\cdot \vec{\tau}\psi
+ g_\delta\overline{\psi}\vec{\delta}\cdot \vec{\tau}\psi, 
\label{lag1} \\
\mathcal{L}_\sigma &=& \frac{1}{2}(\partial^\mu \sigma \partial_\mu \sigma 
- m^2_\sigma\sigma^2) - \frac{A}{3}\sigma^3 - \frac{B}{4}\sigma^4,
\\
\mathcal{L}_\omega &=& -\frac{1}{4}F^{\mu\nu}F_{\mu\nu} 
+ \frac{1}{2}m^2_\omega\omega_\mu\omega^\mu 
+ \frac{C}{4}(g_\omega^2\omega_\mu\omega^\mu)^2,
\\
\mathcal{L}_\rho &=& -\frac{1}{4}\vec{B}^{\mu\nu}\vec{B}_{\mu\nu} 
+ \frac{1}{2}m^2_\rho\vec{\rho}_\mu \cdot \vec{\rho}^\mu,
\\
\mathcal{L}_\delta &=& \frac{1}{2}(\partial^\mu\vec{\delta}\partial_\mu\vec{\delta} 
- m^2_\delta\vec{\delta}^2),
\end{eqnarray}
and

\begin{eqnarray}
\mathcal{L}_{\sigma\omega\rho} &=& 
g_\sigma g_\omega^2\sigma\omega_\mu\omega^\mu
\left(\alpha_1+\frac{1}{2}{\alpha_1}'g_\sigma\sigma\right)
+ g_\sigma g_\rho^2\sigma\vec{\rho}_\mu \cdot \vec{\rho}^\mu
\left(\alpha_2+\frac{1}{2}{\alpha_2}'g_\sigma\sigma\right) 
\nonumber \\
&+& \frac{1}{2}{\alpha_3}'g_\omega^2 g_\rho^2\omega_\mu\omega^\mu
\vec{\rho}_\mu\cdot \vec{\rho}^\mu.
\label{lomegarho}
\end{eqnarray}

\end{widetext}

In this Lagrangian density, $\mathcal{L}_{\rm nm}$ represents the kinetic part of 
the nucleons plus the terms standing for the interaction between them 
and mesons $\sigma$, $\delta$, $\omega$, and $\rho$. The term $\mathcal{L}_j$ 
represents the free and self-interacting terms of the meson $j$, where 
$j=\sigma,\delta,\omega,$ and $\rho$. The $\sigma$
self-interaction terms were the first ones to be introduced
\cite{boguta} to correct some of  the values of the nuclear bulk properties.
The last term, $\mathcal{L}_{\sigma\omega\rho}$, accounts for  
crossing interactions between 
the meson fields. The antisymmetric field tensors $F_{\mu\nu}$ and 
$\vec{B}_{\mu\nu}$ are given by 
$F_{\mu\nu}=\partial_\nu\omega_\mu-\partial_\mu\omega_\nu$
and $\vec{B}_{\mu\nu}=\partial_\nu\vec{\rho}_\mu-\partial_\mu\vec{\rho}_\nu
- g_\rho (\vec{\rho}_\mu \times \vec{\rho}_\nu)$. The nucleon mass is $M$ and the 
meson masses are $m_j$. 

In a mean field approximation, the meson
fields are treated as classical fields and the equations of motion are
obtained via Euler-Lagrange equations.  Translational and rotational
invariance are assumed. The equations of motion
are then solved self-consistently and the 
energy momentum tensor, Eq. (\ref{tem}) is used in the calculation of
the EOS. The calculations follow the steps shown in Section
\ref{nuclear} and \ref{sigmaomega}. The interested reader can also
check them, for instance, in
\cite{walecka} and \cite{relat}.  Nevertheless, some of
the important steps are mentioned in what follows. Within a RMF
approximation, the common substitution mentioned below is again performed:

\begin{widetext}
  
\begin{eqnarray}
\sigma\rightarrow \left<\sigma\right>\equiv\sigma_0, \quad
\omega_\mu\rightarrow \left<\omega_0\right>\equiv\omega_0, \quad
\vec{\rho}_\mu\rightarrow \left<\vec{\rho}_0\right>\equiv \bar{\rho}_{0(3)}, \quad 
\vec{\delta}\rightarrow\,\,<\vec{\delta}>\equiv\delta_{(3)},
\label{meanfield}
\end{eqnarray}
and the equations of motion read:
\begin{eqnarray}
m^2_\sigma\sigma_0 &= g_\sigma\rho_s - A\sigma_0^2 - B\sigma_0^3 
+g_\sigma g_\omega^2\omega_0^2(\alpha_1+{\alpha_1}'g_\sigma\sigma)
+g_\sigma g_\rho^2\bar{\rho}_{0(3)}^2(\alpha_2+{\alpha_2}'g_\sigma\sigma)\,\mbox{,}\quad 
\label{sigmaacm}\\
m_\omega^2\omega_0 &= g_\omega\rho - Cg_\omega(g_\omega \omega_0)^3 
- g_\sigma g_\omega^2\sigma_0\omega_0(2\alpha_1+{\alpha_1}'g_\sigma\sigma_0)
- {\alpha_3}'g_\omega^2 g_\rho^2\bar{\rho}_{0(3)}^2\omega_0, 
\label{omegaacm}\\
m_\rho^2\bar{\rho}_{0(3)} &= \frac{g_\rho}{2}\rho_3 
-g_\sigma g_\rho^2\sigma_0\bar{\rho}_{0(3)}(2\alpha_2+{\alpha_2}'g_\sigma\sigma_0)
-{\alpha_3}'g_\omega^2 g_\rho^2\bar{\rho}_{0(3)}\omega_0^2, 
\label{rhoacm} \\
m_\delta^2\delta_{(3)} &= g_\delta\rho_{s3}, 
\label{deltaacm}\\
\end{eqnarray}

\end{widetext}

and

\begin{equation}
 [ i\gamma^\mu \partial_\mu -\gamma^0 V_\tau  - (M+S_\tau)] \psi = 0,
\label{diracacm}
\end{equation}
where

\begin{equation}
  \rho_s =\left<\overline{\psi}\psi\right>={\rho_s}_p+{\rho_s}_n,
\end{equation}
\begin{equation}
\rho_{s3}=\left<\overline{\psi}{\tau}_3\psi\right>={\rho_s}_p-{\rho_s}_n,
\label{rhos_tot}
\end{equation}

\begin{equation}
  \rho =\left<\overline{\psi}\gamma^0\psi\right> = \rho_p + \rho_n,
 \end{equation}
\begin{equation} 
\rho_3=\left<\overline{\psi}\gamma^0{\tau}_3\psi\right> = \rho_p - \rho_n=(2y_p-1)\rho,
\label{rho_tot}
\end{equation}
with

\begin{equation}
{\rho_s}_{p,n} = \frac{\gamma M^*_{p,n}}{2\pi^2}\int_0^{{k_F}_{p,n}}
\frac{k^2dk}{\sqrt{k^2+M^{*2}_{p,n}}} ,
\label{rhospn}
\end{equation}

\begin{equation}
\rho_{p,n} = \frac{\gamma}{2\pi^2}\int_0^{{k_F}_{p,n}}k^2dk =
\frac{\gamma}{6\pi^2}{k_F^3}_{p,n},
\label{rhopn}
\end{equation}

\begin{equation}
V_{\tau} =g_\omega\omega_0 +
\frac{g_\rho}{2}\bar{\rho}_{0(3)}\tau_3 ,\qquad
S_{\tau} =-g_\sigma\sigma_0 -g_\delta\delta_{(3)}\tau_3,
\end{equation}
with $\tau_3=1$ and $-1$ for protons and neutrons respectively and $\gamma=2$ to account for the spin degeneracy.
The proton and neutron effective masses read:
$$
  M_p^*=M-g_\sigma\sigma_0-g_\delta\delta_{(3)},
 $$
\begin{equation}
M_n^*=M-g_\sigma\sigma_0+g_\delta\delta_{(3)}.
\label{effectivemasses}
\end{equation}

Due to translational and rotational invariance, only the zero
components of quadrivectors remain. From the energy-momentum tensor, 
the following expressions are obtained:

\begin{widetext}
\begin{eqnarray}
\mathcal{E} &=& \frac{1}{2}m^2_\sigma\sigma_0^2 
+ \frac{A}{3}\sigma_0^3 + \frac{B}{4}\sigma_0^4 - \frac{1}{2}m^2_\omega\omega_0^2 
- \frac{C}{4}(g_\omega^2\omega_0^2)^2 - \frac{1}{2}m^2_\rho\bar{\rho}_{0(3)}^2
+g_\omega\omega_0\rho+\frac{g_\rho}{2}\bar{\rho}_{0(3)}\rho_3
\nonumber \\
&+& \frac{1}{2}m^2_\delta\delta^2_{(3)} - g_\sigma g_\omega^2\sigma\omega_0^2
\left(\alpha_1+\frac{1}{2}{\alpha_1}'g_\sigma\sigma_0\right) 
- g_\sigma g_\rho^2\sigma\bar{\rho}_{0(3)}^2 
\left(\alpha_2+\frac{1}{2}{\alpha_2}' g_\sigma\sigma_0\right) \nonumber \\
&-& \frac{1}{2}{\alpha_3}'g_\omega^2 g_\rho^2\omega_0^2\bar{\rho}_{0(3)}^2
+ \mathcal{E}_{\mbox{\tiny kin}}^p + \mathcal{E}_{\mbox{\tiny kin}}^n,
\label{denerg}
\end{eqnarray}
\end{widetext}
with
\begin{eqnarray}
\mathcal{E}_{\mbox{\tiny kin}}^{p,n}&=&\frac{\gamma}{2\pi^2}\int_0^{{k_F}_{p,n}}k^2
(k^2+M^{*2}_{p,n})^{1/2}dk 
\label{decinnlw}
\end{eqnarray}
and
\begin{widetext}
\begin{eqnarray}
P &=& - \frac{1}{2}m^2_\sigma\sigma_0^2 - \frac{A}{3}\sigma_0^3 -
\frac{B}{4}\sigma_0^4 + \frac{1}{2}m^2_\omega\omega_0^2 
+ \frac{C}{4}(g_\omega^2\omega_0^2)^2 + \frac{1}{2}m^2_\rho\bar{\rho}_{0(3)}^2
+ \frac{1}{2}{\alpha_3}'g_\omega^2 g_\rho^2\omega_0^2\bar{\rho}_{0(3)}^2
\nonumber \\
&-&\frac{1}{2}m^2_\delta\delta^2_{(3)} + g_\sigma g_\omega^2\sigma_0\omega_0^2
\left(\alpha_1+\frac{1}{2}{\alpha_1}'g_\sigma\sigma_0\right) 
+ g_\sigma g_\rho^2\sigma\bar{\rho}_{0(3)}^2 
\left(\alpha_2+\frac{1}{2}{\alpha_2}' g_\sigma\sigma\right) \nonumber \\
&+& P_{\mbox{\tiny kin}}^p + P_{\mbox{\tiny kin}}^n,\qquad
\label{pressure}
\end{eqnarray}
\end{widetext}
with

\begin{equation}
P_{\mbox{\tiny kin}}^{p,n} = 
\frac{\gamma}{6\pi^2}\int_0^{{k_F}_{p,n}}\frac{k^4dk}{(k^2+M^{*2}_{p,n})^{1/2}} .
\end{equation}

\begin{widetext}
\begin{table} 
\caption{Parameter sets used in this
  section - all meson masses and $A$ are given in MeV,
 $\Lambda_v={\alpha_3}'/2$ and $M=939$ MeV.}
\label{tab:ctes}
\begin{tabular}{|c|c|c|c|c|c|c|c|c|c|c|}
\hline
Model & $m_\sigma$ & $m_\omega$ & $m_\rho$ & $g_\sigma$ & $g_\omega$ & $g_\rho$ & $A$ & $B$ & $C$ & $\Lambda_v$ \\
\hline
NL3 & 508.194 & 782.501 & 763 & 10.217 & 12.868 & 8.948 & 2.055 $\times 10^{-3}$ & -2.65 $\times 10^{-3}$& 0 & 0 \\
NL3$\omega \rho$ & 508.194 & 782.501 & 763 & 2.192 & 12.868 & 11.276 & 2.055 $\times 10^{-3}$ & -2.65 $\times 10^{-3}$ & 0 & 0.03 \\
IUFSU & 491.5 & 782.5 & 763 & 9.971 & 13.032 & 13.590 & 1.80 $\times 10^{-3}$& 4.9 $\times 10^{-5}$ & 0.18 & 0.046 \\
\hline
\end{tabular}
\end{table}
\end{widetext}

\subsection{Too many relativistic models} 
\label{choice}

In \cite{relat}, a large amount of relativistic models were
confronted with two sets of nuclear bulk properties, one more and one
less restrictive. The interested reader should check the chosen 
ranges of properties in both sets and the respective values of 363
models. In what follows, I will restrict myself to 3 parameter sets:
NL3 \cite{PRC55-540}, NL3$\omega\rho$ \cite{Horo}, which is an
extension of the NL3 parameter set with the introduction of a
vector-isovector interaction and IUFSU \cite{PRC82-055803}. These
models are chosen because they are frequently used in various
applications in the literature. Moreover, NL3$\omega\rho$ and IUFSU satisfy
all nuclear matter bulk properties, but it will be seen along the text
that recent astrophysical observations are not completely satisfied by
them. The inclusion of NL3 and its
comparison with NL3$\omega\rho$  help the understanding of the
importance of the $\omega-\rho$ interaction. Other parameter sets
shown along the next sections are GM1 \cite{Glen2}, GM3 \cite{Glen},
TM1 \cite{tm1} and FSUGZ03 \cite{fsugz03}. All of them are
contemplated in \cite{relat} and the interested reader can check
their successes and failures in satisfying the main nuclear bulk properties.
Notice that none of the parameter sets explicitly mentioned in the
present work includes the $\delta$
meson, which distinguishes protons and neutrons, and consequently, the
effective masses given in eq.(\ref{effectivemasses}) are identical. 
The mesonic crossing terms weighted by
the parameters $\alpha_1$, ${\alpha_1} '$, $\alpha_2$, ${\alpha_2}'$
are not included either.  In Table  \ref{tab:ctes}, the
parameter values for the three parametrisations mostly used are
presented and in Table \ref{tab_sat}, their main
nuclear properties are shown.

\begin{figure}
\begin{tabular}{cc}
\includegraphics[width=7.cm]{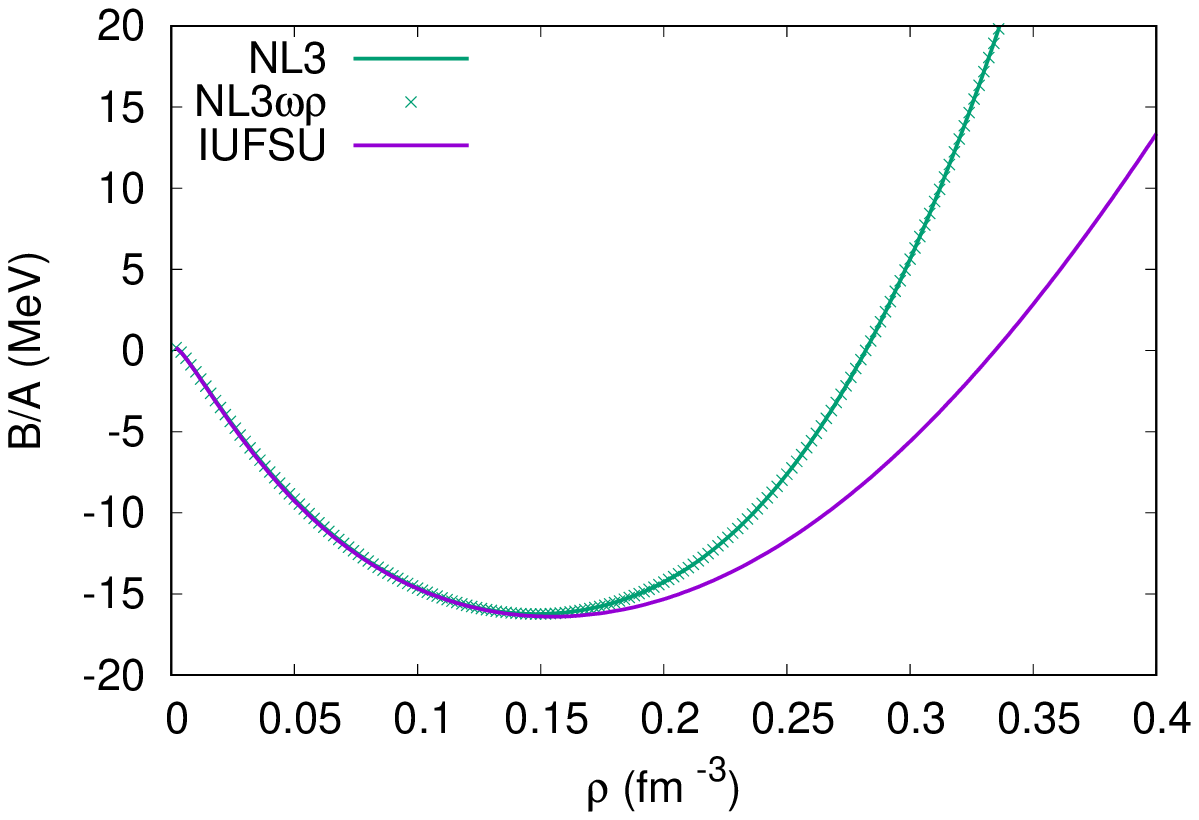} \\
\includegraphics[width=7.cm]{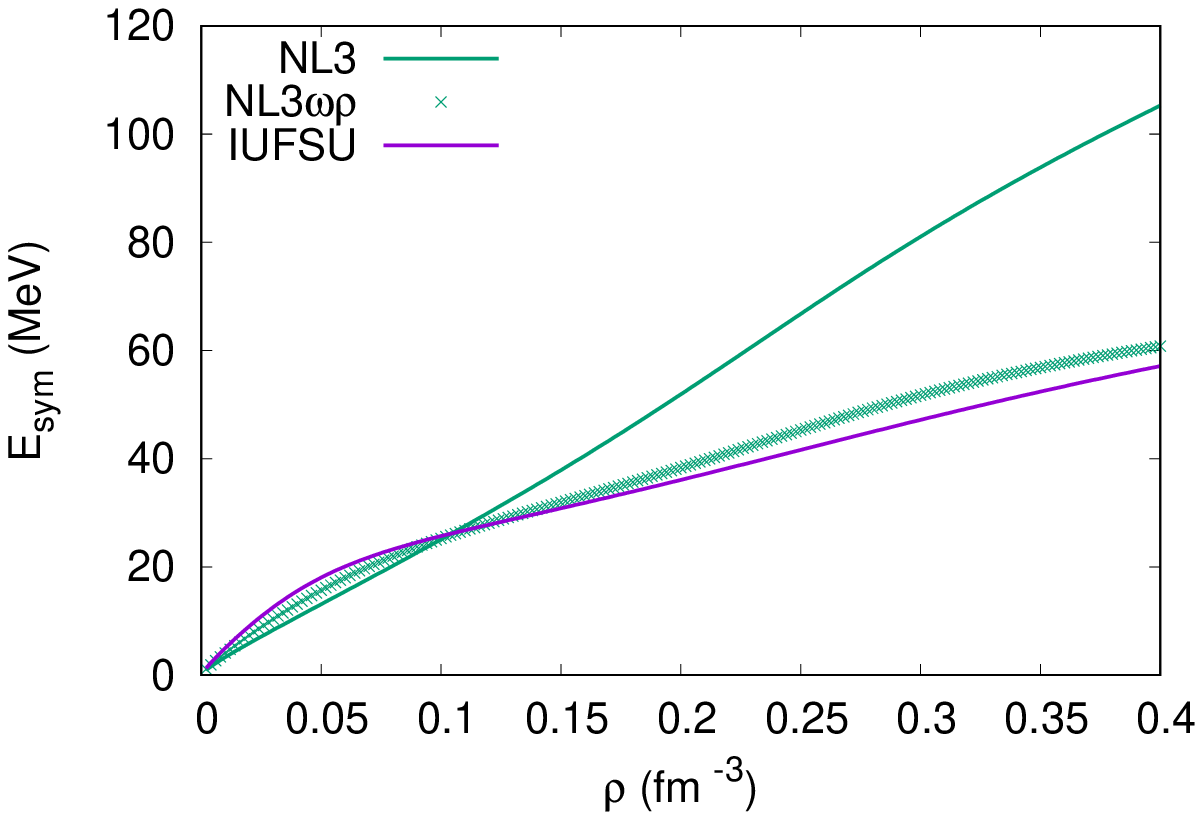} \\
\end{tabular}
\caption{top) Binding energy and bottom) symmetry energy
as a function of the baryonic density for the
 three parameter sets used in this section.} 
\label{binding3}
\end{figure}

In Figure \ref{binding3} top, I plot the binding energy per nucleon for
the three parameter sets and one can clearly see the slightly different
saturation densities and binding energy values. Notice that the
$\omega-\rho$  channel does not influence the binding energy of
symmetric nuclear matter, but plays an important role in asymmetric matter. 
In Figure \ref{binding3} bottom the symmetry energy is depicted and it is
easy to see that they are very similar at sub-saturation densities, but
completely different at larger densities. As a consequence of what is
seen in Figures \ref{binding3},  the
incompressibility, the slope and the curvature of the three models are different, as
shown in Table \ref{tab_sat}.

\begin{widetext}
\begin{table} 
\caption{Saturation and stellar properties. These values are commented along the text.}
\begin{tabular}{|c|c|c|c|c|c|c|c|c|c|c|}
\hline
Model & $\rho_0$ & $B/A$ & $K_0$ & $M^*/M$ & $J$  & $L$  & $M_{max}/M_\odot$& $R_{1.4 M_\odot}$\\
& fm$^{-3}$ & MeV & MeV & & MeV & MeV & & km \\ 
\hline
NL3 & 0.148 & -16.24 & 271.53 & 0.60  & 37.40 & 118.53 & 2.78 & 14.7 \\ 
NL3$\omega \rho$ & 0.148 & -16.24 & 271.60 & 0.60 &  31.70 & 55.50 & 2.76 & 13.7\\
IUFSU & 0.155 & -16.40 & 231.33 & 0.61 & 31.30 & 47.21 & 1.94 & 12.5  \\
\hline
\end{tabular}
\label{tab_sat} 
\end{table}
\end{widetext}

\section{Stellar Matter}
\label{star}

The idea of this section is to show how the relativistic models
presented so far can be applied to describe stellar matter and, in
this case, we refer specifically to neutron stars. Looking back at
the QCD phase diagram presented in the Introduction, one can see that
neutron stars have internal densities that are 6 to 10 times higher than
the nuclear saturation density and their temperature is low. Actually,
if we compare their thermal energy with the Fermi energy of the
system, the assumption of zero temperature is indeed reasonable.
At these very high densities the onset
of hyperons is expected because their appearance is energetically
favourable as compared with the inclusion of more nucleons in the system.
To deal with this fact, the first term in the Lagrangian density of
eq. (\ref{lag1})
has to be modified to take into account, at least, the eight lightest
baryons and it becomes:

\begin{widetext}
\begin{equation}
\mathcal{L}_{\rm Bm} = \sum_B \overline{\psi_B}(i\gamma^\mu\partial_\mu - M_B)\psi_B 
+ g_{\sigma B}\sigma\overline{\psi_B}\psi_B 
- g_{\omega B}\overline{\psi_B}\gamma^\mu\omega_\mu\psi_B 
- \frac{g_{\rho B}}{2}\overline{\psi_B}\gamma^\mu\vec{\rho}_\mu\vec{\tau}\psi_B.
\label{lag2}
\end{equation}
\end{widetext}

The meson-baryon coupling constants are given by  
\begin{equation}
g_{j B} = \chi_{B j} g_j, 
\label{mesonbaryon}
\end{equation}
where $g_j$ is the
coupling of the meson with the nucleon and $\chi_{j B}$ is a value
obtained according to symmetry groups or by satisfying hyperon
potential values. These are important quantities when hyperons are
included in the system \cite{Glen,su3}. 
We come back to the discussion of these quantities below. 
If we perform once again a RMF approximation and use the
Euler-Lagrange equations to obtain the equations of motion, we find:

\begin{equation}
\sigma_0 =  \sum_B \frac{g_{\sigma B}}{m_\sigma^2}  \rho_{sB} -
\frac{1}{m_\sigma^2} \left( A \sigma_0^2 - B \sigma_0^3 \right),
\label{s9}
\end{equation}

\begin{equation}
\omega_0  =\sum_B \frac{g_{\omega B}}{m_\omega^2} \rho_B 
- \frac{1}{m_\omega^2} \left( 2 \Lambda_v g_\omega^2 g_\rho^2 
\bar{\rho}_{0(3)}^2\omega_0
\right)
, \label{s6}
\end{equation}

\begin{equation}
\bar{\rho}_{0(3)}  = \sum_B \frac{g_{\rho B}}{m_\rho^2} \frac{\tau_{3}}{2} \rho_B
-\frac{1}{m_\rho^2} \left( 2 \Lambda_v g_\omega^2 g_\rho^2 \omega_0^2 \bar{\rho}_{0(3)} \right)
,  \label{s8}
\end{equation}
where $\rho_{sB}$ is the scalar density and $\rho_B$ is the baryon B
density, given by:

\begin{eqnarray}
\rho_{sB} =  \frac{\gamma}{2 \pi^2}\int_0^{k_{fB}} \frac{M_B^*}{\sqrt{k^2 + M_B^{*2}}}
  k^2 dk  \quad  ,\\
\rho_B = \frac{\gamma}{6\pi^2} k_{fB}^3, \quad \mbox{and}  \quad  \rho= \sum_B \rho_B ,
\label{s10}
\end{eqnarray}
where $k_{fB}$ is the Fermi momentum of baryon $B$. 
The terms
$\mathcal{E}_{\mbox{\tiny kin}}^p$ and $\mathcal{E}_{\mbox{\tiny
    kin}}^n$ that appear in eq.  (\ref{denerg}), must now be
substituted by

\begin{equation}
\mathcal{E}_B = \frac{\gamma}{2 \pi^2} \sum_B \int_0^{k_{fB}} \sqrt{k^2 +
  M^{*2}_B} k^2 dk 
\label{s11}
\end{equation}
and

\begin{equation}
M_B^*=M_B-g_\sigma\sigma_0.
\label{mstarB}
\end{equation}

Whenever stellar matter is considered, $\beta$-equilibrium and charge
neutrality conditions have to be imposed and hence, the inclusion of leptons (generally
electrons and muons) is necessary.  These conditions read:

\begin{equation}
\mu_B = \mu_n -q_B\mu_e, \quad \mu_e =\mu_\mu, \quad \sum_B q_B \rho_B +
\sum_l q_l \rho_l = 0, \label{s25}
\end{equation}
where $\mu_B$ and $q_B$ are the chemical potential and the electrical
charge of the  baryons, $q_l$ is the electrical charge of the leptons, 
$\rho_B$ and $\rho_l$ are the 
number densities of the baryons and leptons.

After the supernova explosion, the remnant is, at first, a
protoneutron star. Before deleptonisation takes place, neutrinos are
also present in the system and in this case, the chemical stability
condition becomes

\begin{equation}
\mu_B = \mu_n -q_B(\mu_e-\mu_{\nu_e}), \quad \mu_e =\mu_\mu.
\end{equation}

In this process, entropy is usually fixed at values compatible with
simulations of neutron star cooling and the lepton fractions reach
values of the order of  0.3-0.4. This scenario is not considered in the present
paper, but examples of this calculation can be seen in \cite{trapped}.

To satisfy the above conditions of chemical equilibrium and charge
neutrality, leptons have to be incorporated in
the system and this is done with the introduction of a free Fermi gas, i.e.,

\begin{equation}
 \mathcal{L}_{lep} = \sum_l \bar{\psi}_l [i\gamma^\mu\partial_\mu -m_l]\psi_l , \label{s3}
 \end{equation}
where the sum runs over the electron and the muon and their
eigenenergies are

\begin{equation}
\quad E_l = \sqrt{k^2 + m^{2}_l} , \label{s5}
\end{equation}
so that their energy density becomes

\begin{equation}
\mathcal{E}_l = \frac{\gamma}{2 \pi^2} \sum_l \int_0^{k_{fl}} \sqrt{k^2 +
  m^{2}_l} k^2 dk . 
\label{s12}
\end{equation}

The total pression of the system can be either obtained separately for
its baryonic and leptonic parts as in the previous section or by
thermodynamics:

\begin{equation}
P = \sum_f \mu_f \rho_f - \mathcal{E}_f , \label{s15}
\end{equation}
where $f$ stands for all fermions in the system and it is common to
define the particle fraction (including leptons) as $Y_f = \frac{\rho_f}{\rho}$.

As already mentioned, an important point is how to fix the
meson-hyperon coupling constants $g_{iB},
 i=\sigma,\omega,\rho$. There are two methods generally used in the
 literature. The first one is phenomenological and is based on the
 fitting of the hyperon potentials \cite{Glen}:

\begin{equation}
U_Y =  g_{\omega B}\omega_0 - g_{\sigma B}\sigma_0 , \label{s17}
\end{equation} 
 which, unfortunately, are not completely established. The only well
 known potential is the $\Lambda$ potential depth $U_\Lambda = -28$
 MeV \cite{Glen2}.
 Common values for the $\Sigma$ and $\Xi$ potentials are 
 $U_\Sigma = +30$ MeV and  $U_\Xi = -18$ MeV \cite{schaffner,expoy}, but their
 real values remain uncertain. 
According to \cite{Glen2}, appropriate values for the meson hyperon
coupling constants defined in eq. (\ref{mesonbaryon}) are obtained if
$\chi_{B \sigma} = 0.7$ and $\chi_{B \omega} = \chi_{B \rho}$ is given by
0.772 for NL3 and 0.783 if another common parametrisation, the
GM1, is used. However, in these cases, the value of
$\chi_{B \rho}$ remains completely arbitrary. 
We have mentioned GM1 here because it is very
often used in the description of neutron star matter since it was one
of the first parameter sets with a high effective mass at the
saturation density ($M^*/M=0.7$) as compared with 0.6 given by NL3,
for instance (see Table \ref{tab_sat}).  This high effective mass helps
the convergence of the codes when the hyperons are introduced because 
eq. (\ref{mstarB}) accounts for a large contribution of the $\sigma_0$ field,
which in turn, carries the information of the scalar densities of
eight baryons.
The situation is very different as the one in nuclear matter, where
the effective mass only carries the $\sigma_0$ field coming from the
nucleonic scalar density. This means that whenever the 8 lightest
baryons are included, the negative contribution in eq. (\ref{mstarB})
can make the nucleon mass reach zero very rapidly if the effective mass is too low. 

Other examples of how to fit these couplings based
on  phenomenological potentials can be seen in
\cite{james1,james2}. The second possibility to choose the
meson-hyperon couplings is based on the relations established among
them by different group symmetries, the most common being SU(3)
\cite{Weiss, su3}  and SU(6) \cite{Luiz2021}.

In the present work we have used the following sets of couplings,
for which $U_\Lambda = -28$ MeV, $U_\Sigma = +30$ MeV and  $U_\Xi = -18$ MeV: 
\begin{itemize}
\item for NL3 and NL3$\omega\rho$: 
$\chi_{\Lambda \sigma} = 0.613$, $\chi_{\Sigma \sigma}
  = 0.460$, $\chi_{\Xi \sigma} =0.317$, 
\item for IUFSU: $\chi_{\Lambda \sigma} = 0.611$, $\chi_{\Sigma \sigma}
  = 0.454$, $\chi_{\Xi \sigma} =0.316$, 
\end{itemize}
 and $\chi_{\Lambda \omega} = \chi_{\Sigma \omega} = 0.667$,
 $\chi_{\Xi \omega} =0.333$ in all cases due to the SU(6) symmetry and
$\chi_{B \rho} =1$ for all hyperons in all cases.

\begin{figure}
\begin{tabular}{cc}
\includegraphics[width=7.cm]{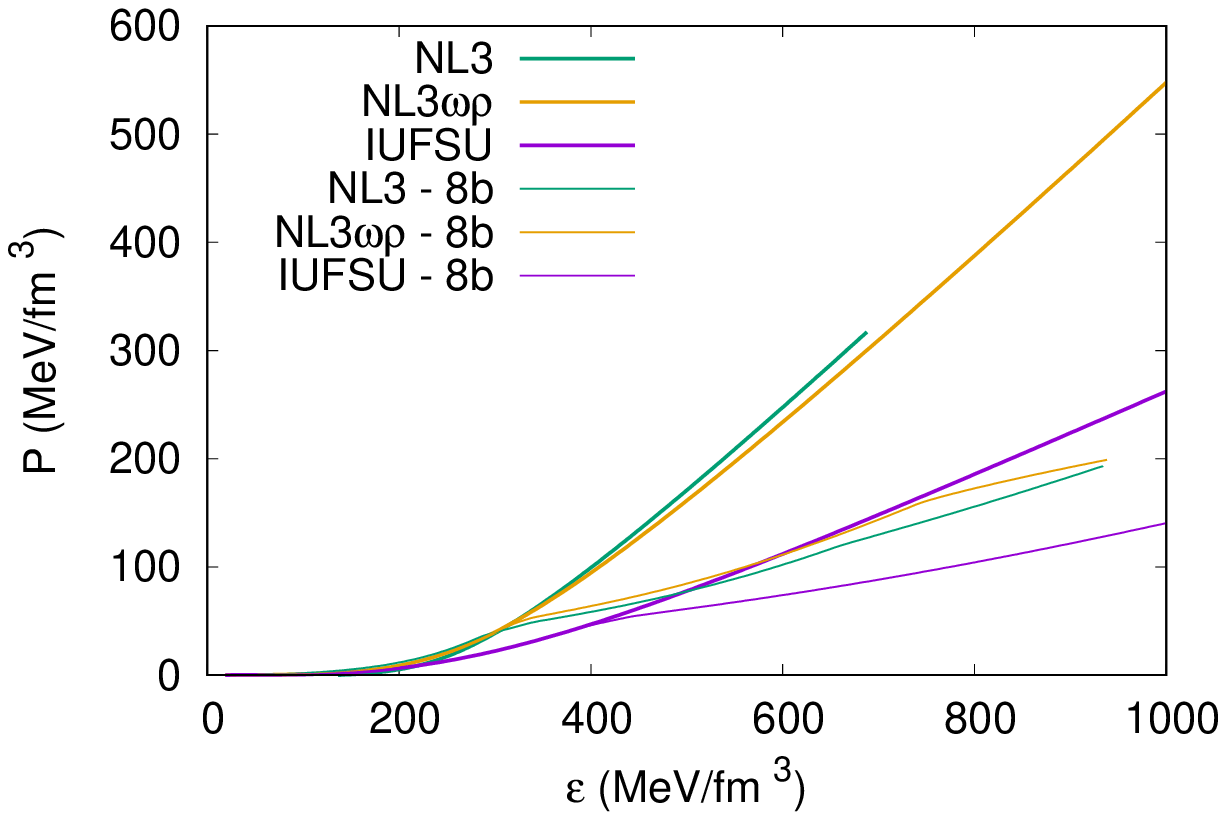} \\
\includegraphics[width=7.cm]{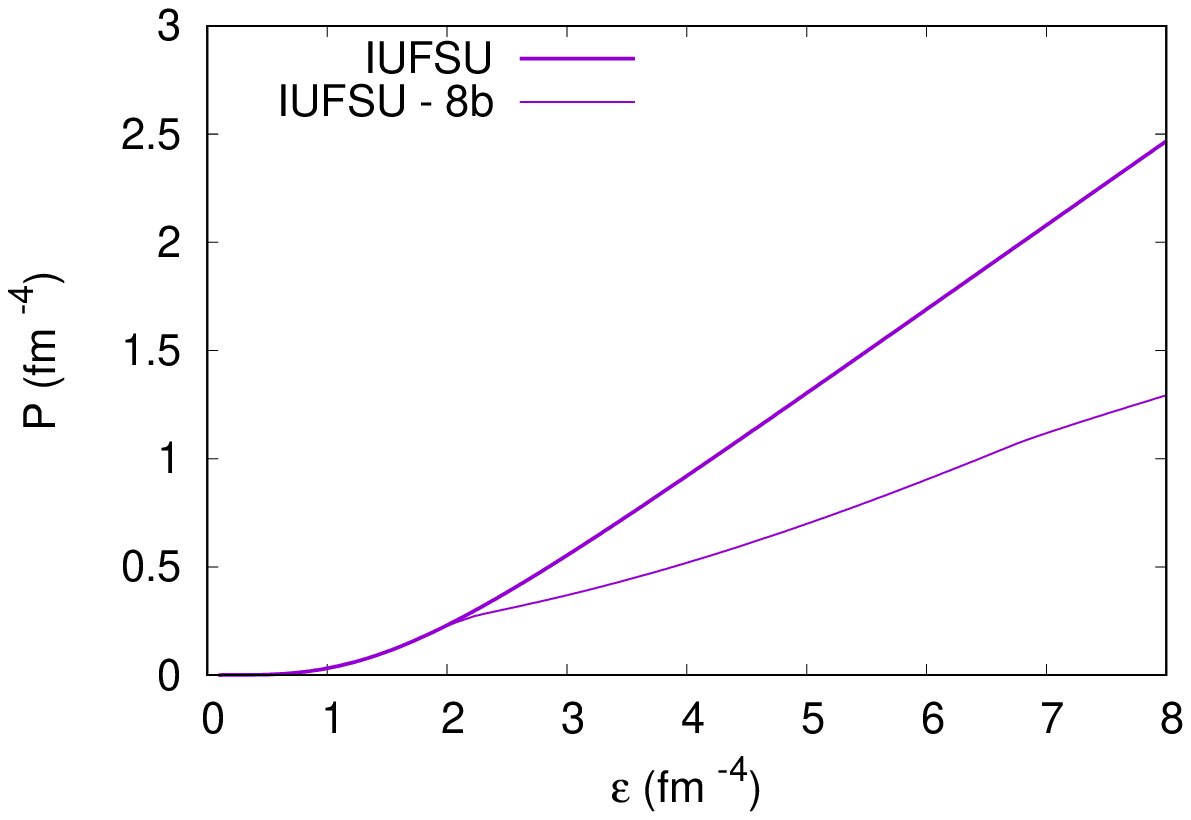} \\
\end{tabular}
\caption{top) Stellar matter EOS obtained with different parametrisations.
  Thick solid lines show EOS with nucleons only  and thin lines represent EOS with
the 8 lightest baryons and
bottom) same EOS for IUFSU with and without hyperons but with
different units for pressure and energy density.} 
\label{eos_estelar}
\end{figure}

In Fig. \ref{eos_estelar} six different EOSs are shown, for the three
parameter sets identified above, with and without the inclusion of the
hyperons. The EOSs for the IUFSU with and without hyperons are
reproduced with different units
(fm$^{-4}$) instead of the more intuitive (MeV/fm$^{3}$) because those
are common units used in stellar matter studies.  Notice that 
$ \hbar c = 197.326$ MeV.fm and Natural units are used in this calculations.

\begin{figure}
\begin{tabular}{cc}
\includegraphics[width=7.cm]{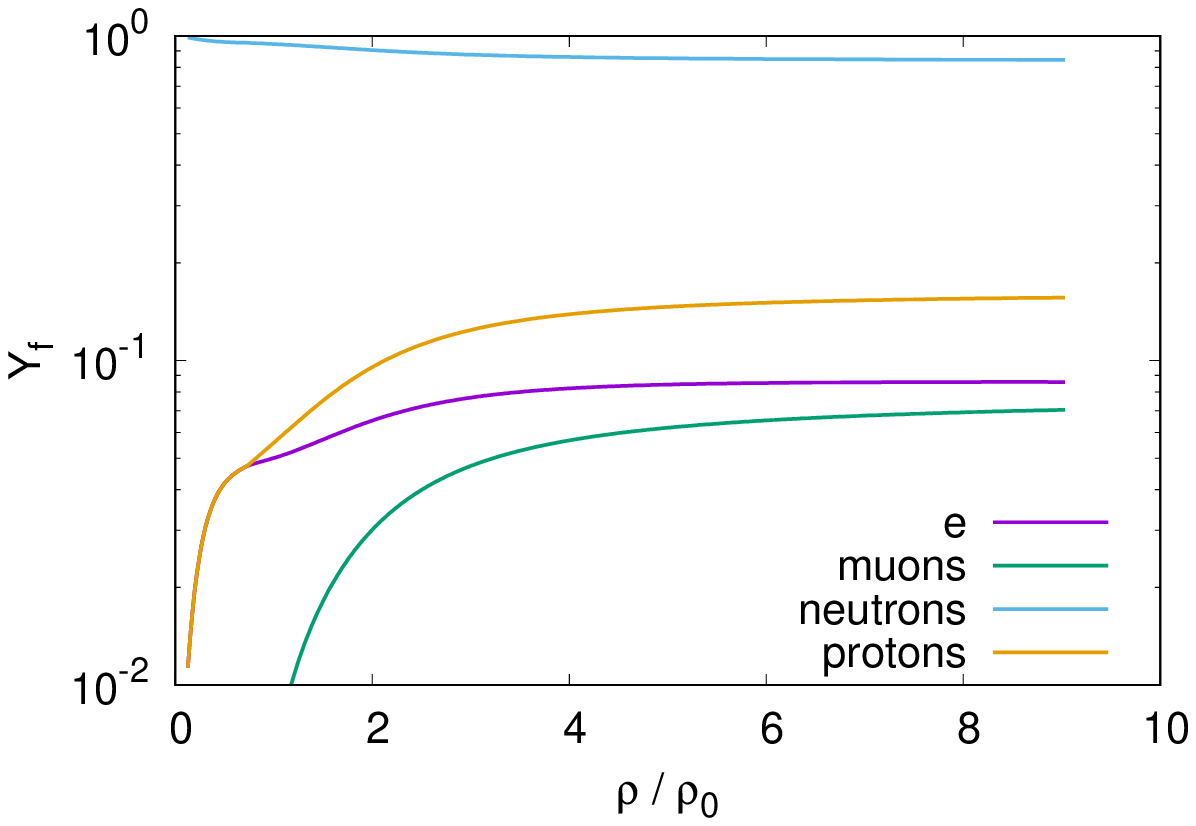} \\
\includegraphics[width=7.cm]{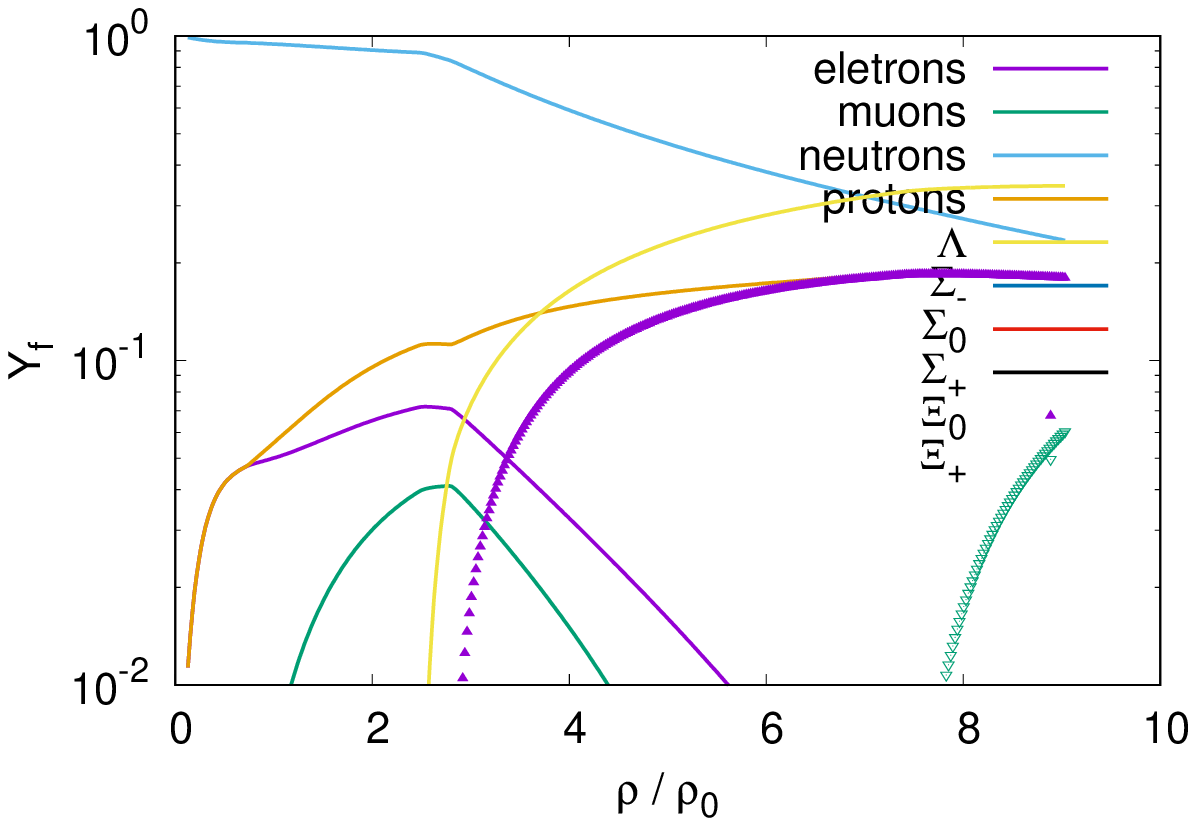} \\
\end{tabular}
\caption{Particle fractions obtained with IUFSU for an EOS with left)
  nucleons only and right) lightest 8 baryons.}
\label{fracao}
\end{figure}

In Fig. \ref{fracao}, the particle fractions obtained with IUFSU are
displayed for the two cases shown in Fig.\ref{eos_estelar} right. 
Notice that when the hyperons are included, these particle
fractions depend on the meson-hyperon couplings discussed above. A
different choice for these couplings would generate different particle
fractions for the same nuclear parametrisation. One
can see that the constituents of the neutron stars change with the
increase of the density, making their core richer in terms of
particles than the region near the crust. From these plots, the
conditions of charge neutrality and chemical equilibrium become clear.

\subsection{The Tolman-Oppenheimer-Volkoff equations}
\label{toveq}

As it was just seen, essential nuclear physics ingredients for astrophysical
calculations are appropriate equations of state (EOS). 
After the EOSs are chosen, they enter as input to the Tolman-Oppenheimer-Volkoff 
equations (TOV) \cite{tov}, which in turn, give as output some macroscopic
stellar properties: radii, masses and central energy densities.
Static properties, as the moment of inertia and rotation rate can
be obtained as well. The EOSs are also necessary in calculations involving the
dynamical evolution of supernova, protoneutron star evolution and
cooling, conditions for nucleosynthesis and stellar chemical
composition and transport properties, for instance.  

The TOV equations were obtained  by Tolman and independently by
Oppenheimer and Volkoff \cite{tov}, as already mentioned, and they read:

\begin{widetext}
\begin{eqnarray}
\frac{dP}{dr}={}&&-G\frac{({\cal E}+P)(M(r)+4\pi
                 P r^3)}{r^2-2M(r)r},\\
={}&& -\frac{G {\cal E} M(r)}{r^2}\left[1 + \frac{P}{{\cal E}}\right]
    \left[1 + \frac{4 \pi r^3 P}{M(r)} \right]
    \left[1-\frac{2GM(r)}{r}\right]^{-1} \nonumber\\
\frac{dM}{dr}={}&&4\pi{\cal E} r^2 ,\nonumber\\
\frac{dM_{Baryonic}}{dr}={}&&4\pi m_n
                   r^2\rho(r)\left[1-\frac{2M(r)}{r}\right]^{-1/2},\nonumber
\label{eqstov}
\end{eqnarray}
\end{widetext}
where $M$ is the gravitational mass, $M_{Baryonic}$ the baryonic mass, $m_n$
is the nucleon mass and $r$ is the radial coordinate and also the
  circumferential radius.
Be aware that
$M_{Baryonic}$ refers to the baryonic mass of the star and it is not
the same as the $M_B$, the individual baryonic masses used to compute the EOS.

The first
differential equation is also shown in such a way that the corrections
obtained from special and general relativity are clearly separated.

\begin{figure}
\begin{center}
\includegraphics[width=0.9\linewidth]{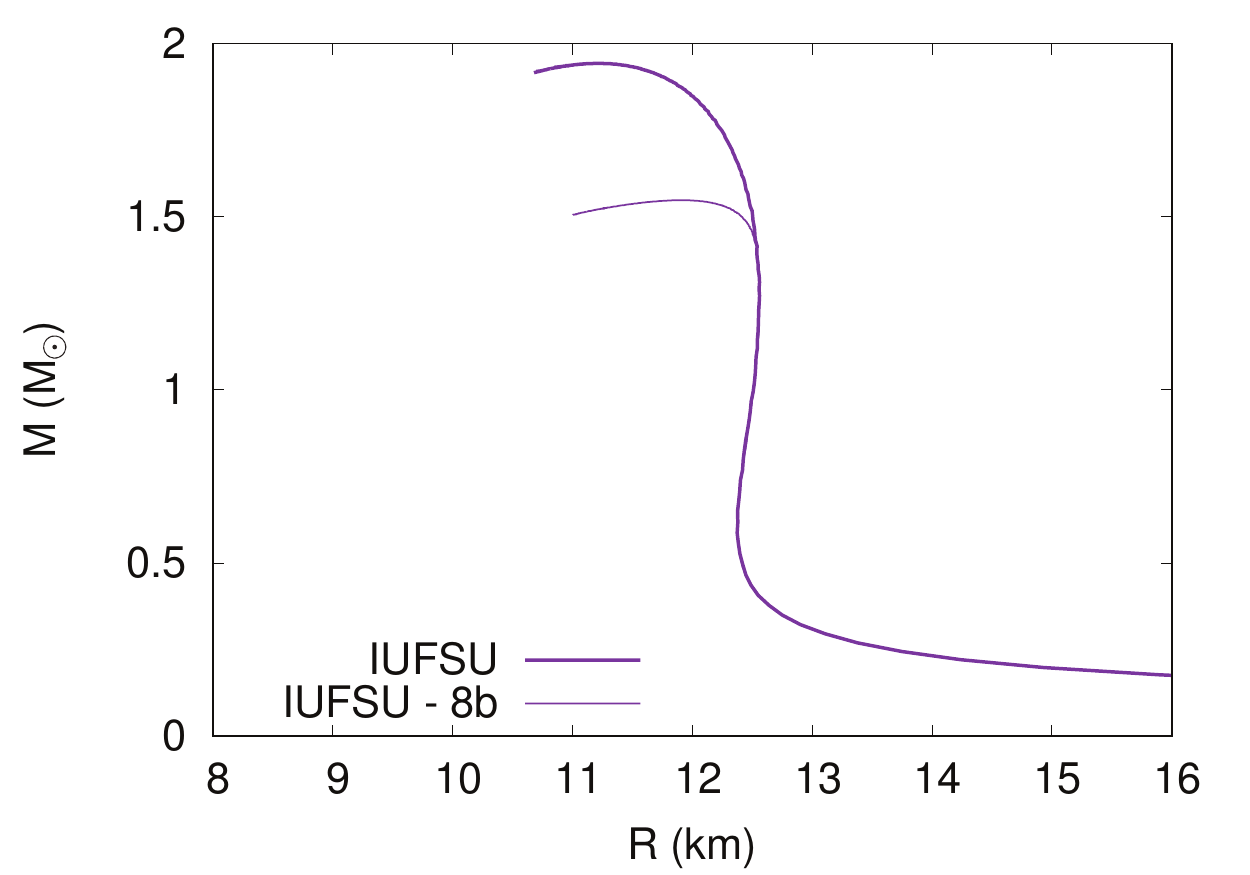} 
\caption{Mass-radius diagram obtained with the IUFSU parametrisation for hadronic matter with (8b) and without hyperons.} 
\label{tov_IUFSU}
\end{center}
\end{figure}

The EOSs shown on the r.h.s. of Fig. \ref{eos_estelar}  are then used as input to the
above TOV equations and the corresponding mass-radius diagram is shown
in Fig. \ref{tov_IUFSU}. Each curve represents a family of stars,
being the maximum point of the curves related to the maximum stellar
mass of the family. By comparing the curves shown in
Fig. \ref{eos_estelar} and Fig. \ref{tov_IUFSU}, one can clearly see
that the harder EOS yields higher maximum mass. Hence, the inclusion
of hyperons makes the EOS softer, as expected, but results in lower
maximum masses. As there is no reason to believe that the hyperons are
not present, this connection of softer EOS with lower neutron star mass 
gave rise to what is known as the {\it hyperon puzzle}. I will go back to
this debate in the next section.

I would like to call the attention of the reader for the values of the
symmetry energy slope  ($L_0$), which has been extensively discussed in
the last years. Although its true value is still a matter
of debate, most studies indicate that it has non-negligible
implications on the neutron star macroscopic
properties~\cite{Rafa,Lopes2014,Tsang,
  Micaela2017,Lattimer2014,Pais2016,Dex2019,Prov2019}.
The slope can be controlled by the inclusion of the $\omega-\rho$
interaction, as can be seen in Table \ref{tab_sat}. In general, the
larger the value of the interaction, the lower the values of the
symmetry energy and its slope \cite{Rafa}. As a general trend, it is
also true that the lower the value of the slope, the lower the radius
of the canonical star, the one with 1.4 $M_\odot$. In Table
\ref{tab_sat} the values of the maximum stellar masses obtained
without the inclusion of hyperons and the radii of the canonical stars
are displayed. Notice, however, that the value of the radius of the
canonical stars depends on the EOS of the crust. To obtain the values
shown in Table \ref{tab_sat}, I used the BPS EOS \cite{bps} for the outer crust
and interpolated the inner crust. As far as the maximum mass is
concerned, the crust barely affects it, since the involved densities
are too low. I will discuss this subject further
when discussing the pasta phase in Section \ref{pasta}. 
Another interesting correlation noticed in \cite{isaac2014} is that
the onset of the charged (neutral) hyperons takes place at lower
(larger) densities for smaller values of the slope.

\subsection{Structure  of neutron stars and observational constraints}
\label{constrain}

Although the internal constitution of a neutron star cannot be
directly tested, it is  reasonably well understood. A famous picture of the NS
internal structure was drawn by Dany Page and can be seen in
\cite{Lattimer2004}.  Close to the surface of the star,
there is an outer and an inner crust and towards the center, 
an outer and an inner core are believed to exist. 
The solid crust is expected to be formed by nonuniform neutron rich
matter in $\beta$-equilibrium. This
inhomogeneous phase is known as pasta phase and calculations predict
that it exists at densities lower than 0.1 fm$^{-3}$, where 
nuclei can coexist with a gas of electrons and neutrons which have dripped out. 
The center of the star is composed of hadronic matter and
the true constituents are still a matter of debate, as one can
conclude from the results presented in the last section.
The fact that the core should contain hyperons is widely accepted,
although this possibility excludes many EOS that become too soft to
explain the existing massive stars, namely,
MSP J0740+6620, whose mass range lies at $2.07 \pm 0.08~M_{\odot}$
~\cite{Cromartie, NICER_news},  PSR J0348+0432 with mass of
$2.01 \pm 0.04~M_{\odot}$~\cite{Antoniadis} and
PSR J1614-2230, which is also a massive neutron
star \cite{Demorest}. Until around 2005, these massive NS had
not been detected and
practically all EOS could satisfy a maximum 1.4 $M_\odot$ star.

\begin{figure}
\begin{center}
\includegraphics[width=0.7\linewidth, angle=270]{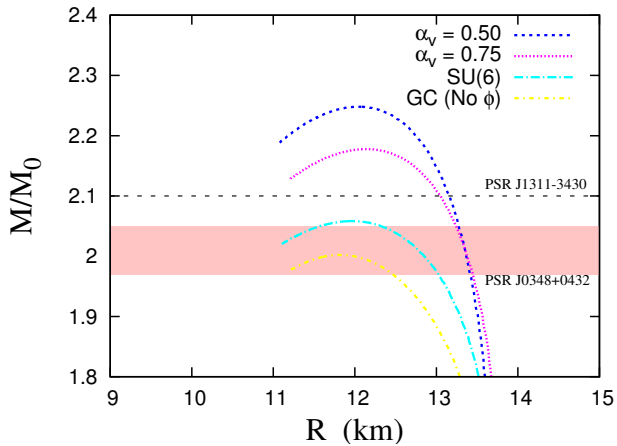}
\caption{\label{figtov} Mass-radius curve for the GM1 parametrisation 
based on \cite{su3}. }
\end{center}
\end{figure}

Since the appearance of hyperons is energetically favorable, different
possibilities were considered in the literature so that the EOS would
be stiffer, as the tuning of the unknown meson-hyperon
coupling constants, for an example. 
Another mechanism that increases the maximum
mass of neutron stars with hyperons in their core is the inclusion of an
additional vector meson that mediates the hyperon-hyperon
interaction~\cite{su3,Weiss}. In Fig. \ref{figtov}, mass-radius curves
 are shown for different hyperon-meson coupling constants of the GM1
 parametrisation \cite{su3}. One can
 see that all choices produce results with high maximum masses,
satisfying the new massive star constraints. I refer the reader to 
\cite{su3} and references therein for explanations on the introduction
of the strange meson channel on the Lagrangian density and the
corresponding strange meson-hyperon couplings. 

As already mentioned in Sec. \ref{astro}, the observation of the
binary neutron star system GW170817 \cite{GW170817}  by the LIGO-Virgo
scientific collaboration and also in the X-ray, ultraviolet, optical,
infrared, and radio bands gave rise to the new era of multi-messenger
astronomy \cite{multi}. The detection of the corresponding
gravitational wave helped the establishment of additional constraints to
the physics of neutron stars. This subject is better discussed next, but
  at this point, I would like to mention that a series of papers based
  on the these constraints imposed restricted values for the
  neutron star radius \cite{Annala2018, Ozel2018, Riley2019, Miller2019,  
Capano2020}, not always compatible among themselves.

The dimensionless tidal deformability, also called tidal
polarisability  and its associated Love number are related to the
induced deformation that a neutron star undergoes by the influence of
the tidal field of its neutron star companion in the binary system.
The idea is analogous to the tidal response of our seas on Earth as a
result of the Moon gravitational field. 
The theory of Love numbers emerges naturally from the theory of tidal
deformation and the first model was proposed in 1909 by Augustus
Love \cite{firstLove} based on Newtonian theory. The relativistic
theory of tidal effects was deduced in 2009 
\cite{Damour,Binn} and since then 
the computing of Love numbers of neutron stars has become a field of
intense investigation.

As different neutron star EOS and related composition have different
responses to the tidal field, the  tidal polarisability  can be used
to discriminate between different
equations of state. A complete overview on the theory of Love numbers
in both Newtonian and General Relativity theories can be found in
\cite{cesar2020}. Here, I show next only the main equations for the
understanding of the constraints on NS.

The second order Love number $k_2$ is given by

\begin{align}
&k_2 =\frac{8C^5}{5}(1-2C)^2[2+C(y_R-1)-y_R]\times
\nonumber\\
&\times\Big\{2C [6-3y_R+3C(5y_R-8)]
\nonumber\\
&+4C^3[13-11y_R+C(3y_R-2) + 2C^2(1+y_R)]
\nonumber\\
&+3(1-2C^2)[2-y_R+2C(y_R-1)]{\rm ln}(1-2C)\Big\}^{-1}.
\label{k2}
\end{align}
where $C= M/R$ is  the star compactness, $M$ and $R$ are the total
mass and total circumferential radius of the star respectively
and $y_R = y(r = R)$, which
is obtained from

\begin{equation}
r \frac{dy}{dr} + y^2 + y F(r) + r^2Q(r)=0.
\label{ydef}
\end{equation}
\noindent Here the coefficients  are given by

\begin{equation}
F(r) = [1 - 4\pi r^2({\cal E} - P)]/E
\end{equation}
\noindent and

\begin{align}
Q(r)=& 4\pi \left[5{\cal E} + 9P + ({\cal E} + P)\left(\frac{\partial 
P}{\partial{\cal E}}\right)-\frac{6}{4\pi r^2}\right]/E 
\nonumber\\ 
&- 4\left[ \frac{m+4\pi r^3 P}{r^2 E} \right]^2,
\end{align}

\noindent where $E = 1-2m/r$, ${\cal E}$ and $P$ are the energy
density and pressure profiles inside the star. Notice that
Eq. (\ref{ydef}) has to be solved coupled to the TOV equations.

Finally, one can obtain the dimensionless tidal deformability
$\Lambda$, which is connected to the compactness parameter $C$ through 

\begin{equation}
\Lambda= \frac{2k_2}{3C^5}.
\label{dtidal}
\end{equation}

\begin{figure}
\begin{tabular}{cc}
\includegraphics[width=7.cm]{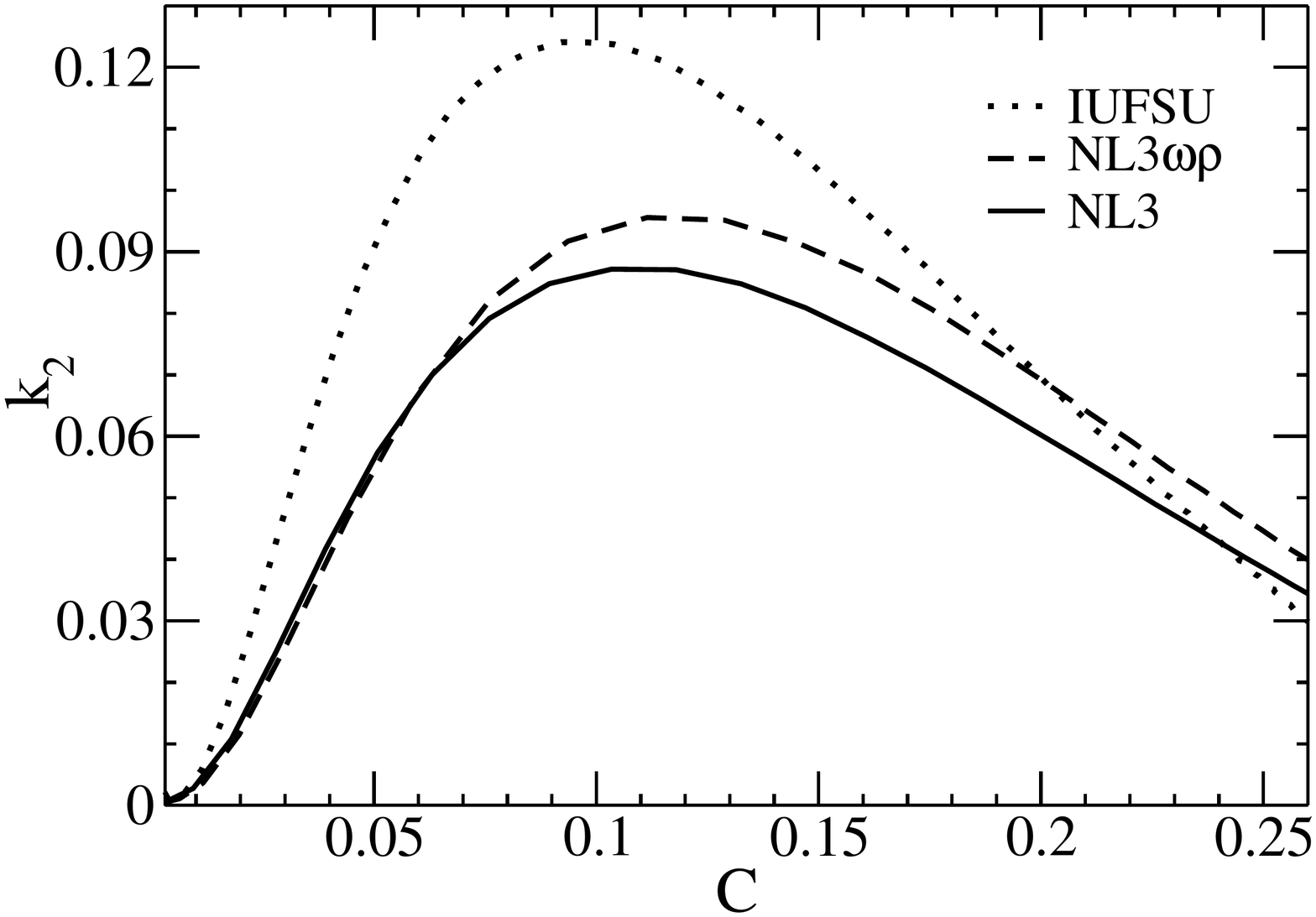} \\
\includegraphics[width=7.cm]{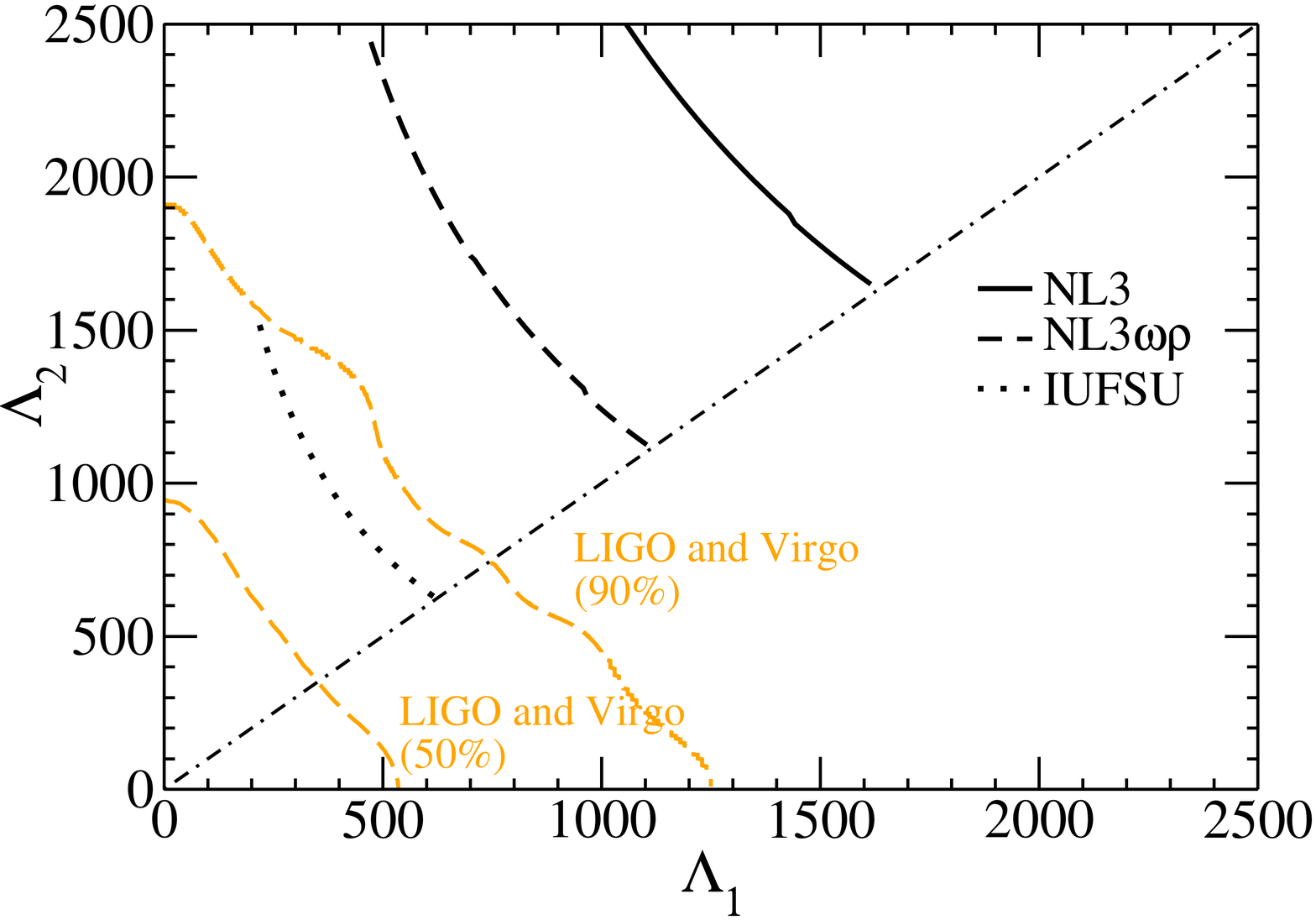} \\
\end{tabular}
\caption{top) Love number as a function of the compactness and bottom)
Tidal deformabilities of both NS in the binary system before the merger.}
\label{tidal}
\end{figure}

In Fig. \ref{tidal}, the second order Love number as a function of the
compactness is shown for the 3 equations of state discussed in Section
\ref{rmf}, as well as the corresponding  tidal deformabilities
$(\Lambda_1,\Lambda_2)$ for the binary system ($M_1,M_2$),
with $M_1 > M_2$. The plots are calculated from the equation for the chirp mass

\begin{equation}
M_{chirp} = (M_{1}M_{2})^{3/5}(M_{1}+M_{2})^{-1/5},
\end{equation}

\noindent and the diagonal dotted line corresponds to the case $M_1 = M_2$.
The lower and upper dashed lines correspond to LIGO/Virgo collaboration 
50$\%$ and 90$\%$ confidence
limits respectively, which are obtained from the GW170817 event.
The EOS used to obtain these curves do not include hyperons to avoid
the uncertainties related to the meson-hyperon couplings. It is
important to mention the matching procedure used to compute the Love
number and the tidal polarisabilities. The outer crust is a BPS EOS,
the inner crust is a polytropic function which interpolates between
the outer crust and the core. A detailed explanation is given in
\cite{skyrmeGW}, sections 2.2 and 2.3. More advanced crustal EOSs
  are available \cite{Pearson2020, Fantina} and I discuss the
  sensitivity of some results on the crust model later on.
One can see from these figures
that the Love numbers are very different for the three models and so
are the tidal polarisabilities, the NL3 and NL3$\omega\rho$ not being
able to reproduce the GW170817 data satisfactorily. Actually, this
behaviour of the NL3 and NL3$\omega\rho$ had already been observed in 
\cite{constanca2018}, but one should notice that in
\cite{constanca2018}, the confidence lines were taken from a
preliminary version of the LIGO/Virgo data \cite{GW170817} while in the
present paper, they are taken from \cite{GW170817_new},  where the
consideration of massive stars was neglected. 

Another important constraint concerns the radii of the canonical
stars, the ones with $M=1.4$  M$_\odot$. According to the LIGO/Virgo
collaboration, the tidal polarisability of canonical stars should lie
in the range $70 \leq \Lambda_{1.4} \leq 580$
\cite{GW170817_new}, a restriction that imposed a constraint
to the radii of the
corresponding stars, which should lie in the range $10.5$ km $ \leq
R_{1.4 M_{\odot}} \leq 13.4$ km. This constraint, which does not take
into account a maximum stellar mass of 1.97 M$_\odot$, only excludes the
NL3 parameter set from the ones we are analysing (see Table
\ref{tab_sat}), exactly the one that was shown not to describe nuclear
bulk properties well enough.  But...the history has become more
complicated: a recently published paper concludes that the canonical
neutron star radius cannot be larger than 11.9 km~\cite{Capano2020}. 
If such small radius is confirmed, it could imply in
a revision of the EOSs or of the gravity theory itself, as done in
\cite{Clesio}. 
Notice, however, that this small radius is
in line with older works that predicted that
  the maximum mass of a canonical star should be 13.6
  km \cite{Annala2018} and \cite{Ozel2018}, whose authors claimed that
any NS, independently of its mass, should bear a radius smaller than
13 km. Moreover, the new information
sent by NICER \cite{NICER_news} supports the evidence that the detected
massive PSR J0740+6620 has a radius of the order of $12.35 \pm 0.75$ km 
and that a star with a mass compatible with a canonical star, J0030+0451,
has a radius of the order of  $12.45 \pm 0.65$ km
\cite{NICER_canonical},  or $12.71^{+1.14}_{1.19}$ km
  \cite{Riley2019} or even $13.02^{+1.24}_{1.06}$ km
  \cite{Miller2019}, depending on the analysis performed.
These recent detections point to the fact that the radii of canonical
and massive stars are of the same order and this feature is not easily
reproduced by most EOSs.
On the other hand, one of the analysis of the results from the PREX
 experiment implies that
$13.25$ km $ \leq R_{1.4 M_{\odot}} \leq 14.26$ km corresponding to
a tidal polarisability in the range
$642  \leq \Lambda_{1.4 M_{\odot}} \leq 955$ \cite{PiekarewiczPREX},
  also much higher than the above mentioned value obtained from
  GW170817 data. Notice, however, that the recent PREX results seem to
  contradict previous understandings on the softness of the symmetry
  energy \cite{Piekarewiczpola}. Hence, the sizes of these objects are still a source
  of debate. One of the conclusions in \cite{PiekarewiczPREX} is that
  a precise knowledge of the crust of these compact objects may help
  to minimize the systematic uncertainties of these results.

A detailed analysis of the relativistic mean field models shown to be
consistent with all nuclear bulk properties in \cite{relat} according
to the masses and radii they yield when applied to describe NS,
can be found in \cite{NS_RMF}. Thirty four models were analysed and only
twelve were shown to describe massive stars with maximum masses in the
range $1.93 \le M/M_\odot \le 2.05$ without the inclusion of
hyperons. In another paper \cite{GW_RMF}, the very same models were
confronted with the constraints imposed by the LIGO/Virgo
collaboration. In this case, 24 models were shown to satisfy
them. However, only 5 models could, at the same time describe massive
stars and constraints from GW170817. These studies did not use EOSs
with hyperons, what poses an extra degree of complication due to the
uncertainties on the meson-hyperon coupling constants. Looking at the
three sets used in the present work, one can clearly see the
difficulty. The two models that can describe massive stars are outside
the range of validity of the GW170817 tidal deformabilities. On the
other hand, IUFSU gives a mass a bit lower than desired, a deficiency that can be correct
with some tuning.

Another aspect that deserves to be mentioned refers to the inclusion
of $\Delta$ baryons in the EOS. If they are  considered as a
possible constituent of neutron stars, at least with the parametrisations
studied (GM1 and GM1$\omega\rho$), no ``$\Delta$ puzzle''  is
observed \cite{Veronica_Kauan}.

\subsection{The importance of the inner  and outer crusts}
\label{pasta}

When examining neutron star merger, the coalescence time is determined
by the tidal polarisability, which as already explained, is a direct response of the tidal
field of the companion that induces a mass quadrupole. This scenario suggests that
the neutron star crust should play a role in this picture. If one
looks at the famous figure drawn by Dany Page \cite{Lattimer2004},
one can see that the crust is divided in two pieces, the outer and the
inner crust, the latter being the motive for the present section. It
may include a {\it pasta phase}, the result of a frustrated
system in which there is a unique competition between the Coulomb and
the nuclear interactions,  possible at very low baryonic
densities. In the simplest interpretation of the
geometries present in the pasta phase, they are known
as droplets (3D), rods (2D) and slabs (1D) 
and their counterparts (bubbles, tubes and slabs) are also possible.
 Much more sophisticated geometries such as
waffles, parking garages and triple periodic minimal surface  have
been proposed \cite{schneider2018, newton2009,schuetrumpt2019},
but I next describe only the more traditional picture.

The pasta phase is the dominant matter configuration if its free
energy  (binding energy at $T=0$) is lower than its corresponding 
homogeneous phase. Depending on the model, the used parametrisation and the
temperature \cite{pasta}, typical pasta densities lie between 0.01 and 0.1
fm$^{-3}$. Different approaches are used to compute the pasta phase structures:
the coexisting phases (CP) method, the Thomas-Fermi approximation, numerical
simulations, etc. For detailed calculations, one can look at
\cite{pasta}, \cite{maruyama}, \cite{Schneider2016}, for instance. In
what follows, I only show the main equations used to build the pasta
phase with the CP method.

According to the Gibbs conditions, both pasta phases have the
same pressure and chemical potentials for proton and neutron and, at a
fixed temperature, the following equations must be solved simultaneously:

\begin{equation}
P^I=P^{II},
\label{p1e2}
\end{equation}

\begin{equation}
\mu_p^I=\mu_p^{II},
\end{equation}

\begin{equation}
\mu_n^I=\mu_n^{II},
\end{equation}
\begin{equation}
f(\rho^{I}_p-\rho_e)+(1-f)(\rho^{II}_p-\rho_e)=0.
\end{equation}
where $I$ ($II$) represents the high (low) density region, $\rho_p$ is
the global proton density, $\rho_e$ is the electron density taken as
constant in both phases
and $f$ is the volume fraction of the phase $I$, that reads

\begin{equation}
f=\frac{\rho-\rho^{II}}{\rho^I-\rho^{II}}.
\end{equation} 
The total hadronic matter energy reads:
\begin{eqnarray}
{\cal E}_{matter}=f{\cal E}^I+(1-f){\cal E}^{II}+{\cal E}_e,
\label{energymat}
\end{eqnarray}
where ${\cal E}^I$ and ${\cal E}^{II}$ are the energy densities of
phases $I$ and $II$ respectively and  ${\cal E}_e$ is the energy
density of the electrons, included to account for charge neutrality. 
The total energy can be obtained by adding the surface and Coulomb terms to the matter energy in Eq. (\ref{energymat}),
\begin{eqnarray}
{\cal E}={\cal E}_{matter}
  +{\cal E}_{surf}+{\cal E}_{Coul}.
\label{totalener}
\end{eqnarray}
Minimizing ${\cal E}_{surf}+{\cal E}_{Coul}$ with respect to the size of the droplet/bubble, cylinder/tube or slabs, we obtain  \cite{maruyama}
${\cal E}_{surf}=2 {\cal E}_{Coul}$ where

\begin{equation}
{\cal E}_{Coul}=\frac{2\alpha}{4^{2/3}}(e^{2}\pi \Phi)^{1/3}\left[ \sigma^{surf} D(\rho^{I}_p 
-\rho^{II}_p)\right] ^{2/3},
\end{equation}
with $\alpha=f$ for droplets, rods and slabs, and $\alpha=1-f$ for
tubes and bubbles. The quantity $\Phi$ is given by 
\begin{eqnarray}
\Phi=\left\lbrace\begin{array}{c}
\left(\frac{2-D \alpha^{1-2/D}}{D-2}+\alpha\right)\frac{1}{D+2}, \quad
                  D=1,3 \\
\frac{\alpha-1-\ln \alpha}{D+2},  \quad D=2 \quad 
\end{array} \right. 
\end{eqnarray}
where, $\sigma^{surf}$ is the surface tension, which measures the
energy per area necessary to create a planar interface between the two
regions. The surface tension is a crucial quantity in the pasta
calculation and it is normally parametrised with the help of more
sophisticated formalisms. Another important aspect is that the
 pasta phase is only present at the low-density regions of the neutron
 stars, and in this region, muons are not present, although they are
 present in the EOS that describes the homogeneous region.

\begin{figure}
\begin{center}
\includegraphics[width=0.7\linewidth]{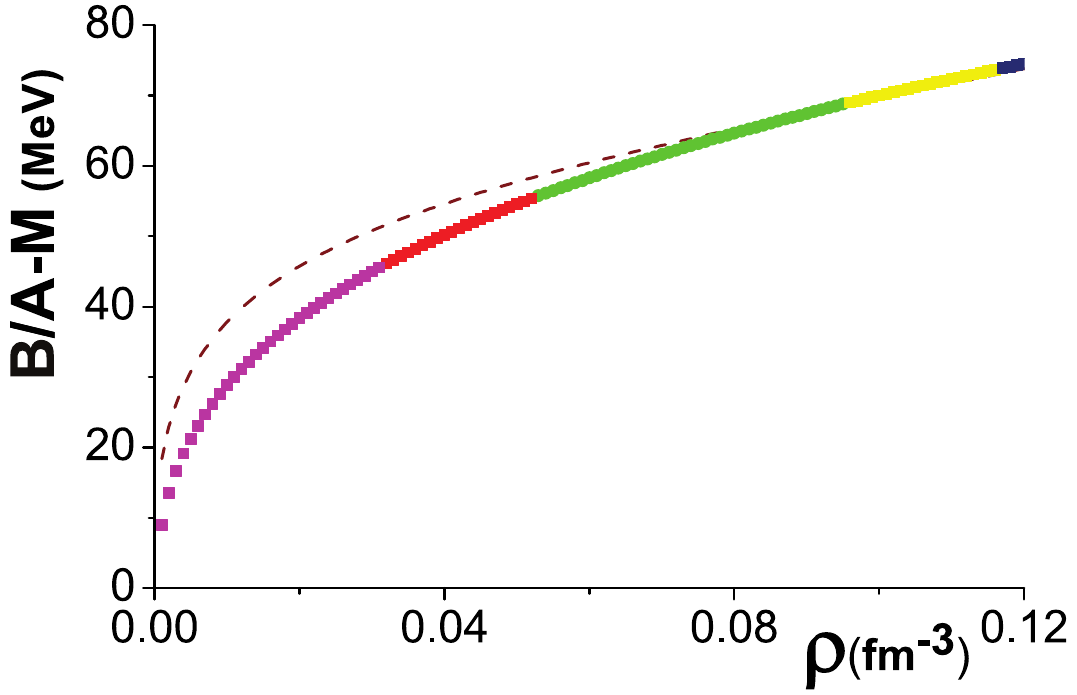}
\caption{\label{fig8} $npe$ matter binding energy obtained with the CP
method and NL3 parametrisation \cite{PRC55-540}. Figure taken from \cite{pasta}. }
\end{center}
\end{figure}

\begin{figure}
\begin{center}
\begin{tabular}{cc}
\includegraphics[width=0.5\linewidth]{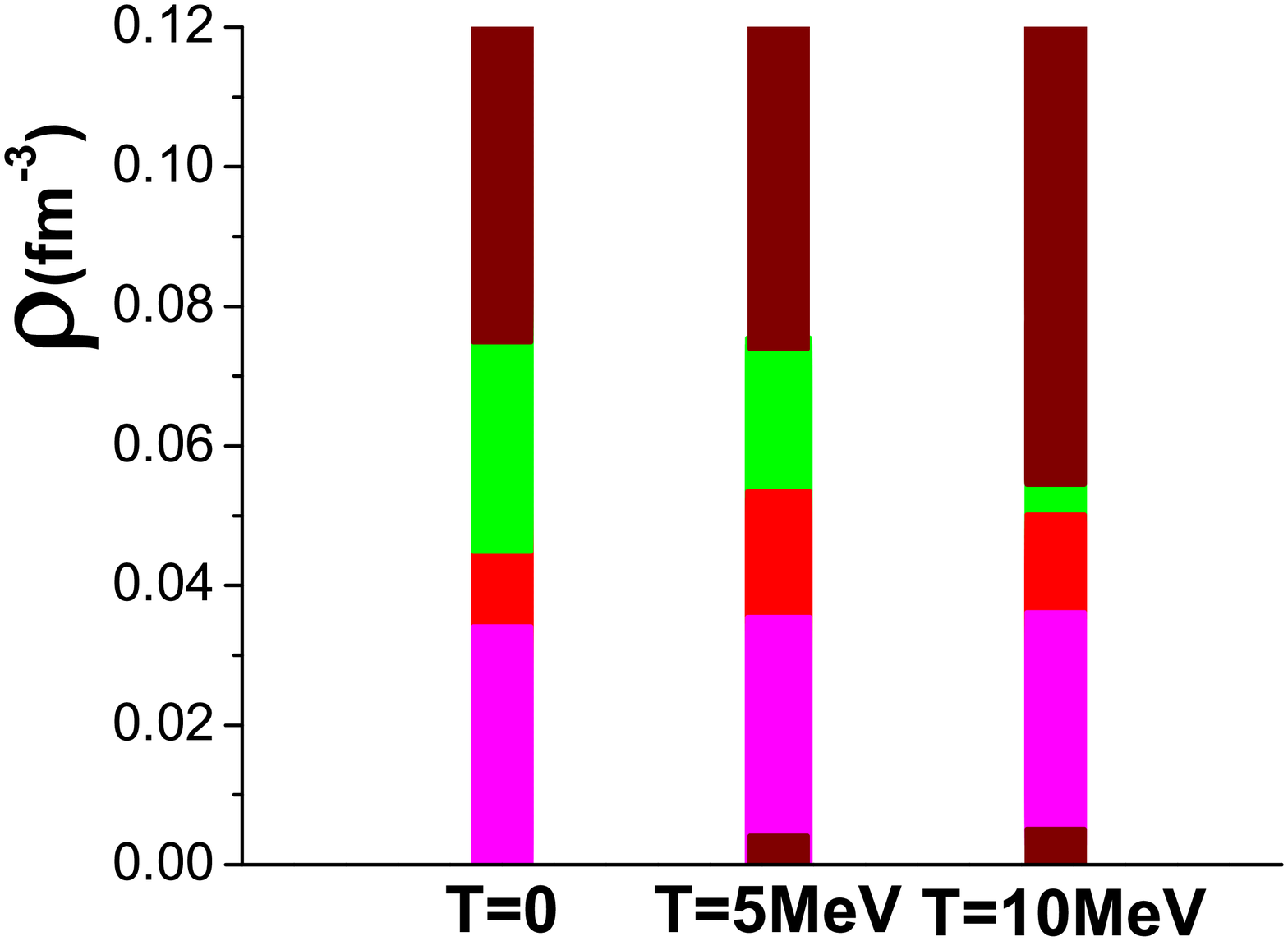} &
\includegraphics[width=0.5\linewidth]{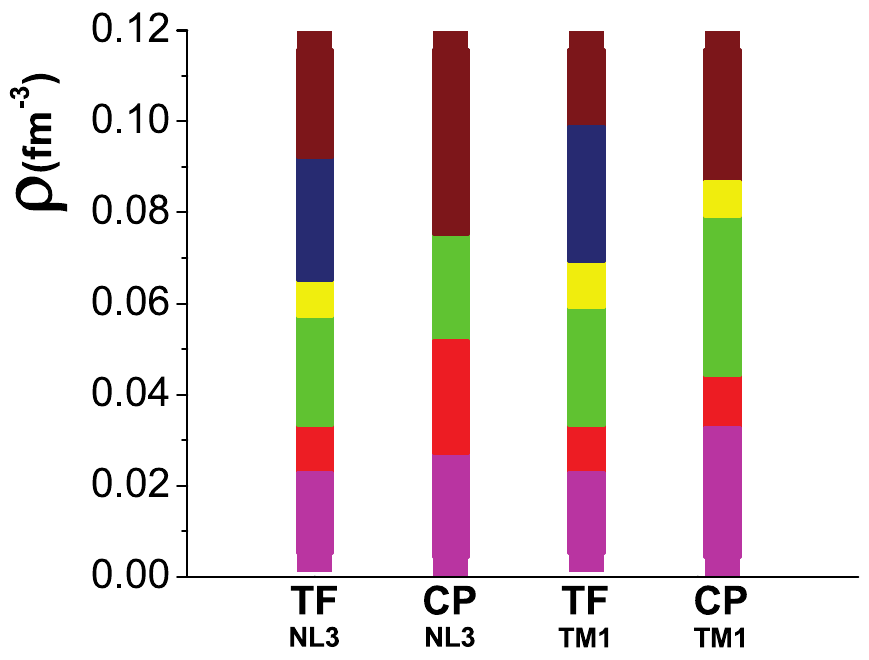} \\
\end{tabular}
\caption{\label{fig9} Phase diagrams obtained with Left) NL3 
  parametrisation and CP method for $Y_p=0.5$. From bottom to top the colors
  represent homogeneous phase ($T=5$ and 10 MeV only), droplets, rods and homogeneous phase.
Right) NL3 and TM1 parametrisations with CP and TF methods for
$Y_p=0.3$ and $T=0$. 
Figures taken from \cite{pasta}.}
\end{center}
\end{figure}

In Fig. \ref{fig8}, I plot the binding energy of the homogeneous
matter (dashed line) as compared with the pasta phase
binding energy (solid line with different colours representing the
different structures). One can see that the pasta phase binding energy
is lower up to a certain density,
when the homogeneous phase becomes the preferential state. 
In Fig. \ref{fig9}, I show various phase diagrams obtained with the CP and TF
methods for  fixed proton fractions at different
temperatures. As the temperature increases, the pasta phase shrinks.
Here I have mentioned the TM1 parameter set \cite{tm1}, 
not used before in the present work,
but also quite common in the literature. The purpose is only to show
that different approximations and different parametrisations
result in different internal structures with different
transition densities from one phase to another.

But then, what is the influence of the pasta phase on the calculation of the tidal
polarisability and if this structure is not well determined, how much
its uncertainty contributes to the final calculations? This problem
was tackled in \cite{GW_QMC} and, as the model used in that paper is quite different
from the RMF models we use in the present work, we do not include any
figures, but it is fair to say that the contribution is indeed
minor. In \cite{GW_QMC}, the BPS EOS was used for the outer crust. For
the inner crust, two possibilities were considered: the existence of
the pasta phase and a simple interpolation between the outer crust and
the core.  It was observed that, although the explicit inclusion of
the pasta phase affected the Love number in a visible way, it almost did not change
the tidal polarisabilities, a result that corroborated the findings in
\cite{piekarewicz}. These results can be explained by the fact that,
for a fixed compactness, even if the Love number is sensitive to the
inner crust structure, the tidal polarisability scales with the fifth
power of $C$ and hence, the influence is small.

\begin{figure}
\begin{center}
  \begin{tabular}{cc}
    \includegraphics[width=0.9\linewidth]{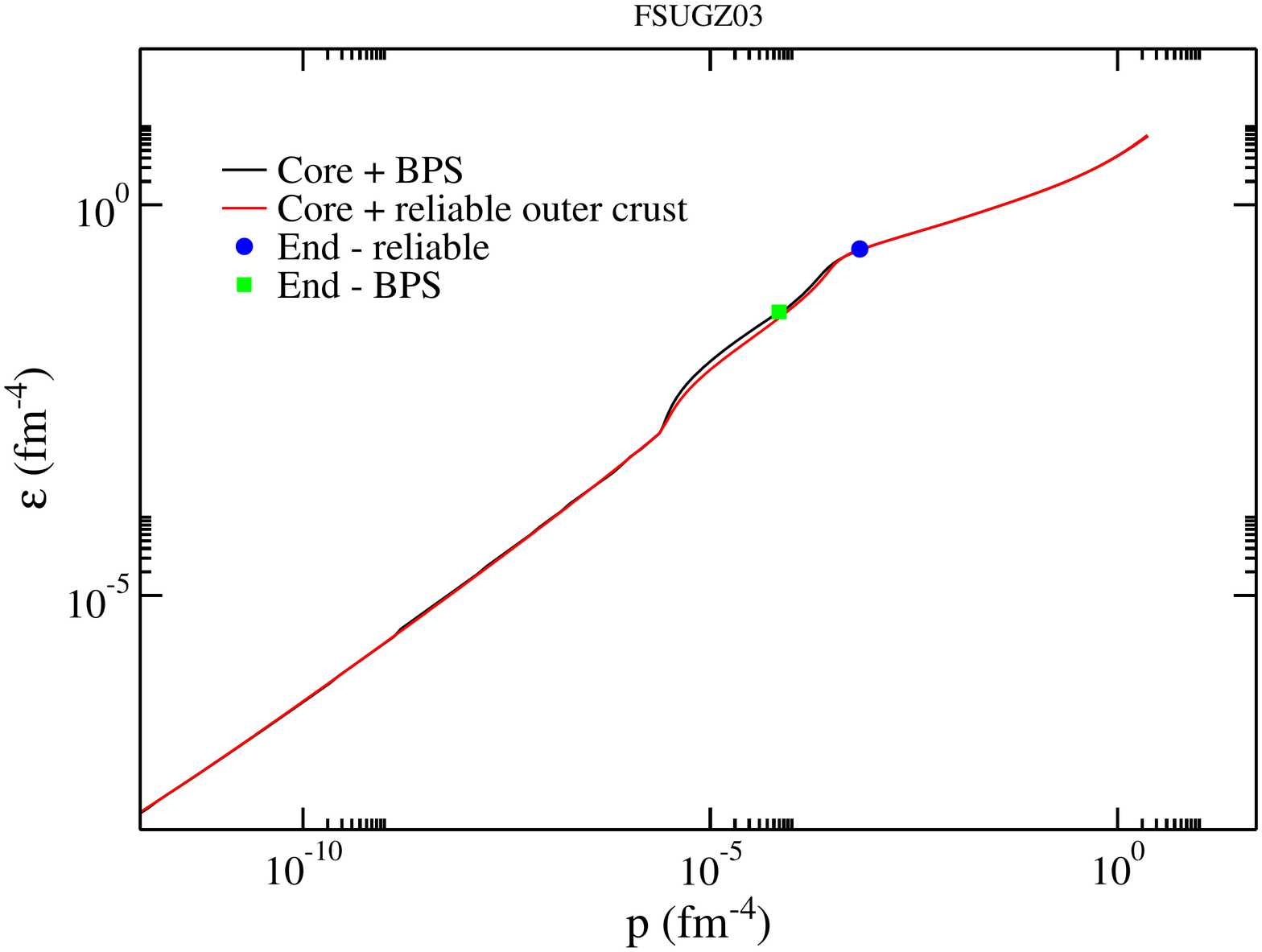} \\
    \includegraphics[width=0.9\linewidth]{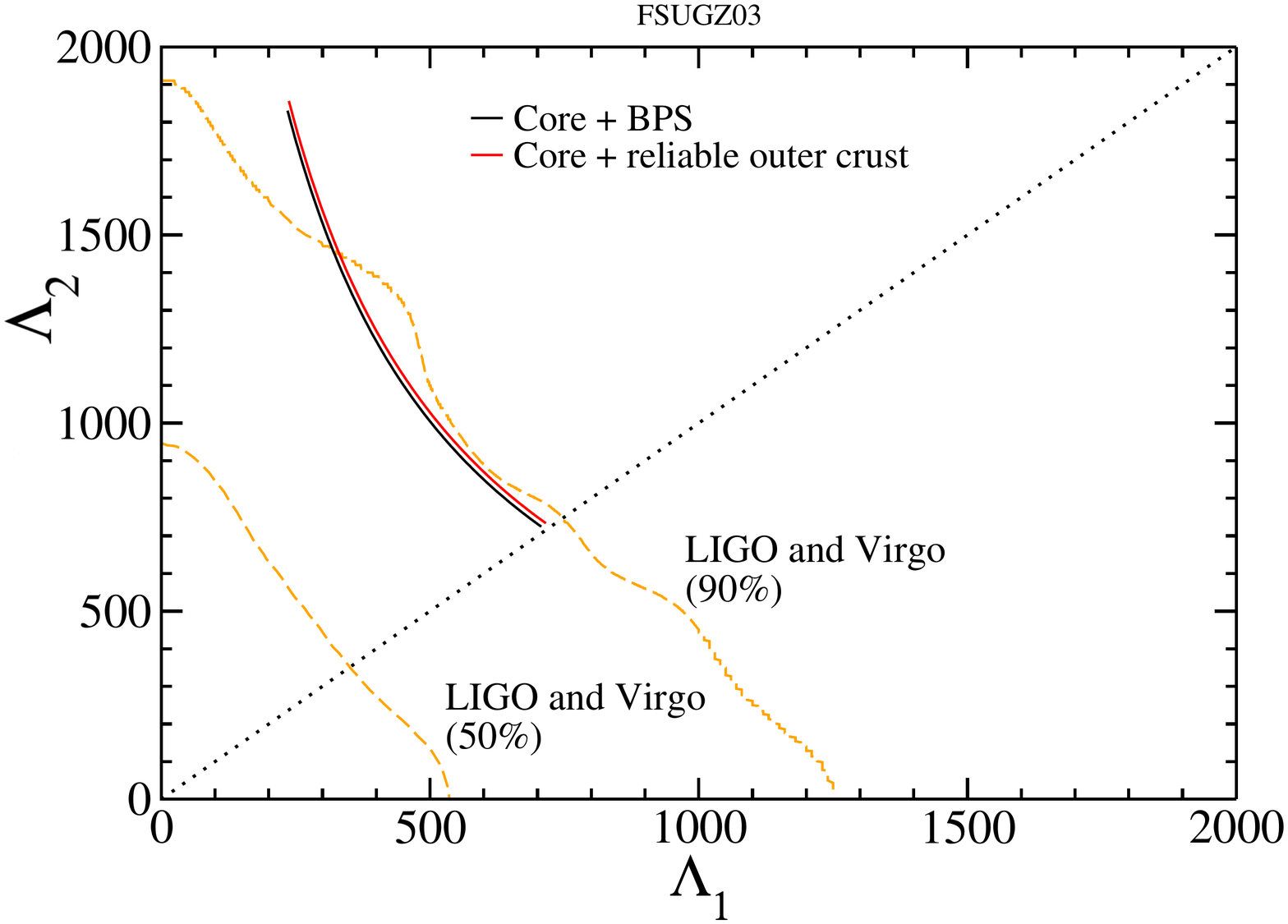}\\
  \end{tabular}
  \caption{top) EOSs obtained with the outer crust described by BPS
    and by a reliable model \cite{Fantina}; bottom) Tidal
    polarisabilities for both NSs with the EOSs shown on the right.}  
  \label{outer_test}
\end{center}
\end{figure}

And what about the outer crust? Indeed, in this case, the tidal
effects should be even more sensitive. In what follows, I test
how much the use of a modern EOS for the outer crust, which we call
{\it reliable} \cite{Fantina}
change the results as compared with the BPS
generally used and mentioned below. A modified version of the IUFSU
model known as FSUGZ03 \cite{fsugz03} was used to plot 
Figure \ref{outer_test} and we trust the
qualitative results would be the same for any other
parametrisation. In this Figure, the outer crust is linked directly to
the core EOS, as seen on the top. Log-scale is used because the
differences cannot be seen in linear scale.  Then,
the different prescriptions are
used to compute the tidal polarisabilities shown on the
bottom. Once again, one can see that the influence is very small. 

Although we have seen that  neither the outer nor the inner crust
alter significantly the tidal polarisabilities, they do have an
impact, which was quantified in \cite{marcio1,marcio2}. The authors of
these works concluded that the impact of the crust EOS is not larger
that 2\%, but the matching procedure (crust-core) can account for a 5\% difference 
on the determination of the low mass NS radii and up to 8\% on the
tidal deformability.
In another recent work \cite{Luiz_crust}, the inner crust was
parametrised in terms of a polytropic-like EOS and the sound velocity
and canonical star radii were computed. EOS for the inner crust with different sound
velocities produced radii with up to 8\% difference when the same EOS
was used for the core.

\begin{figure}
\begin{center}
\includegraphics[width=0.9\linewidth]{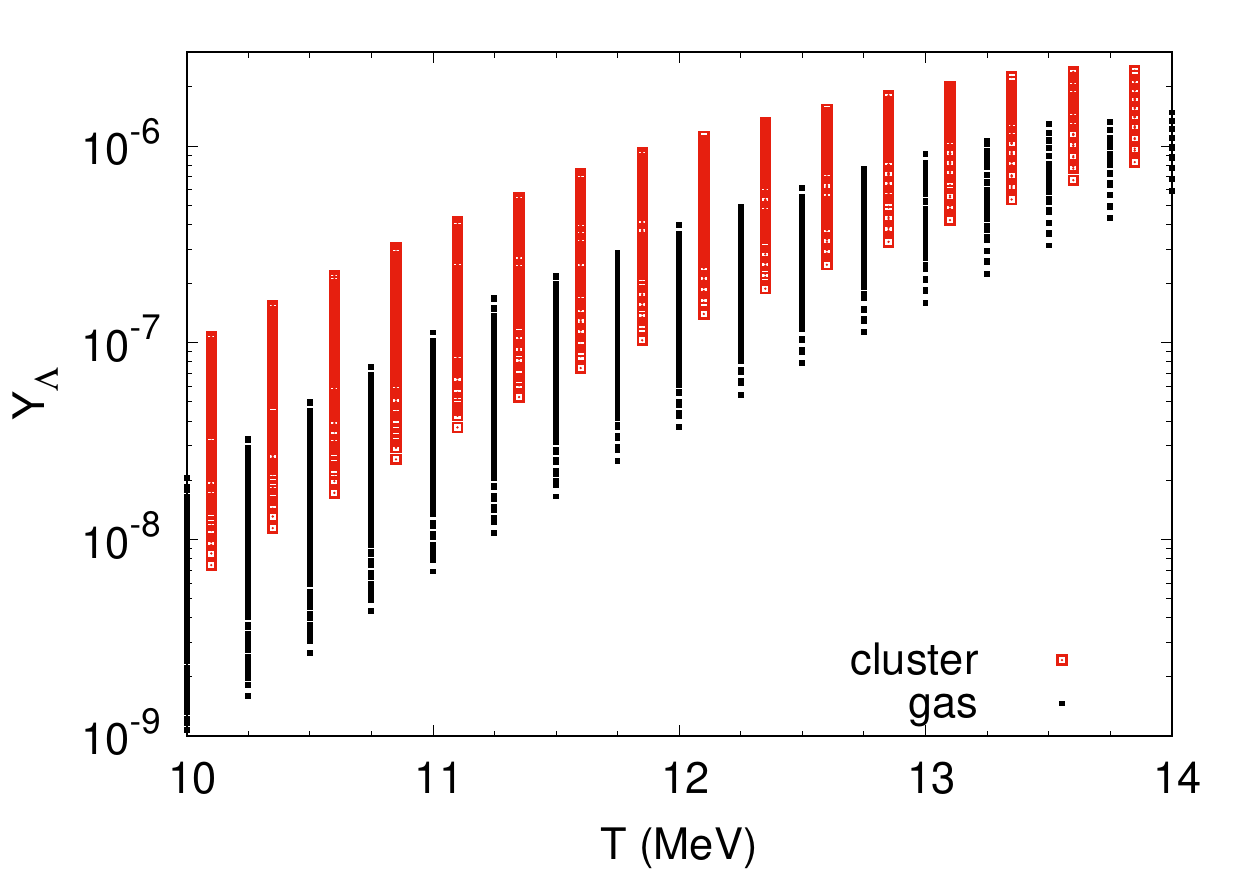}
\caption{ $\Lambda$ fraction as a function of temperature in the
  cluster and gas phases with the NL3$\omega\rho$ parametrisation \cite{Horo}.
Figure taken from \cite{hyperons}. }
\label{hyp}
\end{center}
\end{figure}

Albeit the fact that present results show that the inclusion of  the
pasta phase is not essential when the above discussed
macroscopic properties of NSs are computed, it may indeed be important
for the thermal \cite{thermal} , magnetic evolution
\cite{magnetic1,magnetic2} and neutrino diffusion of NS
\cite{opacity1, opacity2},
processes that take place at different epochs.
Hence, being able to handle properly the pasta phase structure is
still a matter or concern. The first issue worth discussing is the possible existence of 
baryons that are more massive than nucleons and carry strangeness in
the pasta phase. In \cite{hyperons}, it was verified that the $\Lambda$ hyperons, 
can indeed be present, although in small amounts, as seen in
Fig.\ref{hyp}, where the $\Lambda$ fraction is shown as a function of
temperature in phases $I$ (clusters) and $II$ (gas). For the
parametrisation used, NL3$\omega\rho$, the pasta phase disappears at
T=14.45 MeV, the $\Lambda$s being present for electron fractions
ranging from 0.1 to 0.5 and in quantities larger than $10^{-11}$ for
$T>7$ MeV. The $\Xi^-$ can also be found, but in much
smaller amounts, of the order of  $10^{-12}$.

\begin{figure}
\begin{center}
  \begin{tabular}{cc}
    \includegraphics[width=0.8\linewidth]{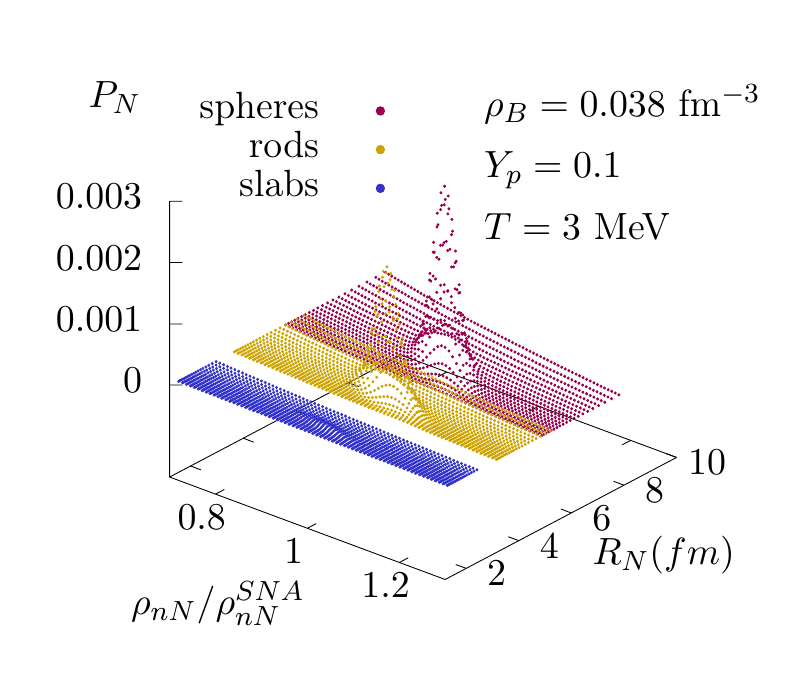} \\
    \includegraphics[width=0.8\linewidth]{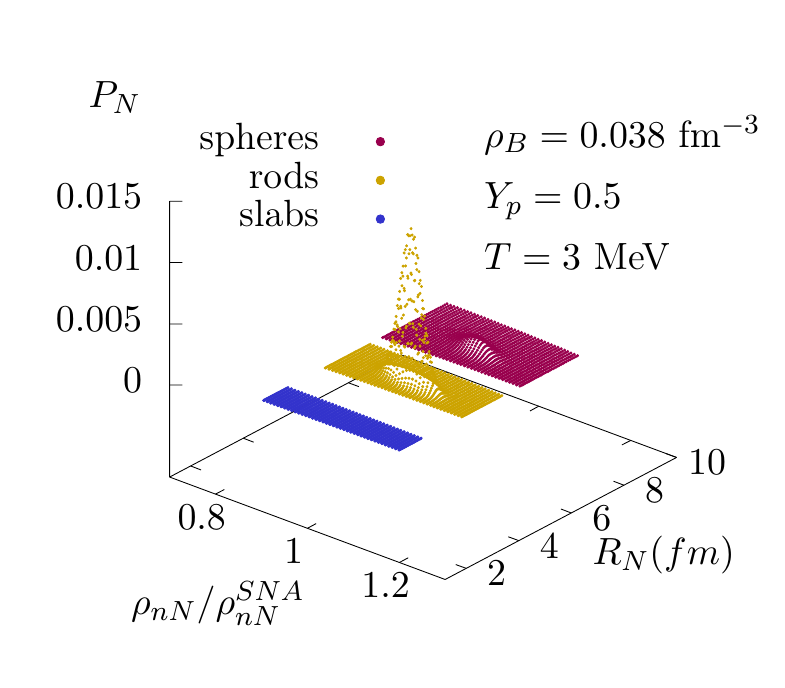}\\
  \end{tabular}
  \caption{3D probability distribution as a function of the pasta linear
    dimension and the normalised cluster density with different
    geometries obtained with the IUFSU parametrisation for different
    proton fractions. Notice the vertical scale differences.}  
  \label{fig_flu}
\end{center}
\end{figure}

The second important point refers to the fact that the CP method just
presented and also another commonly used method, the Thomas-Fermi
approximation \cite{TF}, can only provide one specific geometry for each
density, temperature and proton (or electron) fraction but it is well
known that this picture is very naive. In fact, different geometries
can coexist at thermal equilibrium
\cite{maruyama2013,Schneider2016,Fattoyev2017}. The problem with these
more sophisticated approaches is that the computational cost is
tremendous, making them inadequate to be joined to other expensive
computational methods that may be necessary to calculate neutrino
opacities and transport properties, for instance.  In a recent paper,
a  prescription with a very low computational cost was presented
\cite{flu}. In that paper, fluctuations are taken into account in a
reasonably simple way by the introduction of a rearrangement term in
the free energy density of the cluster.  A simple result can be seen
in Figure \ref{fig_flu}, where one can see that different geometries
can coexist at a certain temperature for a fixed density. If different
proton fractions are considered, the dominat geometry changes as in
the CP or TF method, but the other geometries can still be present. 
The complete formalism has been revised and extended to asymmetric
matter and can be found in \cite{flu2}.

\section{Hybrid stars} 
\label{hybrid}

So far, I have discussed the possibility that hadronic matter exists
in the core of a neutron star and that nuclear physics underlies the
models that describe it. The idea of a hybrid star, containing a 
hadronic outer core that has a different composition than the inner
core, which could be composed of deconfined quarks was first proposed
by Ivanenko and Kurdgelaidze \cite{firsthybrid} in the late 60's. In their papers, they
have even foreseen that a transition to a superconducting phase would
be possible. This idea has gained credibility lately.
A model-independent analysis based on the sound velocity in hadronic
and quark matter points to the fact that the existence of quark cores
inside massive stars should be considered the standard pattern \cite{nature_2020}.
In this case, one would be dealing with what is known as hybrid star
and, from the theoretical point of view, its description requires a
sophisticated recipe: a reliable model for the outer hadronic core
and another model for the inner quark core. The ideal picture would be
a chiral model that could describe both matters as density increases,
but those models are still rarely used
\cite{veronica1,veronica2,Lena2016, Clebson19}. 
Generally, what we find in
the literature are Walecka-type models as the ones presented in Section \ref{rmf} 
or density-dependent models, whose density dependence is introduced on
the meson-baryon couplings as in \cite{DDHD,Typel} for the hadronic
matter and the MIT bag model \cite{mit} or the Nambu-Jona-Lasinio (NJL)
model \cite{njl} for the quark matter. While the MIT bag model is very
simplistic, the NJL model is more robust and accounts for the expected chiral
symmetry but cannot satisfy the condition of absolutely stable strange
matter that will be discussed next.  The MIT bag model EOS is
simply the EOS calculated for a free Fermi gas in Section
\ref{nuclear}, where the masses are the ones of the $u,d,s$ quarks, generally
taken as $m_u=m_d=5$ MeV and $m_s$ varying from around 80 to 150 MeV
and the inclusion of a bag constant $B$ of arbitrary value, which is responsible for confining
the quarks inside a certain surface.  $B$ enters with a negative sign
in the pressure equation and consequently a positive one in the energy
density equation.  The NJL EOS is more complicated
and besides accounting for chiral symmetry breaking/restoration, also
depends on a cut-off parameter. The derivation of the EOS can be
obtained in the original papers \cite{njl}, in an excellent review
article \cite{buballa_review} or in one of the papers I have co-authored as \cite{DC2003}, for
instance and I will refrain from copying the equations here. Contrary
to the MIT bag model, the NJL model does not offer the possibility of
free parameters. All of them are adjusted to fit the pion mass, its
decay constant, the kaon mass and the quark condensates in the vacuum. 
There are different sets of parameters for describing the SU(2) (only
considers $u$ and $d$ quarks)  and the SU(3) versions of the model. 

When building the EOS to describe hybrid stars, two constructions are comonly
made: one with a mixed phase (MP) and another without it, where
the hadron and quark phases are in direct contact. In the first case, 
neutron and electron chemical potentials are continuous throughout the stellar
matter, based on the standard thermodynamical rules for phase coexistence known
as Gibbs conditions. In the second case, the electron chemical potential 
suffers a discontinuity and only the neutron chemical potential is
continuous. This condition is known as Maxwell construction. 
The differences between  stellar structures obtained with both constructions
were discussed in many papers \cite{maruyama07,vos2, paoli} and I just
reproduce the main ideas next.

In the mixed phase, constituted of hadrons and quarks, charge neutrality is not
imposed locally but only globally, meaning that quark and hadron
phases are not neutral separately. Instead, the system rearranges itself so that
$$\chi \rho_c^{QP}+(1-\chi)\rho_c^{HP}+\rho_c^l=0,$$
where $\rho_c^{iP}$ is the charge density of the phase $i=H,Q$, $\chi$ is the volume fraction occupied by the quark phase,
and $\rho_c^l$ is the electric charge density of leptons. 
The Gibbs conditions for phase coexistence impose that \cite{Glen}:
$$\mu^{HP}_n=\mu^{QP}_n,\quad \mu^{HP}_e=\mu^{QP}_e \quad {\rm and} \quad P^{HP}=P^{QP},$$
and consequently,

\begin{equation}
\langle {\cal E} \rangle = \chi {\cal E}^{QP}+(1-\chi){\cal E}^{HP}+{\cal E}^l
\end{equation}

and

\begin{equation}
\langle \rho \rangle = \chi \rho^{QP}+(1-\chi)\rho^{HP}.
\end{equation}

The Maxwell construction is much simpler than the case above and it is
only necessary to find the transition point where

$$ \mu^{HP}_n=\mu^{QP}_n \quad {\rm and} \quad P^{HP}=P^{QP},$$
and then construct the EoS. 

\begin{figure}
\begin{center}
  \begin{tabular}{cc}
    \includegraphics[width=0.8\linewidth]{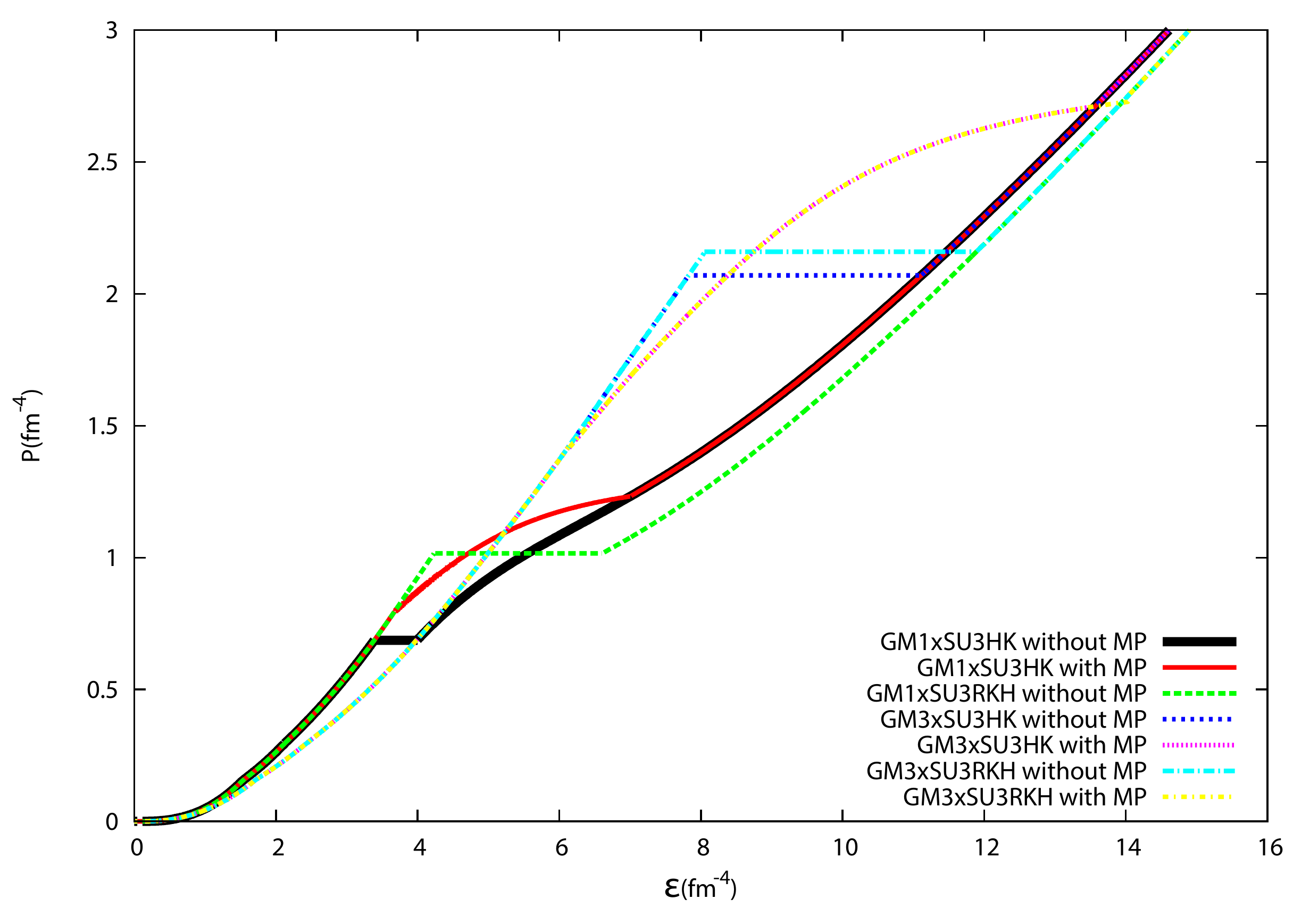} \\
    \includegraphics[width=0.8\linewidth]{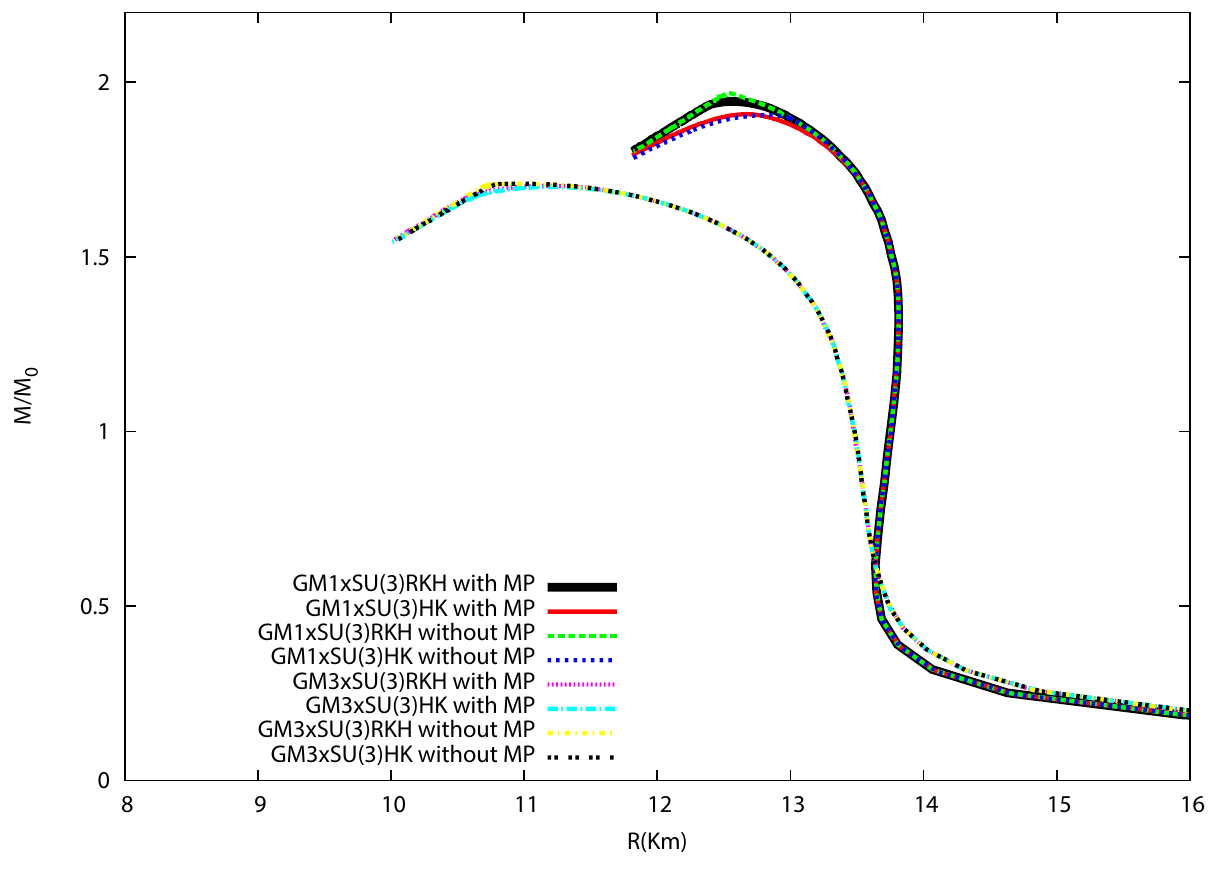}\\
  \end{tabular}
  \caption{top) EOS built with Maxwell (without MP) and Gibbs (with
    MP) constructions;  bottom) Corresponding mass-radius diagram.
    Figure based on the results presented in \cite{paoli}.}  
  \label{fig_hybrid}
\end{center}
\end{figure}

In Fig.\ref{fig_hybrid}, different EOS are built with both
constructions and the respective mass radius curves are also shown. In
all cases, the hadronic matter was described with either GM1 \cite{Glen2}
or GM3 parametrisations \cite{Glen}  and the quark phase with the two
most common parametrisations for the NJL model (HK \cite{HK} and RKH
\cite{RKH}). On the top one can see that under the Maxwell
construction, the EOS presents a step at fixed pressure and under
the Gibbs construction, the EOS is continuous. It is then easy to see
that for both constructions, the mass radius curves are indeed very
similar and yield almost indistinguishable results for gravitational masses and
radii. In these cases, the differences of the hadronic EOSs dominate over the
differences of quark EOSs. Hence, the maximum mass is mostly
determined by the hadronic part. It is also important to stress that
the quark core is not always present in the star even if it the quark
matter EOS is included in the EOS. This fact is noticed when one
compares the density where the onset of quarks takes place with the
star central density.  If the star central density is lower than the
quark onset, no quark core exists.

A more recent analysis of the dependence of the macroscopic properties of hybrid stars
on meson-hyperon coupling constants and on the vector channel added to
the NJL model can be seen in \cite{Luiz2021}.  

In 2019, the LIGO/Virgo collaboration detected yet another
gravitational wave, the GW190814
\cite{GW190814}, resulting from the merger of a 23 $M_\odot$ black
hole and another object with $2.59^{+0.08}_{-0.09} M_\odot$, which
falls in the mass-gap category, i.e., too light to be a black hole and
too massive to be a NS.  In \cite{veronica2}, a
chirally invariant model was used to describe hybrid stars with a
variety of different vector interactions and this compact object could
be explained as a massive rapidly rotating NS. 
A comprehensive discussion on ultra-heavy NS (masses larger than
  2.5 $M_\odot$) and the possibility that they are hybrid objects can
  be found in \cite{Veronica_Jacquelyn2021}.

If the reader is interested in understanding the effects of
different quark cores that also include trapped neutrinos at
fixed entropies, reference \cite{trapped} can be consulted.

\section{Quark stars} 
\label{quark}

\begin{figure}
\begin{center}
    \includegraphics[width=0.9\linewidth]{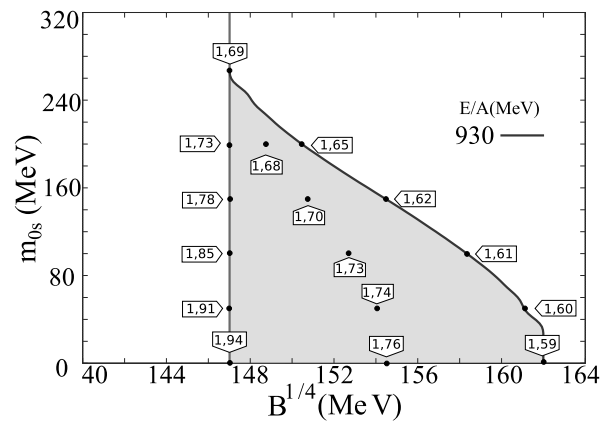}
  \caption{ Stability window shown in the shaded area. The flags
    indicate the maximum stellar masses obtained with various $B$ and
    strange quark mass values. Figure taken from \cite{James2013}.}  
  \label{window_mit}
\end{center}
\end{figure}

All experiments that can be realised in laboratories show that hadrons
are the ground state of the strong interaction. Around 50 years ago,
Itoh \cite{Itoh} and Bodmer \cite{Bodmer}, in separate studies,
proposed that under specific
circumstances, as the ones existing in the core of neutron stars,
strange quark matter (SQM) may be the real ground state.
This hypothesis, later on also investigated by Witten, became known as
the Bodmer-Witten conjecture and it is theoretically tested with the search of a
stability window, defined for different models
in such a way that a two- flavour quark matter (2QM) must be unstable
(i.e., its energy per baryon has to be larger than 930 MeV, which is
the iron binding energy) and SQM (three-flavour quark matter) must be
stable, i.e., its energy per baryon must be lower than 930 MeV 
\cite{Bodmer,Witten}. 
As shown in the previous section, although the 
Nambu-Jona-Lasinio (NJL) model \cite{njl} can be used to describe the core
of a hybrid star \cite{Clebson19,DC2003}, it cannot be used in the
description of absolutely stable SQM 
as shown in \cite{buballa_stability,NJLmag,NJLmag2,James_Veronica}. 
 The most common model, the
MIT bag model \cite{mit} satisfies the Bodmer-Witten conjecture, but
cannot explain massive stars  J0348+0432 \cite{Antoniadis},  J1614-2230
\cite{Demorest} and J0740+6620) \cite{Cromartie, NICER_news}, as can be seen in
Fig.\ref{window_mit}, from where one can observe that the maximum attained
mass is 1.94 $M_\odot$ obtained for a non-massive strange quark.

Hence, we next mention another quark
matter model that satisfies de Bodmer-Witten conjecture at the same
time that can describe massive stars and canonical stars with small
radii, the density dependent quark mass (DDQM) proposed in
\cite{Peng2000, Xia2014} and investigated in \cite{Betania21}. 
In the DDQM model, the quark masses depend on two arbitrary parameters
and are given by

\begin{equation}
    m_{i} = m_{i0} + m_I \equiv m_{i0} + \frac{D}{\rho_b^{1/3}} + C \rho_{b}^{1/3},
    \label{masses}
\end{equation}
where the index $I$ stands for the medium corrections and the baryonic
density is written in terms of the quark densities as

\begin{equation}
    \rho_{b} = \frac{1}{3} \sum_i \rho_i , \quad \rho_i  = \frac{g_i\nu_i^3}{6\pi^2},
    \label{baryon-number-density}
\end{equation}
and $\nu_i$ is the Fermi momentum of quark $i$, which reads:

\begin{equation}
\nu_i = \sqrt{\mu_i^{*2} - m_i^2} 
   \label{eq:fermimom}
\end{equation}
and $\mu_i^*$ is the $i$ quark effective chemical potential.
The energy density and pressure are respectively given by

\begin{equation}
    {\cal E} = \Omega_0 - \sum_i \mu_i^* \frac{\partial\Omega_0}{\partial\mu_i^*},
    \label{energy-density}
\end{equation}
and 

\begin{equation}
    P = -\Omega_0 + \sum_{i,j} \frac{\partial\Omega_0}{\partial m_j} n_i \frac{\partial m_j}{\partial \rho_i},
    \label{pressure_ddqm}
\end{equation}
where 
$\Omega_0$ stands for the thermodynamical potential of a free system
with particle masses $m_i$ and effective chemical potentials $\mu_i^*$
\cite{Peng2000}:

\begin{equation}
    \Omega_0 = -\sum_i \frac{g_i}{24 \pi^2} \left[ \mu_i^*\nu_i \left( \nu_i^2 - \frac{3}{2}m_i^2 \right) + \frac{3}{2} m_i^4 \ln \frac{\mu_i^* + \nu_i}{m_i} \right],
    \label{thermodynamic-potential-free-system}
\end{equation}
with $g_i$ being the degeneracy factor 6 (3 (color) x 2 (spin)) and 
the relation between the chemical potentials and their effective
counterparts is simply

\begin{equation}
    \mu_i = \mu_i^* + \frac{1}{3}\frac{\partial m_I}{\partial n_b} \frac{\partial \Omega_0}{\partial m_I} \equiv \mu_i^* - \mu_I,
    \label{real-effective-chemical-potentials}
\end{equation}

\begin{figure}
\begin{center}
  \begin{tabular}{cc}
\includegraphics[width=0.6\linewidth, angle=270]{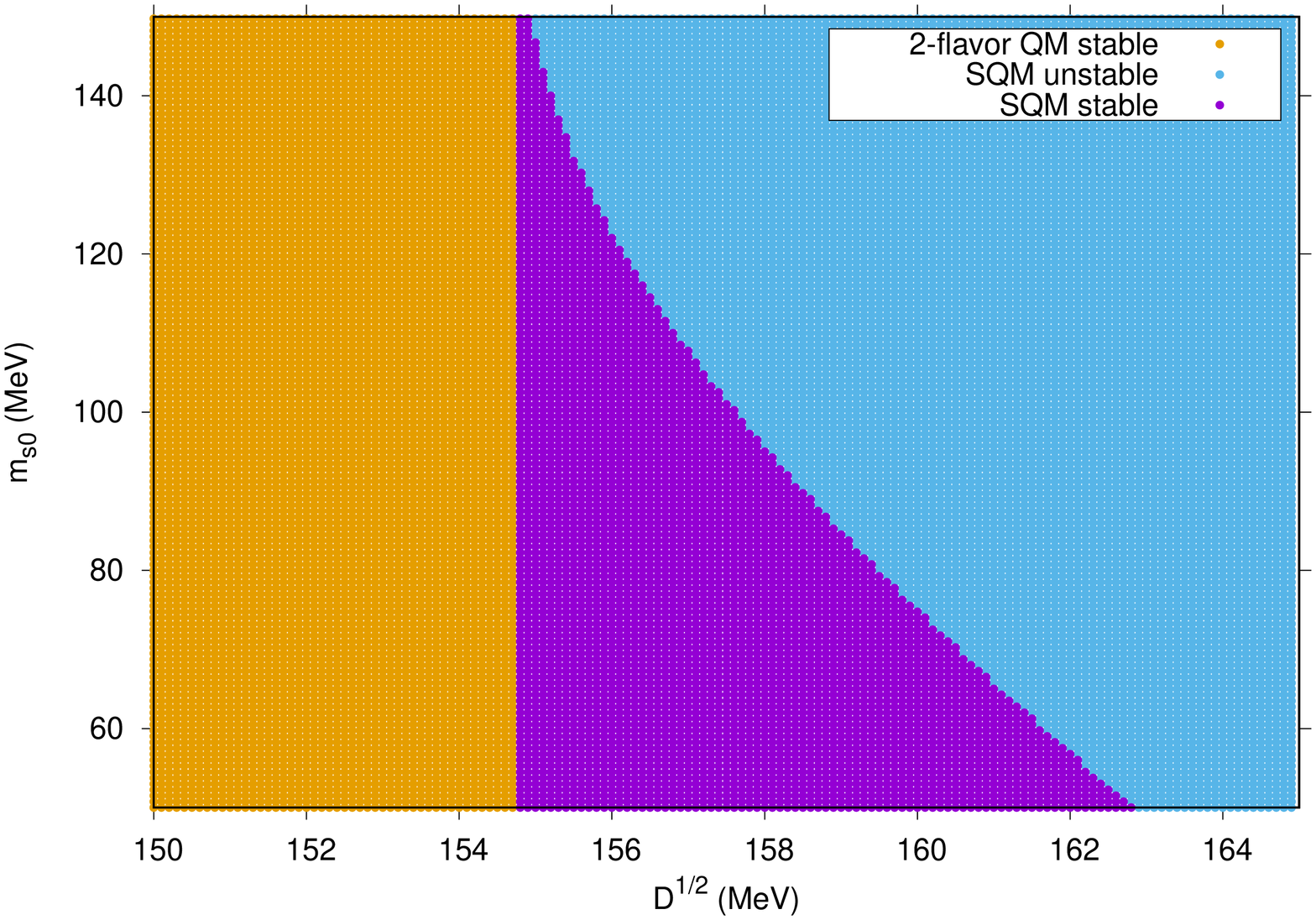}\\
\includegraphics[width=0.6\linewidth, angle=270]{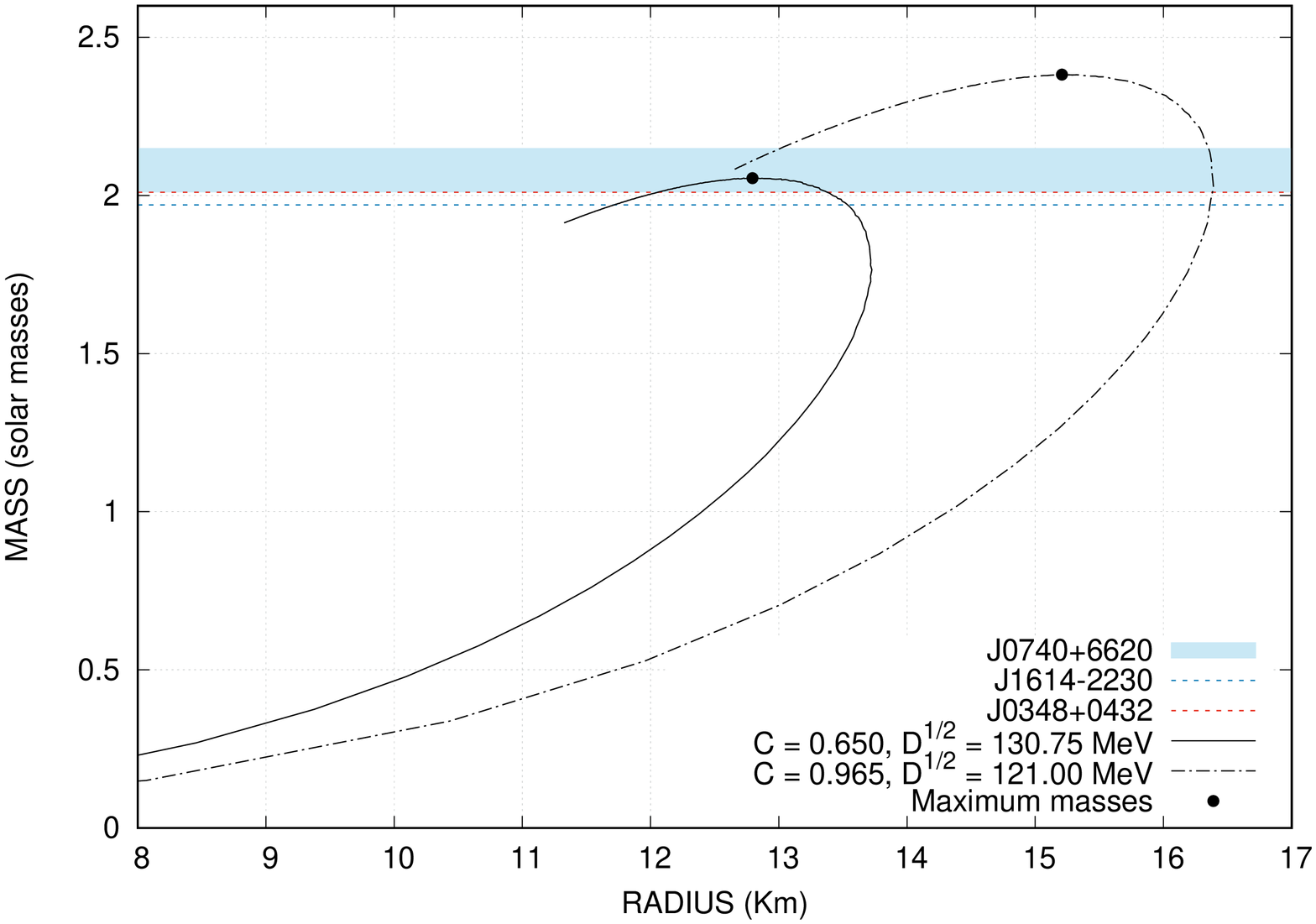}\\
  \end{tabular}
  \caption{top) Stability window for fixed $C=0$, bottom) Mass-radius
    diagram for different values of $C$ and $\sqrt D$.
    Figures based on the results shown in \cite{Betania21}.}  
  \label{ddqm}
\end{center}
\end{figure}

On the left of Fig. \ref{ddqm} the stability window is plotted for a fixed value
of $C$, so that it displays a shape that can be compared with
Fig. \ref{window_mit}. For other values of the constants, more
stability windows are shown in \cite{Betania21}.  On the right of
Fig. \ref{ddqm}, different mass-radius curves are shown and one can
see that very massive stars can indeed be obtained. At this point, it
is worth mentioning that quark stars are believed to be bare (no
crust is supported) and for this reason, the shape of the curves shown
in Fig.\ref{ddqm} is so different from the ones obtained for hadronic
stars and shown in Figs. \ref{tov_IUFSU}, Fig.\ref{figtov} and for
hybrid stars, as seen in  Fig.\ref{fig_hybrid}.

There is still another very promising model, an extension of the MIT
bag model based on the ideas of the QHD model. In this extended
version, the Lagrangian density accounts for the free Fermi gas part
plus a vector interaction and a self-interaction mesonic field and
reads \cite{Carline1}:

\begin{widetext}
\begin{equation}
\mathcal{L} = \sum_{u,d,s}\{ \bar{\psi}_q  [ \gamma^{\mu}
(i\partial_\mu - g_{qqV}V_\mu) - m_q ]\psi_q - B
\}\Theta(\bar{\psi_q}\psi_q)
+\frac{1}{2}m_\omega^2V_\mu V^\mu
+  b_4 \frac{(g_{uuV} ^2V_\mu V^\mu)^2}{4} 
, \label{v1}
\end{equation}
\end{widetext}
where the quark interaction is mediated by the vector channel $V_\mu$
representing the $\omega$ meson, in the same way as in QHD models
~\cite{walecka}. The relative quark-vector field interaction is fixed
by symmetry group and results in

\begin{equation}
  g_{ssV} = \frac{2}{5}g_{uuV} = \frac{2}{5}g_{ddV} ,
\end{equation}
with adequate redefinitions given by

\begin{equation}
(g_{uuV}/m_V)^2 = G_V, \quad
  X_V =  \frac{g_{ssV}}{g_{uuV}}  
\end{equation}
and $b_4$ taken as a free parameter. Using a mean field approximation and
solving the Euler-Lagrange equations of motion, the following
eigenvalues for the quarks and $V_0$ field can be obtained:  

\begin{equation}
  E_q =  \mu = \sqrt{m_q^2 + k^2} + g_{qqV}V_0,
\end{equation}
\begin{equation}
g_{uuV} V_0  + \bigg ( \frac{g_{uuV}}{m_\omega} \bigg)^2 \bigg (  b_4
  (g_{uuV}V_0)^3 \bigg )=
  \nonumber 
 \bigg (\frac{g_{uuV}}{m_\omega} \bigg ) \sum_{u,d,s} \bigg
  (\frac{g_{qqV}}{m_\omega} \bigg )n_q .
  \label{MITv}
\end{equation}

\begin{figure}
  \begin{center}
\begin{tabular}{cc}
\includegraphics[width=5.cm,angle=270]{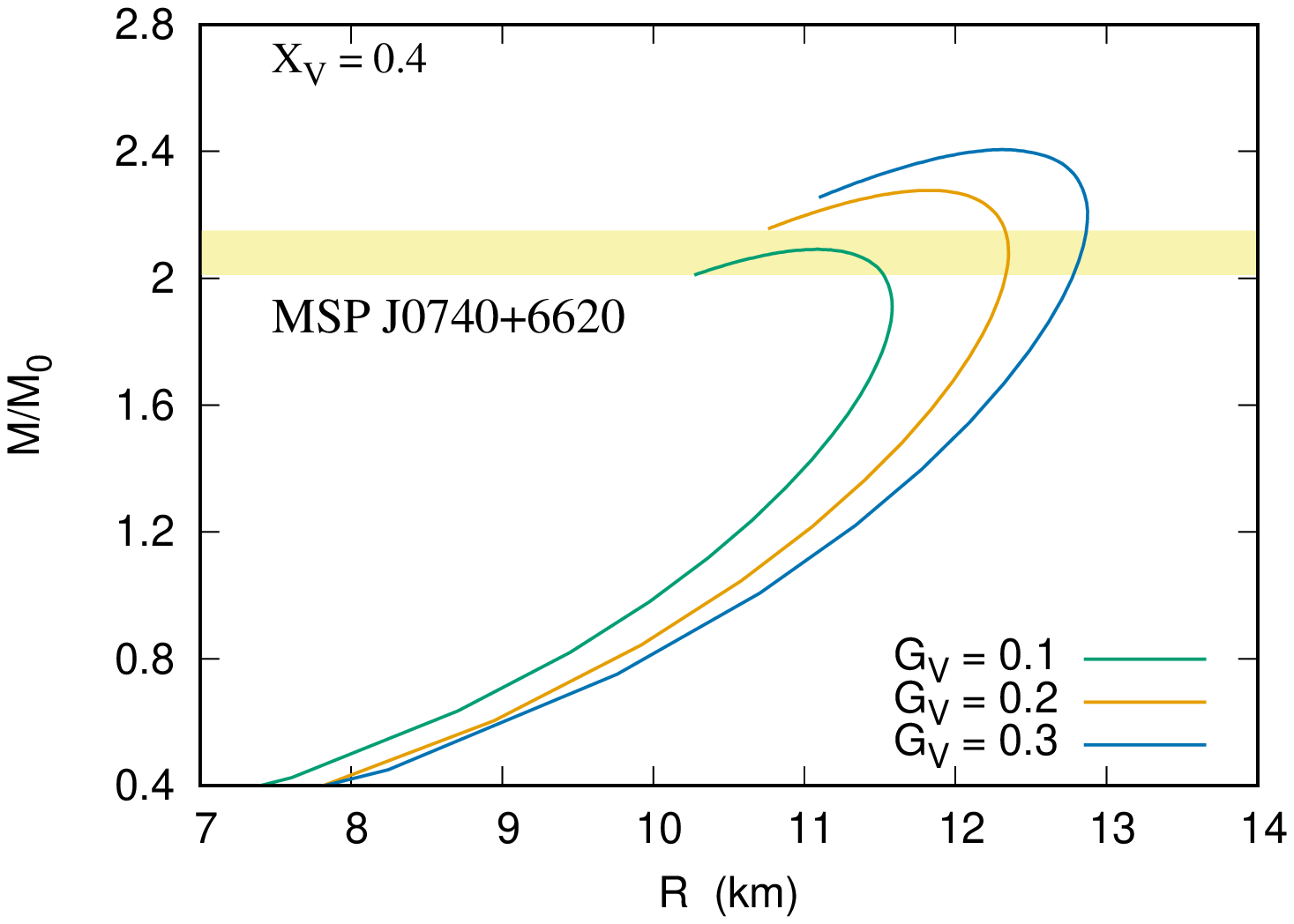} \\
\includegraphics[width=5.cm,angle=270]{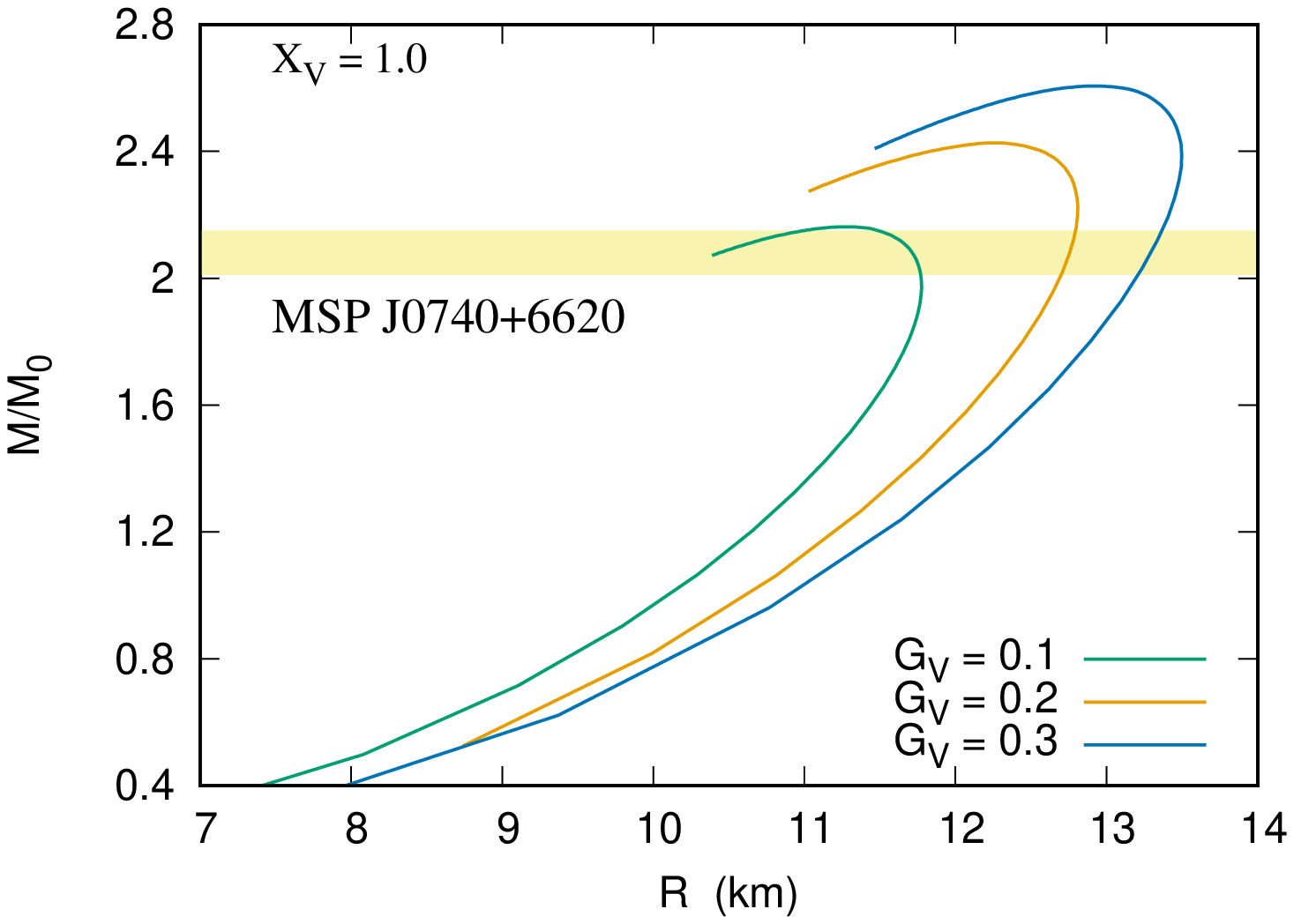} \\
\end{tabular}
\caption{Mass-radius diagram obtained with the minimum value of the
bag pressure that produces stable quark stars with different values of
$G_v$ and top) $X_v=0.4$, bottom) $X_v=1.0$. Figure based on the
results presented in \cite{Carline1}.}
\label{mitvec1}
\end{center}
\end{figure}

With this new approach, when the self-interaction vector channel is
turned off,  the stability window increases and a
2.41$M_\odot$ quark star that satisfies all astrophysical constraints is
obtained. The self-interaction vector channel does not change the
stability window, but allow even more flexibility in the calculation
of the tidal polarisability and the canonical star radius due to the
inclusion of the free parameter $b_4$. In this case, a 2.65 $M_\odot$
quark star corresponding to a 12.13 km canonical star radius and a
tidal polarisability within the expected observed range is
obtained along many other results which satisfy many of the presently 
known astrophysical
constraints. Some of the results are displayed in figure
\ref{mitvec1}.
After all the discussion on the radii of NS constrained with the help of
  gravitational wave observation and neutron skin thickness
  experimental results presented in section \ref{constrain} and on the
  uncertainty of these values, I just would like to add one comment: 
  contrary to what is obtained for a family of hadronic stars (maximum mass stars
  are generally associated with a smaller radii than their canonical
  star counterparts), a family of quark stars may produce canonical stars
  with radii that can be approximately the same as the maximum mass
  star radii, depending on the model used \cite{Carline1}
and this feature could accomodate the recent NICER detections 
for J0030+0451 and J0740+6620 .

\begin{figure}
 \begin{center} 
\includegraphics[width=7.cm]{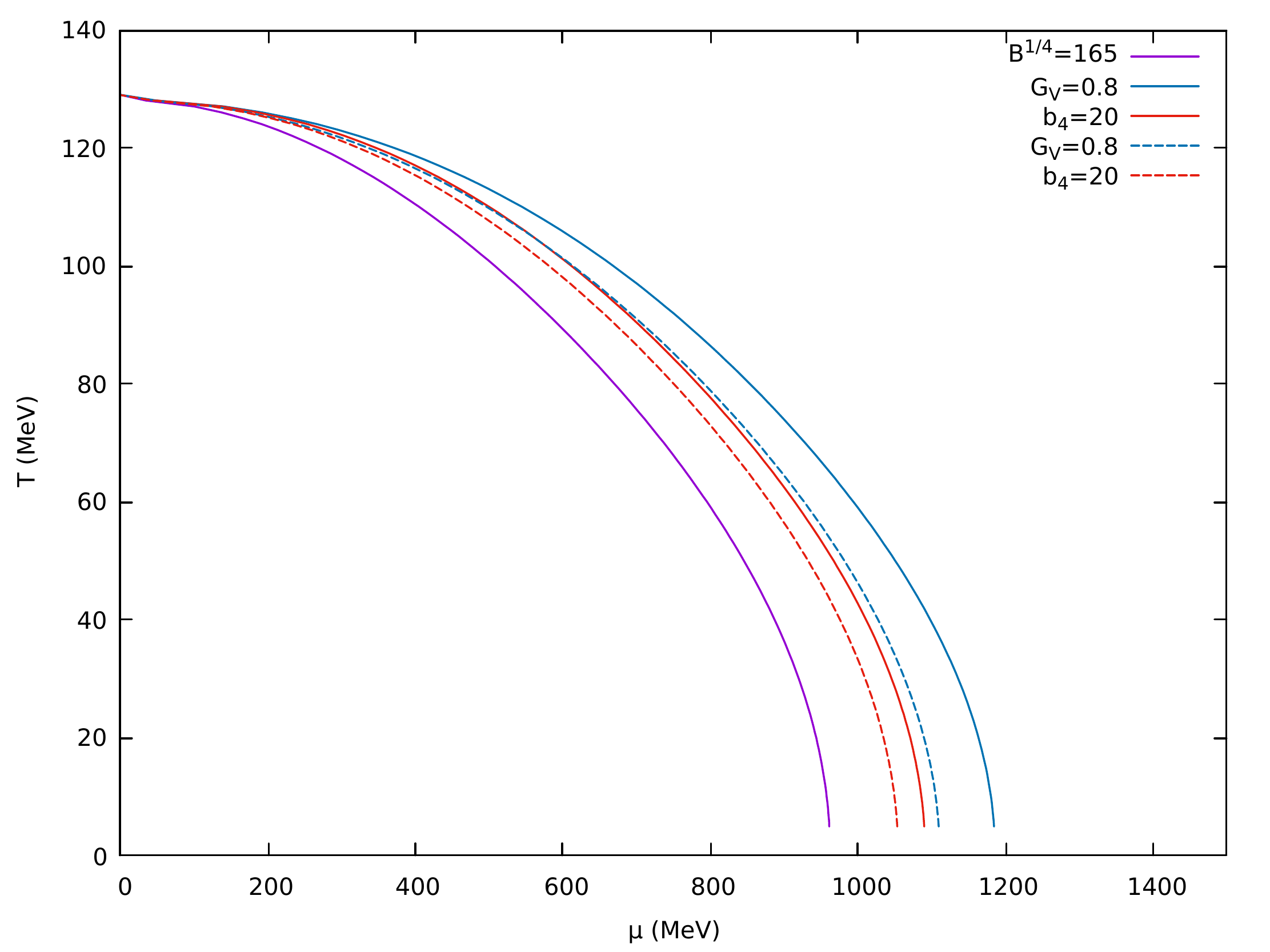}
\caption{Phase diagram for neutral matter in $\beta$-equilibrium with
a temperature dependent bag pressure $B(T)$. Dashed lines stand for
$X_v=0.4$ and solid lines for $X_v=1.0$. Figure based on the results
presented in \cite{Carline2}. }
\label{mitvec2}
\end{center}
\end{figure}

This modified MIT bag model has also been used to
investigate the finite temperature systems and to obtain the QCD phase
diagram in \cite{Carline2} with the help of a temperature dependent
bag $B(T)$, as discussed in the Introduction of the
paper. Some of the possible phase diagrams are shown in figure \ref{mitvec2}.

I have outlined the main aspects concerning the internal
  structure of quark stars, but the discussion about their bare surface
  \cite{Melrose, quarkstars} is not completely settled
  \cite{Yakovlev} and important problems as its high plasma frequency and
neutrinosphere are out of the scope of the present work but should not
be disregarded.

\section{Magnetars: crust-core transitions and oscillations}
\label{magnetars}

I could not end this review without mentioning magnetars \cite{Duncan,
  Pons}, a special class of neutron stars with surface magnetic fields 
three orders of magnitude (reaching up to $10^{15}$ G at the
surface) stronger than the ones present in standard neutron stars ($10^{12}$ G at the
surface). Most of the known magnetars detected so far are isolated
objects, i.e., they are not part of a binary system and manifest
themselves as either transient X-ray sources, known as
soft-$\gamma$ repeaters or persistent anomalous x-ray pulsars.
They are also promising candidates for the recent discovery of
  fast radio bursts \cite{bochenek}.
So far, only about 30 of them have been clearly identified \cite{catalog} but more
 information is expected from NICER \cite{NICER} 
 and ATHENA \cite{ATHENA}, launching foreseen to take place in 2030.
So far, NICER has already pointed to the fact that the beams of radiation
emitted by rapidly rotating pulsars may not be as simple as often
supposed: the detection of two hot spots in the same hemisphere
suggests a magnetic field
configuration more complex than perfectly symmetric dipoles \cite{NICER_news}.

From the theoretical point of view, there is no reason to believe that
the structure of the magnetars differs from the ones I have mentioned
in this article. Thus, they can also be described as hadronic objets
\cite{Lopes2012,Rudinei2014, Lopes2015, Debi2019,Lopes2020,cesar2020},
as quark stars \cite{NJLmag, Veronica2014, Lopes2015, Lopes2016, Lopes2020}
or as hybrid stars \cite{Rudinei2014,Lopes2020}.

At this point, it is fair to claim that the best approach to calculate
macroscopic properties of magnetars is the use of the LORENE code
\cite{LORENE}, which takes into account Einstein-Maxwell equations and equilibrium
solutions self consistently with a density dependent magnetic
field. LORENE avoids discussions on anisotropic effects and violation
of Maxwell equations as pointed out in \cite{Lopes2016}, for
instance. However, at least two important points involving matter subject to
strong magnetic fields can be dealt with even without the LORENE
code. The first one is the crust core transition density discussed in
\cite{Debi2019} and \cite{Constanca2017}. Although the magnetic fields
at the surface of  magnetars are not stronger than $10^{15}$ G, if the
crust is as large as expected (about 10\% of the size of the star), at
the transition region the magnetic field can  reach $10^{17}$ G.
The transition density can then be estimated by computing the
spinodal sections, both dynamically and
thermodynamically. The point where the EOS crosses the spinodal 
defines the transition density \cite{older}. An interesting aspect is that the
spinodals of magnetised matter are no longer smooth curves. Due to
the filling of the Landau levels, more than one crossing point is
possible \cite{Debi2019,Constanca2017}, what introduces an extra
uncertainty on the calculation.

The second aspect refers to possible oscillations in magnetars caused
by the violent dynamics of a merging binary system. One
has to bear in mind that so far, all observed magnetars are isolated
compact objects, but there is no reason to believe that binary systems
do not exist. In this case, the perturbations on the metric can couple
to the fluid through the field equations \cite{Thorne, Hinderer}. For
a comprehensive discussion on the equations involved, please refer to
\cite{cesar2020}. The gravitational wave frequency of the
fundamental mode is expected to be detected in a near future by detectors
like  the  Einstein Telescope. In \cite{cesar2020}, the effect of
strong magnetic fields on the fundamental mode was
investigated. From the results presented in that paper, one can
  clearly see that  magnetars bearing masses below 1.8 M$_\odot$ present
practically the same frequencies. Nevertheless, more massive stars
present different frequencies depending on their constitution: 
nucleonic stars present frequencies lower than their hyperonic
counterparts, a feature which may define the internal constitution of magnetars.

The DDQM described in Section \ref{quark} was also investigated under
the effects of strong magnetic fields and the main expressions can be
found in \cite{Betania_Kauan}. This may be an interesting model for future
calculations of the fundamental models.

\section{Final Remarks}

From the existence of a massive ordinary star that lives due to nuclear
fusion, to its explosive ending and its aftermath, I have tried to
tell the neutron star history. All these stages can be explained
thanks to nuclear physics and I have revisited the main aspects and
models underlying each one. 

I have also tried to emphasise that nuclear models are generally
parameter dependent and a plethora of models have been proposed in the
last decades, but it is unlikely that the very same models can be
used to describe different aspects of nuclear matter and, at the
same time, all macroscopic properties of neutron stars. I  do not
advocate that the models I have chosen to use are the best ones, but
the main idea is to show that different models should be used at
the discretion of the people who employ them. I have not used density
dependent hadronic models as the ones proposed, for instance, in \cite{Typel, ddme2}
to avoid extra theoretical complications, but they are indeed very
good options, since they can describe well nuclear matter, finite
nuclei and NS properties, as seen in \cite{relat,NS_RMF,GW_RMF}.

As far as detections of gravitational waves are concerned, a
  window was opened in 2015 and many observations will certainly be
  disclosed even before I finish writing this paper. Besides the ones
  already mentioned, I would like to comment on the GW190425
  \cite{GW190425}, GW200105 and GW200115 \cite{GW200115}. The first
  one was used in conjunction with a chiral effective field theory to
  constrain the NS equation of state \cite{Dietrich2020}. The authors
  obtained a radius equal to $11.75^{+0.86}_{0.81}$ for a canonical
  star, also quite small as compared with the ones obtained from the PREX
 experiment. The other two probably refer to neutron star - black hole
 mergers, systems that have been conjectured for a long time and will
 probably contribute to the understanding of NS EOS. 

 Before concluding, I would like to mention that many aspects
  regarding either isolated NS or binary systems have not been tackled
  in this manuscript and, in my opinion, rotation is the most important one.
 A better understanding of these compact objects depends
  on many rich features, including thermal and magnetic
  evolution. Different observation manifestations as pulsars,
  accreting X-ray binaries, soft $\gamma$-repeaters and anomalous
  X-ray pulsars, also deserve an attentive investigation. Hence, this
  review is just one step towards the incredible exotic world of neutron
  stars.

 As far as the QCD phase diagram is concerned, many aspects have 
been extensively studied and are
well understood: matter at zero temperature, symmetric nuclear and pure
neutron matter, low density matter, including clusterisation and the
pasta phase, high density matter and matter in $\beta$-equilibrium.
Nevertheless, an EOS that covers the complete QCD phase diagram parameter
space in $(T, \mu_B)$ in a single model is not available yet.
Some of the EOS can be found on the  CompOSE (CompStar
Online Supernovae Equations of State) website \cite{compose}.

\vspace{6pt}

\acknowledgments{This  work is a part of the project INCT-FNA Proc. No. 464898/2014-5. D.P.M. is partially supported by Conselho Nacional de Desenvolvimento Cient\'ifico e  Tecnol\'ogico (CNPq/Brazil) under grant 301155.2017-8.
  Figures \ref{fig0}, \ref{fig3} and \ref{fig4} were drawn by
   Kevin Schroeder, the data used to draw figure \ref{outer_test} with
   the reliable crust were obtained by Thomas Carreau and Francesca
   Gulminelli. Figures \ref{tidal} were drawn by Odilon Louren\c co,
   figures \ref{fig_flu} by Mateus Renke Pelicer, figures \ref{ddqm}
   by Bet\^ania C.T. Backes, figure \ref{mitvec1} by Luiz L. Lopes and
 figure  \ref{mitvec2} by Carline Biesdorf. I thank them all and also
 Prof. Constan\c ca Provid\^encia for useful comments and suggestions.}


\end{document}